\documentclass{jfm}
\usepackage{graphicx}
\usepackage{newtxtext}
\usepackage{natbib}
\usepackage[hidelinks]{hyperref}
\usepackage{placeins}
\usepackage{natbib}
\usepackage{amsmath,mathtools}
\usepackage{color} 
\usepackage{amssymb}
\usepackage{tabularx}
\usepackage{longtable}
\usepackage{listings}
\usepackage{booktabs}
\usepackage{float}
\usepackage{makecell}
\usepackage{epstopdf}
\usepackage{pifont} 
\usepackage{wasysym}
\usepackage{tikz}
\usepackage{upgreek}
\usepackage[justification=justified,singlelinecheck=false]{caption}

\usepackage[normalem]{ulem}

\usepackage{flafter}  
\usepackage{ulem}
\usepackage{slashbox}
\usepackage{multirow}

\definecolor{darkorange}{rgb}{0.8500 0.3250 0.0980}

\definecolor{darkgreen}{rgb}{0.4660 0.6740 0.1880}

\def\mline{\vrule width4pt height2.5pt depth -2pt}
\def\dashed{\mline\hskip3.5pt\mline\thinspace}

\newcommand{\RomanNumeralCaps}[1]
\linenumbers

\usepackage{lmodern,pdftexcmds,amsmath}

\DeclareMathAlphabet{\mathsfbi}{OT1}{\sfdefault}{bx}{sl}
\DeclareMathVersion{sfletters}
\SetSymbolFont{letters}{sfletters}{OML}{ntxsfmi}{b}{it}

\makeatletter
\newcommand{\mathbfsbilow}[1]{%
  \text{\mathversion{sfletters}$\m@th#1$}%
}
\makeatother

\usepackage{etoolbox}

\title{Fixed-flux Rayleigh-B\'enard convection in doubly periodic domains: generation of large-scale shear}

\author{Chang Liu\aff{1,3}
  \corresp{\email{chang\_liu@uconn.edu}}, Manjul Sharma\aff{2}, Keith Julien\aff{2} \and 
  Edgar Knobloch\aff{1}
 }
\affiliation{
\aff{1}Department of Physics, University of California, Berkeley, CA 94720, USA
\aff{2}Department of Applied Mathematics, University of Colorado, Boulder, CO 80309, USA
\aff{3}School of Mechanical, Aerospace, and Manufacturing Engineering, University of Connecticut, CT 06269, USA
}

\begin{document}
\maketitle

\begin{abstract}
  This work studies two-dimensional fixed-flux Rayleigh-B\'enard convection with periodic boundary conditions in both horizontal and vertical directions  and analyzes its dynamics using numerical continuation, secondary instability analysis and direct numerical simulation. The fixed-flux constraint leads to time-independent elevator modes with a well-defined amplitude. Secondary instability of these modes leads to tilted elevator modes accompanied by horizontal shear flow. For $Pr=1$, where $Pr$ is the Prandtl number, a subsequent subcritical Hopf bifurcation leads to hysteresis behavior between this state and a time-dependent direction-reversing state, followed by a global bifurcation leading to modulated traveling waves without flow reversal. Single-mode equations reproduce this moderate Rayleigh number behavior well. At high Rayleigh numbers, chaotic behavior dominated by modulated traveling waves appears. These transitions are characteristic of high wavenumber elevator modes since the vertical wavenumber of the secondary instability is linearly proportional to the horizontal wavenumber of the elevator mode. At a low $Pr$, relaxation oscillations between the conduction state and the elevator mode appear, followed by quasiperiodic and chaotic behavior as the Rayleigh number increases. In the high $Pr$ regime, the large-scale shear weakens, and the flow shows bursting behavior that can lead to significantly increased heat transport or even intermittent stable stratification. 
\end{abstract}


\begingroup
\allowdisplaybreaks

\section{Introduction}
\label{sec:introduction}

Fixed-flux temperature boundary conditions describing adiabatic boundaries appear in a wide range of geophysical and astrophysical applications including convection in the Earth's mantle \citep{mckenzie1974convection,hewitt1980large,chapman1980long,sakuraba2009generation,long2020scaling} and models of solar supergranulation \citep{vieweg2021supergranule,vieweg2022inverse,kaufer2023thermal}. Such boundary conditions also model Rayleigh-B\'enard convection between poorly conducting horizontal plates, suggesting an explanation for discrepancies between experimentally measured heat transport and numerical simulations \citep{verzicco2004effects,verzicco2008comparison}. Fixed-flux temperature or mass boundary conditions are also used to model the free surface in ocean circulation models \citep{huck1999interdecadal,abernathey2011dependence} and for understanding nutrient productivity in the oceans \citep{pasquero2005impact}. A low enough constant injection rate of CO$_2$ concentration in saline aquifers can also be modeled as a fixed-flux problem \citep{amooie2018solutal}. Moreover, fixed-flux boundary conditions are also relevant for concentration transport within an enclosure with impermeable boundaries \citep{mamou1998double,mamou1999thermosolutal}.  

Rayleigh-B\'enard convection (RBC) provides a canonical set-up for understanding the effect of fixed-flux conditions, see e.g., the monograph by \citet{goluskin2016internally}. With fixed-flux at both the top and bottom boundaries, the critical horizontal wavenumber at the onset of convective instability vanishes \citep{sparrow1964thermal,hurle1967solution,chapman1980nonlinear}, and the resulting near-onset evolution is described by the Cahn-Hilliard equation \citep{novick2008cahn,miranville2019cahn}. In the weakly supercritical Rayleigh number regime, each convection cell is thus unstable to perturbations with ever longer wavelength in a process referred to as coarsening \citep{chapman1980long,chapman1980nonlinear}. This large scale manifests itself in the turbulent regime of fixed-flux RBC in three dimensions (3D) and organizes the resulting flow. For example, recent 3D DNS with aspect ratio $\Gamma= 60$ show that convection cells gradually aggregate into a single large cell that eventually fills the whole domain, thereby providing insight into the aggregation of granules into a supergranule in the solar convection zone \citep{vieweg2021supergranule,vieweg2022inverse}, a process also confirmed in experiments \citep{kaufer2023thermal}. There is also evidence that fixed-flux boundary conditions influence the generation and reversals of large-scale shear. For example, experimental studies found that a configuration with constant flux at the bottom and constant temperature at the top exhibits less frequent reversals of the large-scale circulation than in a configuration with constant temperature on both surfaces \citep{huang2015comparative}. 

Heat transport in fixed-flux RBC has also been widely analyzed. In 2D domains of modest aspect ratio, fixed flux and fixed temperature RBC display essentially identical heat transport, overall flow dynamics and mean temperature profiles at Rayleigh number $Ra_T=10^{10}$ based on temperature difference \citep{johnston2009comparison} despite possible differences in the dominant scale. In 3D cylindrical geometry \citet{stevens2011prandtl} investigated the difference in heat transport introduced by replacing the bottom plate by a fixed-flux condition and showed that this difference decreases with increasing Rayleigh number. Different heat transport scaling predictions may be realized depending on the details of the thermal forcing \citep{lepot2018radiative,bouillaut2019transition} as shown in experiments using radiative heating in a thermally insulating container to control the heat flux. However, the current upper bound on the Nusselt number with fixed-flux boundary conditions displays the same scaling law with the Rayleigh number based on temperature difference as the fixed temperature configuration with either no-slip \citep{otero2002bounds,wittenberg2010bounds} or stress-free velocity boundary conditions \citep{fantuzzi2018bounds}. 

Fixed-flux temperature boundary conditions are also studied in related set-ups. For example, the difference between Neumann and Dirichlet boundary conditions on the buoyancy field in moist convection decreases as the Rayleigh number increases \citep{weidauer2012moist} while fixed heat flux and fixed temperature boundary conditions are shown to be asymptotically equivalent in rapidly rotating convection \citep{calkins2015asymptotic}. For rotating convection with no-slip boundaries, the Nusselt number increases significantly when a fixed heat flux is imposed instead of a fixed temperature difference \citep{kolhey2022influence}.

The top and bottom boundaries are often absent in geophysical applications, suggesting that periodic boundary conditions in the vertical are more appropriate. The resulting {\it homogeneous} RBC problem driven by a constant temperature gradient has been employed to understand bulk RBC \citep{borue1997turbulent,lohse2003ultimate,calzavarini2005rayleigh,calzavarini2006exponentially,ng2018bulk,pratt2020lagrangian,barral2023asymptotic}. Similar homogeneous configurations are also commonly employed to analyze double-diffusive convection \citep{radko2013double,stellmach2011dynamics,garaud2018double} and different shear flows \citep{rogers1987structure,sekimoto2016direct}. Periodic boundary conditions in the vertical within these homogeneous set-ups have the benefit of eliminating inessential but computationally expensive thermal or viscous boundary layers. Within homogeneous RBC, the Nusselt number $Nu$ scales with the Rayleigh number $Ra$ according to the ultimate regime prediction \citep{lohse2003ultimate,calzavarini2005rayleigh}, a prediction supported by experimental evidence from a cylindrical cell \citep{schmidt2012axially}. Moreover, \citet{calzavarini2005rayleigh} showed that $Nu$ scales with the Prandtl number $Pr$ according to mixing length theory \citep{spiegel1963generalization} and attributed this fact to more frequent appearances of elevator modes at high $Pr$. An exponentially growing elevator mode is an exact nonlinear solution of the homogeneous RBC problem, whose growth in DNS is only arrested by secondary numerical noise with a resolution-dependent instability ultimately leading to statistically steady transport \citep{calzavarini2006exponentially}. The appearance of elevator modes also leads to high intermittency in the heat transport \citep{borue1997turbulent,calzavarini2005rayleigh,calzavarini2006exponentially,barral2023asymptotic}, thereby adversely affecting the flow statistics, by increasing the sensitivity to round-off noise and discretization error \citep{calzavarini2006exponentially}. In DNS these polluting elevator modes can be suppressed by explicitly removing the mean flow parallel to gravity at each time step \citep{pratt2020lagrangian}, or by introducing an artificial horizontal buoyancy field \citep{xie2022bolgiano} or large-scale friction \citep{barral2023asymptotic}, but the modes remain a major source of contention.

This work formulates a fixed-flux homogeneous RBC problem that is not only relevant to a wide range of geophysical and astrophysical applications but also avoids the exponentially growing elevator modes that plague homogeneous RBC driven by a constant temperature gradient. Our study is motivated by a recent formulation imposing fixed-flux salinity conditions on homogeneous inertia-free salt-finger convection (IFSC) \citep{xie2020fixed}. This fixed-flux constraint saturates the elevator mode in IFSC and it does so in the present case as well. In both cases the resulting formulation leads to differential-integral equations with time-varying mean salinity  or temperature gradients that adjust the system response. Moreover, fixed-flux conditions result in a more potent restoring mechanism towards the statistically steady state that can be used as a diagnostic to determine whether in situ salt-finger convection is flux-driven or gradient-driven \citep{xie2020fixed}.

This work thus focuses on the underlying dynamics of fixed-flux homogeneous RBC using numerical continuation, secondary instability analysis, and DNS. Secondary instabilities of the elevator mode lead to tilted elevator modes accompanied by horizontal jet formation. At $Pr=1$ this state is in turn unstable to a subcritical Hopf bifurcation displaying hysteresis behavior between this state and the resulting direction-reversing state. A subsequent global bifurcation of Shil'nikov type \citep{shilnikov2007shilnikov} leads to modulated traveling waves without flow reversal. Single-mode equations that severely truncate the horizontal flow structure reproduce this moderate Rayleigh number behavior rather well. 

At high Rayleigh numbers, chaotic flow dominated by modulated traveling waves appears, and resembles no-slip instead of stress-free boundary conditions in RBC with fixed temperature. The vertical wavenumber of the secondary instability of steady elevator modes leading to these transitions is linearly proportional to the horizontal wavenumber of the elevator mode, leading to its suppression when the vertical extent of the domain precludes its presence. Thus the domain aspect ratio requires adjustment as the parameters are varied.

At low Prandtl numbers, relaxation oscillations between the conduction state and the elevator mode appear, followed by quasiperiodic and chaotic behavior as the Rayleigh number increases. Since the secondary and Hopf bifurcation points shift closer to the primary instability as $Pr$ decreases, the single-mode description works well in this regime. In contrast, at high $Pr$ the large-scale shear weakens, and the flow exhibits bursting behavior resembling that in homogeneous RBC driven by a constant temperature gradient \citep{borue1997turbulent,calzavarini2005rayleigh,calzavarini2006exponentially} and resulting in significantly increased heat transport or even intermittent stable stratification.

The remainder of this paper is organized as follows. Section \ref{sec:fixed_flux_setup} formulates the fixed-flux homogeneous RBC problem. Bifurcation analysis at moderate Rayleigh numbers performed via numerical continuation is described in \S \ref{sec:moderate_Ra} and confirmed by DNS. Section \ref{sec:high_Ra} analyzes the high Rayleigh number dynamics arising from the secondary instability of the elevator modes. The effects of changing the Prandtl number are discussed in \S \ref{sec:Pr}. The paper concludes with a summary and suggestions for future work in \S \ref{sec:conclusion}.

\section{Fixed-flux homogeneous Rayleigh-B\'enard convection}
\label{sec:fixed_flux_setup}

We consider a layer of fluid of depth $h$ with a constant upward heat flux $-k q$ through it, where $k$ is thermal conductivity and $q<0$ is the associated vertical temperature gradient. The equation of state, $(\rho_*-\rho_{r*})/\rho_{r*}=-\alpha (T_*-T_{r*})$, is linear in the Boussinesq approximation, with constant expansion coefficient $\alpha$, constant reference density $\rho_{r*}$ and constant reference temperature $T_{r*}$. The subscript $*$ denotes a dimensional variable. In the following, we nondimensionalize the temperature $T_*$ by the temperature gradient $|q|$ associated with imposed constant heat flux, $\underline{T}=T_*/|hq|$. Spatial variables are normalized by the depth $h$ of the layer while time and velocity are normalized using the thermal diffusion time $h^2/\kappa_T$ and the corresponding speed $\kappa_T/h$, respectively. Here, $\kappa_T=k/(\rho_{r*} c_p)$ is the thermal diffusivity with $c_p$ denoting specific heat capacity. In homogeneous double-diffusive convection, lengths are usually normalized by the expected finger width \citep{radko2013double,stellmach2011dynamics}, while here we normalize lengths by the layer depth $h$ for consistency with the usual procedure for Rayleigh-B\'enard convection.

We introduce the velocity field $\boldsymbol{u}:=(u,v,w)$ in Cartesian coordinates $(x,y,z)$ with $z$ in the upward vertical direction. Under the Boussinesq approximation, the system is governed by
\begin{subequations}
\label{eq:NS_tot}
\begin{align}
    \frac{\partial \boldsymbol{u}}{\partial t}+\boldsymbol{u}{\cdot} \boldsymbol{\nabla}\boldsymbol{u}=&-\boldsymbol{\nabla} p+Pr Ra_{T,q}\underline{T}\boldsymbol{e}_z+Pr\nabla^2 \boldsymbol{u}\label{eq:NS_tot_mom},\\
    \boldsymbol{\nabla}{\cdot} \boldsymbol{u}=&0\label{eq:NS_tot_mass},\\
    \frac{\partial \underline{T}}{\partial t}+\boldsymbol{u }{\cdot} \boldsymbol{\nabla}\underline{T}=&\nabla^2 \underline{T}\label{eq:NS_tot_T},
\end{align}
\end{subequations}
where $\boldsymbol{e}_z$ is the unit vector in the vertical. The governing parameters are the flux Rayleigh number $Ra_{T,q}$ and the Prandtl number $Pr$:
\begin{align}
    Ra_{T,q}:=\frac{\alpha g |q| h^4}{\nu \kappa_T},\;\;\;
    Pr:=\frac{\nu}{\kappa_T}.
\end{align}
A similar flux Rayleigh number is also employed by \citet{otero2002bounds}, \citet{johnston2007rayleigh,johnston2009comparison}, \citet{verzicco2008comparison} and \citet{goluskin2016internally}. 

We decompose the total temperature  $\underline{T}(x,y,z,t)$ as 
\begin{align}
    \underline{T}(x,y,z,t)=1+\bar{\mathcal{T}}_{z,q}z+T(x,y,z,t),
    \label{eq:T_decompose}
\end{align}
where $\bar{\mathcal{T}}_{z,q}$ is a spatially and temporally averaged temperature gradient. This decomposition allows us to impose vertically periodic boundary conditions on $T$. The velocity is taken as periodic in the vertical, and periodic conditions in the horizontal are imposed on all variables. We then define volume averaged heat flux
\begin{subequations}
\label{eq:Q_define}
\begin{align}
    Q(t):=&\langle (\boldsymbol{u}\underline{T}-\boldsymbol{\nabla}\underline{T}){\cdot} \boldsymbol{e}_z\rangle_{h,v},\label{eq:Q_define_a}\\
    =&\langle wT\rangle_{h,v}-\bar{\mathcal{T}}_{z,q},
    \label{eq:Q_define_b}
\end{align}
\end{subequations}
where $\langle\cdot \rangle_{h,v}$ is the horizontal and vertical average. The equality in \eqref{eq:Q_define_b} is obtained on assuming a vanishing homogeneous mode, $\langle w\rangle_{h,v}(t)=\langle T\rangle_{h,v}(t)=0$. We further assume that the instantaneous heat flux $Q(t)$ recovers the imposed value $Q_c$ exponentially rapidly at a rate $\beta$:
\begin{align}
    \frac{dQ}{dt}+\beta(Q-Q_c)=0.
    \label{eq:Q}
\end{align}
Here $Q_c=1$ because the temperature is normalized based on the imposed heat flux. 

We can now write equation \eqref{eq:NS_tot_T} in terms of $T$ that is periodic in all spatial directions. This is obtained by substituting the decomposition \eqref{eq:T_decompose} and \eqref{eq:Q_define} into \eqref{eq:NS_tot_T}:
\begin{align}
    \frac{\partial T}{\partial t}+\boldsymbol{u }{\cdot} \boldsymbol{\nabla}T-Qw + w\langle wT\rangle_{h,v}=&\nabla^2 T,\label{eq:NS_T_T}
\end{align}
where $Q(t)$ is governed by \eqref{eq:Q}. Note that \eqref{eq:Q} is taken to be independent of $(\boldsymbol{u},T,p)$. Thus, setting $Q(t=0)=Q_c$ leads to $Q(t)=Q_c=1$. This corresponds to the $\beta\to\infty$ limit in which $Q(t)$ recovers the reference value $Q_c$ instantaneously. We show, moreover, that for $\beta=10^4$ and a random $Q(t=0)$ the results display the same behavior as those for $\beta=\infty$ (Appendix \ref{sec:finite_beta}). Setting $Q(t)=Q_c=1$ and eliminating the hydrostatic pressure, we obtain the governing equations in the form
\begin{subequations}
\label{eq:NS}
\begin{align}
    \frac{\partial \boldsymbol{u}}{\partial t}+\boldsymbol{u}{\cdot} \boldsymbol{\nabla}\boldsymbol{u}=&-\boldsymbol{\nabla} p+Pr Ra_{T,q}T\boldsymbol{e}_z+Pr\nabla^2 \boldsymbol{u},\label{eq:NS_mom}\\
    \boldsymbol{\nabla}{\cdot} \boldsymbol{u}=&0,\label{eq:NS_mass}\\
    \frac{\partial T}{\partial t}+\boldsymbol{u }{\cdot} \boldsymbol{\nabla}T-w + w\langle wT\rangle_{h,v}=&\nabla^2 T.\label{eq:NS_T}
\end{align}
\end{subequations}
The integral term $w\langle wT\rangle_{h,v}$ in \eqref{eq:NS_T} is the new flux feedback term that does not appear in earlier formulations of the homogeneous RBC problem driven by a constant temperature gradient \citep{borue1997turbulent,lohse2003ultimate,calzavarini2005rayleigh,calzavarini2006exponentially,ng2018bulk,pratt2020lagrangian,xie2022bolgiano,barral2023asymptotic}. 

The response parameter is the instantaneous Nusselt number, which measures the ratio of the total convective transport to the conductive heat transport in the vertical:
\begin{align}
    nu(t):=\frac{ \langle (\boldsymbol{u}\underline{T}-\boldsymbol{\nabla}\underline{T}){\cdot} \boldsymbol{e}_z\rangle_{h,v}}{\langle(-\boldsymbol{\nabla}\underline{T}){\cdot} \boldsymbol{e}_z\rangle_{h,v}}=\frac{1}{1-\langle wT\rangle_{h,v}(t)}.
    \label{eq:nusselt_time}
\end{align}
We also define a Nusselt number measuring the time-averaged heat transport as:
\begin{align}
    Nu:=\frac{1}{1-\langle wT\rangle_{h,v,t}},
    \label{eq:nusselt_mean}
\end{align}
where $\langle\cdot \rangle_{h,v,t}$ denotes spatio-temporal averaging. We can also obtain the mean temperature gradient by time-averaging \eqref{eq:Q_define},
\begin{align}
\bar{\mathcal{T}}_{z,q}=&\langle wT\rangle_{h,v,t}-\langle Q\rangle_t\\
=&\langle wT\rangle_{h,v,t}-1,
\end{align}
where we assume $\langle Q\rangle_t=Q_c=1$, as appropriate for long time averages. The Rayleigh number based on the mean temperature gradient is
\begin{align}
    Ra_T:=\frac{\alpha g |\bar{\mathcal{T}}_{z,q*}|h^4}{\nu \kappa_T}=\frac{\alpha g |q|h^4}{\nu \kappa_T}(-\bar{\mathcal{T}}_{z,q})=\frac{Ra_{T,q}}{Nu},
\end{align}
where $\bar{\mathcal{T}}_{z,q*}=q\bar{\mathcal{T}}_{z,q}$ is the dimensional mean temperature gradient. A relation similar to (2.10) was also noted in RBC with a fixed imposed flux \citep{otero2002bounds,johnston2007rayleigh,johnston2009comparison,verzicco2008comparison,goluskin2016internally}. As a result, a scaling law $Nu\sim Ra_{T,q}^{1/3}$ based on imposed flux corresponds to $Nu\sim Ra_T^{1/2}$ based on the mean temperature gradient, a relation similar to that between fixed flux and fixed temperature RBC \citep{otero2002bounds}.

\section{Bifurcation analysis at moderate Rayleigh number}
\label{sec:moderate_Ra}

In this section, we analyze flow structures originating from the primary instability at moderate Rayleigh numbers and their subsequent destabilization by means of analytical calculation, numerical continuation as well as direct numerical simulations. The nonlinear solutions and their stability determined from this analysis provide the pathway towards chaotic behavior or even fully developed turbulent states, which generally visit neighbourhoods of (unstable) steady, periodic or travelling wave solutions, and these visits leave an imprint on the flow statistics; see, e.g. \citep{kawahara2001periodic,van2006periodic} and the reviews by \citet{kawahara2012significance,graham2021exact}. We keep our analytical calculation general as appropriate for three dimensions (3D), although our numerical results are confined to 2D $(x,z)$ configurations.

\subsection{Primary instability and the steady elevator mode}
\label{subsec:primary_instability_elevator_mode}
We start from the primary instability and the steady elevator mode that originates from this instability, which allows analytical progress. We linearize Eq. \eqref{eq:NS} around the conduction base state ($\boldsymbol{u}=\boldsymbol{0}$, $T=0$) by dropping the nonlinear terms. After eliminating the pressure in the vertical momentum equation by applying $-\boldsymbol{e}_z\cdot \boldsymbol{\nabla}\times [\boldsymbol{\nabla}\times (\cdot)]$ to the momentum equation, we obtain
\begin{subequations}
\label{eq:NS_linear_transformed}
\begin{align}
    \frac{\partial \nabla^2 w}{\partial t}=&Pr\nabla^4 w+Pr Ra_{T,q} \nabla^2_\perp T,\label{eq:NS_linear_transformed_mom}\\
    \frac{\partial T}{\partial t}=&w+\nabla^2 T,\label{eq:NS_linear_transformed_T}
\end{align}
\end{subequations}
where $\nabla^2_\perp:=\partial_x^2+\partial_y^2$. We use the normal mode assumption $\phi(x,y,z,t)=\hat{\phi}\exp[\text{i}(k_x x+k_y y+k_z z) +\lambda t]+c.c.$, where $\phi=w,T$, and $k_x$, $k_y$, and $k_z$ are the wavenumbers in the corresponding directions and $\lambda$ is the (necessarily real) growth rate. Here, $\text{i}$ is the imaginary unit and $c.c.$ denotes the complex conjugate. This normal mode assumption yields
\begin{subequations}
\begin{align}
    \lambda \hat{w}=&-PrK^2\hat{w}+Pr Ra_{T,q}\frac{k_\perp^2}{K^2}\hat{T},\\
    \lambda \hat{T}=&\hat{w}-K^2\hat{T},
\end{align}
\end{subequations}
where $K^2:=k_x^2+k_y^2+k_z^2$ and $k_\perp^2:=k_x^2+k_y^2$. Solving this eigenvalue problem gives growth rate
\begin{align}
    \lambda=-\frac{1}{2}(Pr+1)K^2\pm \frac{1}{2}\sqrt{(Pr+1)^2K^4+4Pr(Ra_{T,q}\frac{k_\perp^2}{K^2}-K^4)}.
\end{align}
When $k_z=0$, this growth rate is the same as that associated with the elevator mode in homogeneous RBC driven by a constant temperature gradient \citep[Eq. (9)]{calzavarini2006exponentially}. 

The growth rate $\lambda$ vanishes at the onset of a steady bifurcation, leading to the neutral curve $Ra_{T,q}=\frac{K^6}{k_\perp^2}$. For a given $Ra_{T,q}$, the most unstable mode corresponds to $k_z=0$, i.e. to an elevator mode. The corresponding neutral curve then simplifies:
\begin{align}
    Ra_{T,q}=k_\perp^4,
    \label{eq:critical_Ra_Tq}
\end{align}
again as for the case of constant temperature gradient forcing \citep{calzavarini2006exponentially}. As a result, the critical horizontal wavenumber is $k_{\perp,c}=0$, as in RBC with fixed-flux boundary conditions; see e.g. \citet{sparrow1964thermal,chapman1980nonlinear}. 

The resulting steady elevator mode  ($k_z=0$) plays an important role in the subsequent behavior of the system. The amplitude of the elevator mode is obtained by substituting 
\begin{subequations}
    \begin{align}
    w(x,y,z,t)=&\hat{w}_e \exp[\text{i}(k_x x+k_y y)]+c.c.,\\    T(x,y,z,t)=&\hat{T}_e\exp[\text{i}(k_x x+k_y y)]+c.c.
\end{align}
\end{subequations}
into Eq. \eqref{eq:NS}, which gives:
\begin{align}
    \hat{w}_e=\sqrt{\frac{Ra_{T,q}}{2k_\perp^2}-\frac{k_\perp^2}{2}},
    \quad \hat{T}_e=\frac{k_\perp^2 \hat{w}_e}{Ra_{T,q}}.
    \label{eq:amplitude_elevator_mode}
\end{align}
Note that this is an exact solution of the nonlinear governing equation \eqref{eq:NS} and corresponds to the Nusselt number 
\begin{align}
    Nu=\frac{Ra_{T,q}}{k_\perp^4}.
    \label{eq:Nu_elevator_mode}
\end{align}
The steady elevator mode within the fixed-flux formulation thus has a unique time-independent amplitude \eqref{eq:amplitude_elevator_mode} for each Rayleigh number. In contrast, within homogeneous RBC driven by a constant temperature gradient, the steady elevator mode bifurcates from $Ra_{T,q}^{(p)}$ with arbitrary amplitude and grows exponentially for $Ra_{T,q}>Ra_{T,q}^{(p)}$, leading to intermittent heat transport in DNS \citep{borue1997turbulent,lohse2003ultimate,calzavarini2005rayleigh,calzavarini2006exponentially}.

\subsection{Numerical methods}

For solution branches beyond the steady elevator mode, numerical computations are required. We compute each solution branch and associated bifurcation points by numerical continuation using pde2path \citep{uecker2021numerical,uecker2014pde2path} with horizontal and vertical directions discretized by the Fourier collocation method \citep{weideman2000matlab} following the implementation in \citet{uecker2021pde2path}. We use a streamfunction formulation of the full 2D equations in Eq. \eqref{eq:NS} to reduce the number of variables thereby facilitating computation. The horizontal and vertical directions use $N_x=N_z=32$ grid points, and doubling the number of grid points in each direction does not influence the results. The tolerance of the maximum absolute value of the residual at each vertical location ($L_\infty$ norm) is set to $10^{-6}$. We implement the phase condition associated with horizontal translation symmetry for elevator modes and the phase conditions corresponding to both horizontal and vertical translations for all other 2D solution branches \citep{rademacher2017symmetries}. The stability of each branch is determined by computing a subset of the eigenvalues, and this subset is enlarged as necessary to ensure that instability and bifurcation points are correctly identified. 

We also analyze time-dependent states through direct numerical simulation in Dedalus \citep{burns2020dedalus} using a Fourier spectral method in both horizontal and vertical directions. We set the spatial homogeneous mode associated with $k_x=k_z=0$ to zero, which can be implemented by adding the constraint $\langle T\rangle_{h,v}(t)=0$. The quantity $\langle w\rangle_{h,v}(t)$ is conserved over time, and all of the results here are for $\langle w\rangle_{h,v}(t)=0$. A nonzero $\langle w\rangle_{h,v}(t)$ leads to vertically advected structures whose behavior in the comoving frame is identical to that described below. We use $N_x=N_z=128$ grid points for moderate $Ra_{T,q}$ and set $Pr=1$.

\subsection{Flow structures beyond elevator mode}

Here, we choose the domain size $L_x=0.2\pi$, unless otherwise mentioned, selected to accommodate the secondary instability of the elevator mode. With this domain size, the horizontal wavenumber of a domain-filling elevator mode corresponds to $k_\perp=2\pi/L_x=10$ and thus the critical Rayleigh number is $Ra_{T,q}=10^4$ according to Eq.~\eqref{eq:critical_Ra_Tq}. A larger domain will instead display a stable finite-amplitude elevator mode up to $Ra_{T,q}=10^8$ at $Pr=1$, as demonstrated below through secondary instability analysis (figure \ref{fig:secondary_growth_rate_Lx_ke}) and DNS (figure \ref{fig:DNS_u_h_ke_20_ke_30}).

\begin{figure}
(a) \hspace{0.48\textwidth} (b)
    
    \centering
    \includegraphics[width=0.49\textwidth]{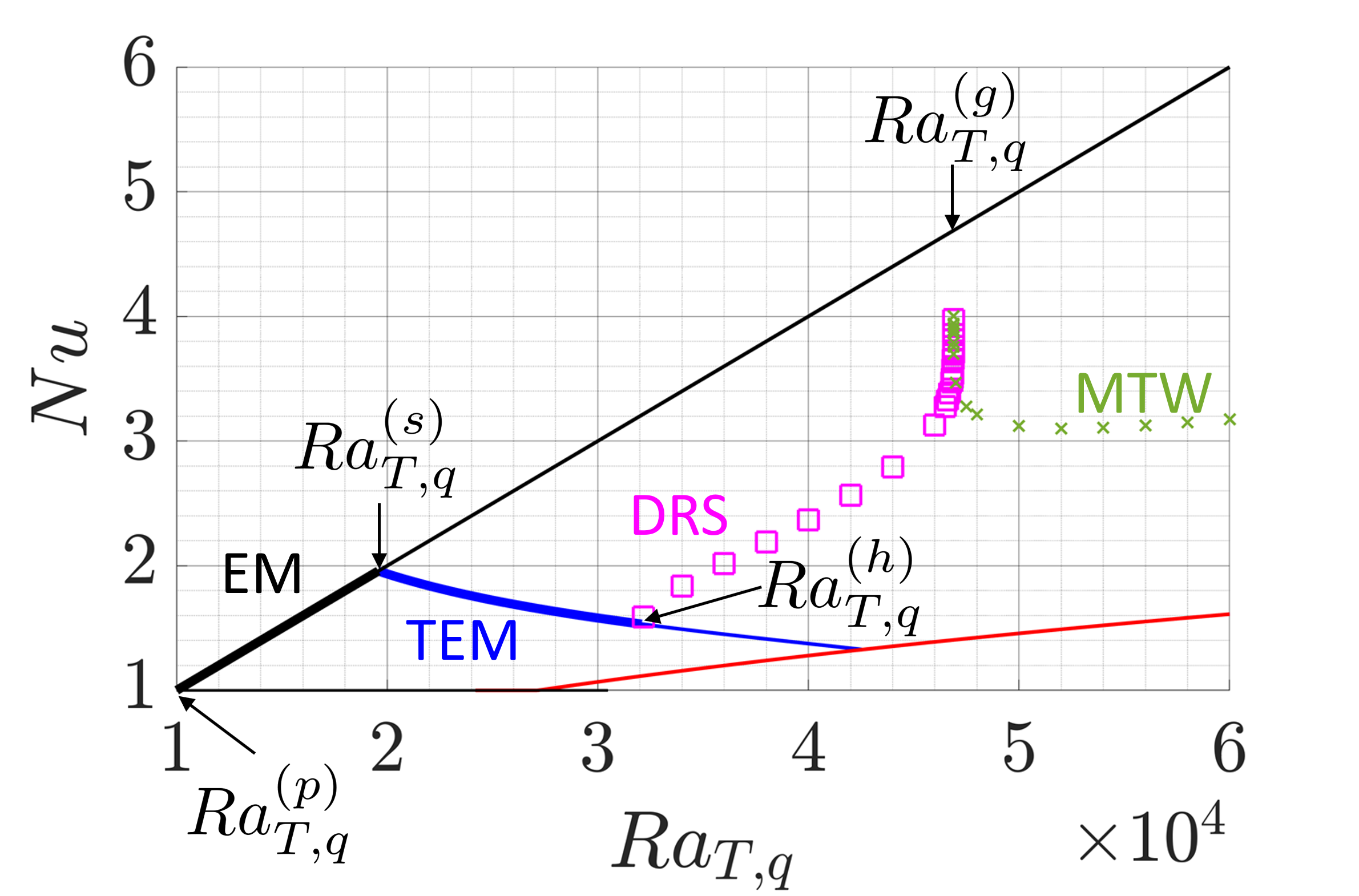}
    \includegraphics[width=0.49\textwidth]{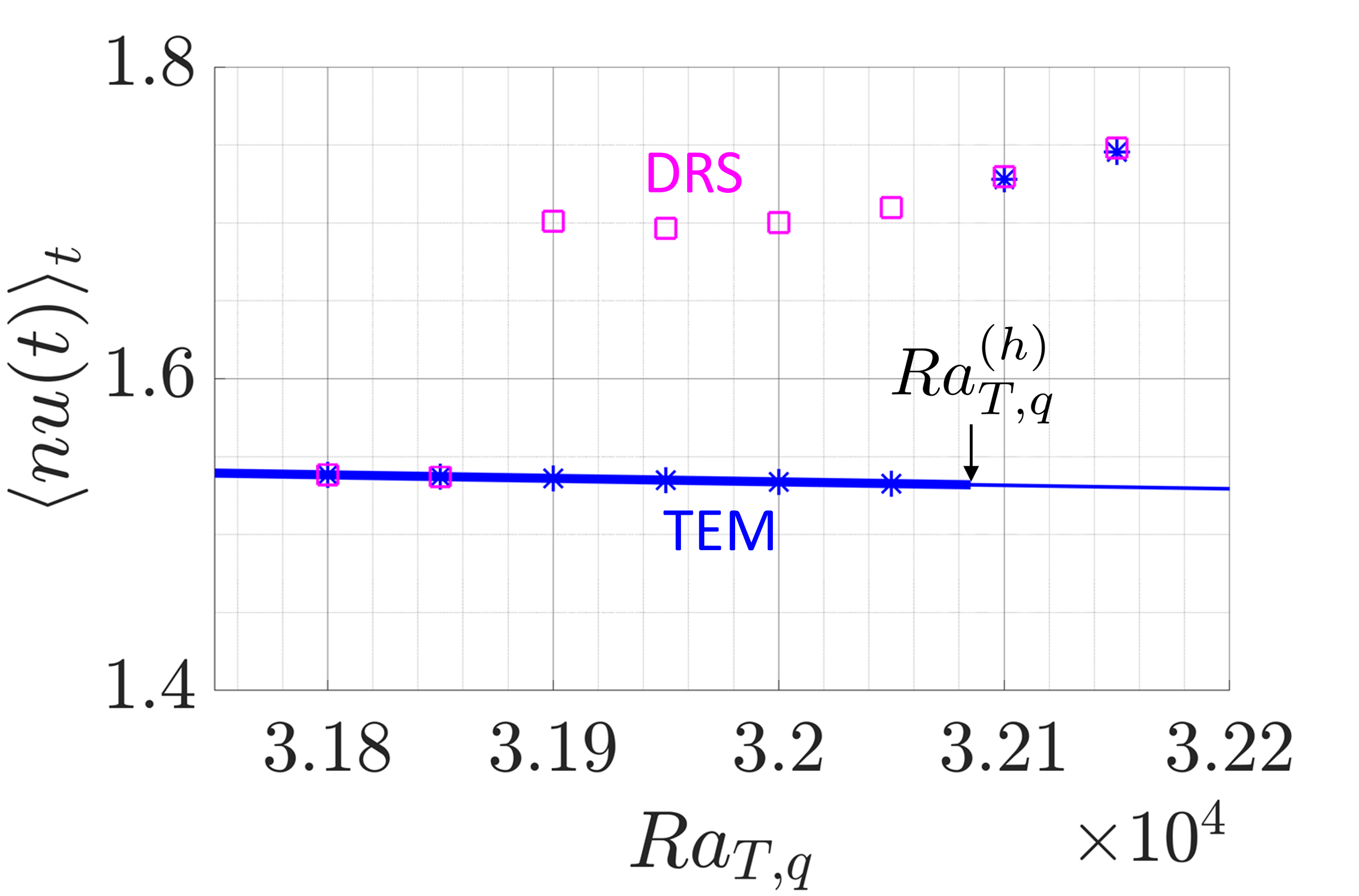}
    \caption{(a) Bifurcation diagram with elevator mode (EM, $\mline$, black), tilted elevator mode (TEM, {\color{blue}$\mline$}, blue), direction-reversing state (DRS, {\color{magenta}$\square$}, magenta) and modulated traveling waves (MTW, {\color{darkgreen}$\times$}, green). The bifurcation points include the primary bifurcation $Ra_{T,q}^{(p)}$, the secondary bifurcation $Ra_{T,q}^{(s)}$, the Hopf bifurcation $Ra_{T,q}^{(h)}$ and a global bifurcation at $Ra_{T,q}^{(g)}$. (b) Hysteresis diagram near the Hopf bifurcation point $Ra_{T,q}^{(h)}$ with TEM initial condition and increasing $Ra_{T,q}$ ({\color{blue}$*$}, blue) or DRS initial conditions and decreasing $Ra_{T,q}$ ({\color{magenta}$\square$}, magenta).}
    \label{fig:bif_diag_full}
\end{figure}

Figure \ref{fig:bif_diag_full} shows the resulting bifurcation diagram using thick (thin) lines for stable (unstable) states obtained from numerical continuation. The markers in figure \ref{fig:bif_diag_full} correspond to the final stable state as obtained from DNS. Figure \ref{fig:bif_diag_full}(a) shows that the elevator mode (EM, black) bifurcates from the primary instability at $Ra_{T,q}^{(p)}=10^4$ consistent with Eq. \eqref{eq:critical_Ra_Tq} and that the Nusselt number of this mode displays a linear relation with $Ra_{T,q}$ according to Eq. \eqref{eq:Nu_elevator_mode}. The secondary instability of the elevator mode at $Ra_{T,q}^{(s)}=19576.3$ leads to a branch of steady tilted elevator modes (TEM, blue) accompanied by large-scale shear. Figure \ref{fig:DNS_Ra_Tq_3e4_Pr_1}(a) shows the evolution of this shear from DNS at $Ra_{T,q}=3\times 10^4$, starting from an unstable EM at this Rayleigh number. The figure shows that the large-scale shear $\langle u\rangle_h(z,t)$ becomes nonzero at $t\approx 2$ and then saturates in a horizontal flow with an approximately sinusoidal profile in the vertical. Figure \ref{fig:DNS_Ra_Tq_3e4_Pr_1}(b) shows that the associated temperature deviation $T(x,z,t)$ at $t=10$ is tilted in the direction corresponding to the generated large-scale shear. This secondary bifurcation resembles the behavior observed in RBC between fixed temperature boundaries, whereby a secondary bifurcation of steady convection rolls leads to tilted rolls accompanied by large-scale shear \citep{howard1986large,rucklidge1996analysis}, as also observed in both experiments \citep{krishnamurti1981large} and DNS \citep{goluskin2014convectively,matthews1996three,von2015generation,wang2020zonal}. This TEM branch  terminates in another unstable steady state (red) that bifurcates from the conduction state without large-scale shear generation and resembles the two-layer (S2) solutions identified in salt-finger convection \citep{liu2022staircase}. 

\begin{figure}
(a) $\langle u\rangle_{h}(z,t)$ \hspace{0.4\textwidth} (b) $T(x,z,t=10)$

    \centering
        \includegraphics[width=0.49\textwidth]{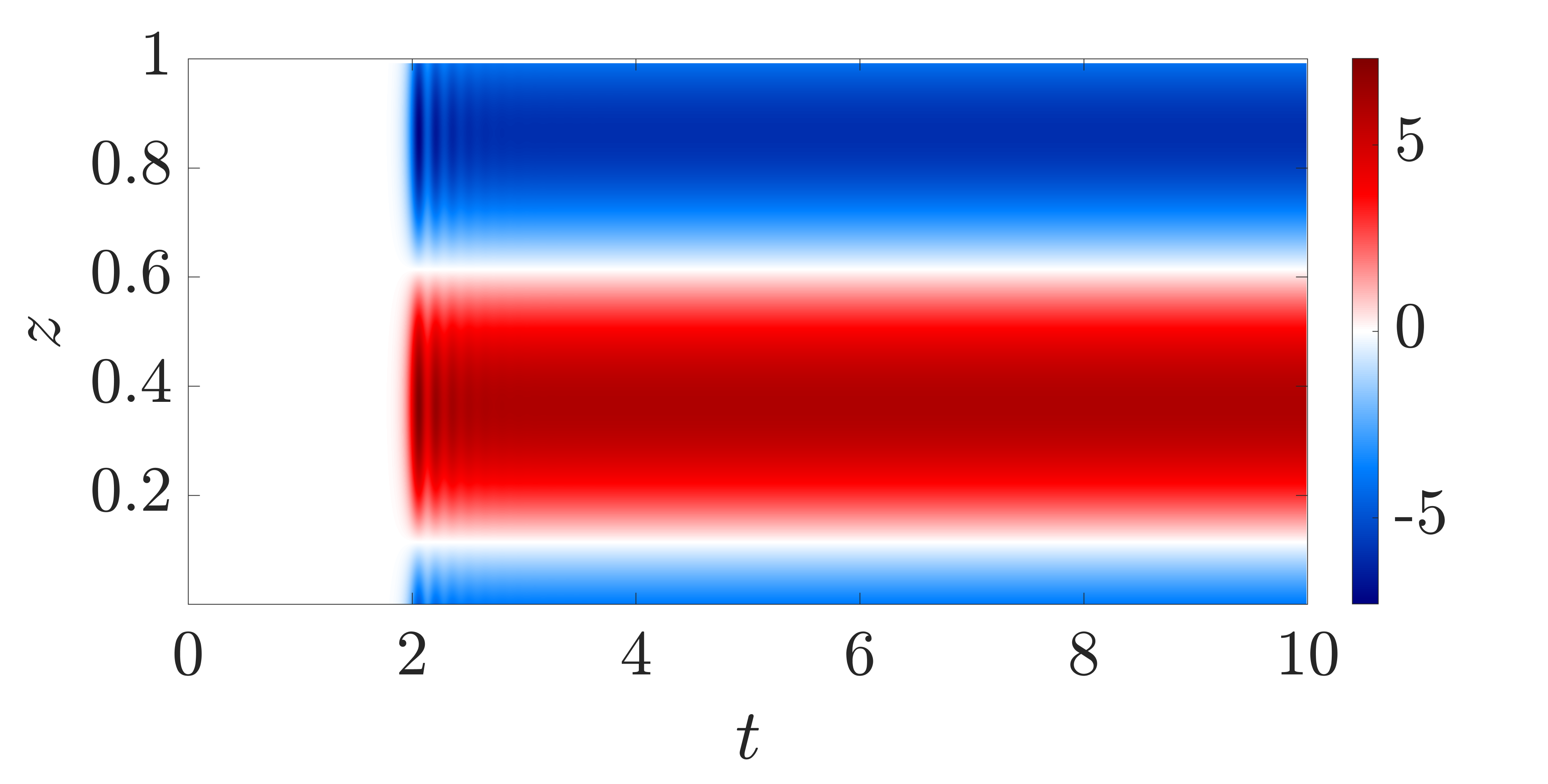}
\includegraphics[width=0.44\textwidth]{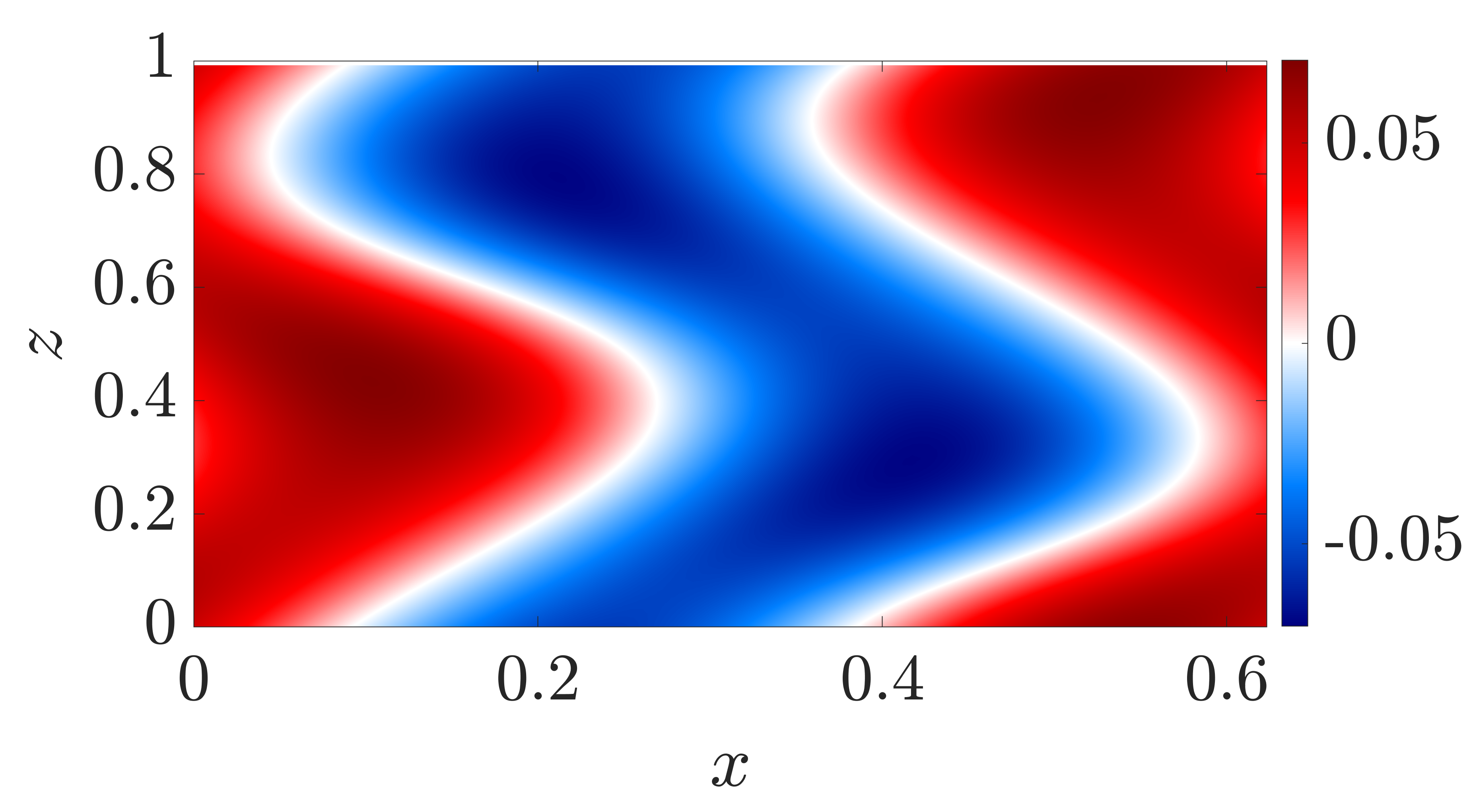}

    \caption{(a) Large-scale shear $\langle u\rangle_h(z,t)$ and (b) temperature deviation $T(x,z,t)$ at $t=10$ for the steady tilted elevator mode (TEM) at $Ra_{T,q}=3\times 10^4$, $Pr=1$ and $L_x=0.2\pi$.}
    \label{fig:DNS_Ra_Tq_3e4_Pr_1}
\end{figure}

\begin{figure}
(a) $\langle u\rangle_{h}(z,t)$ \hspace{0.4\textwidth} (b) 

    \centering
        \includegraphics[width=0.49\textwidth]{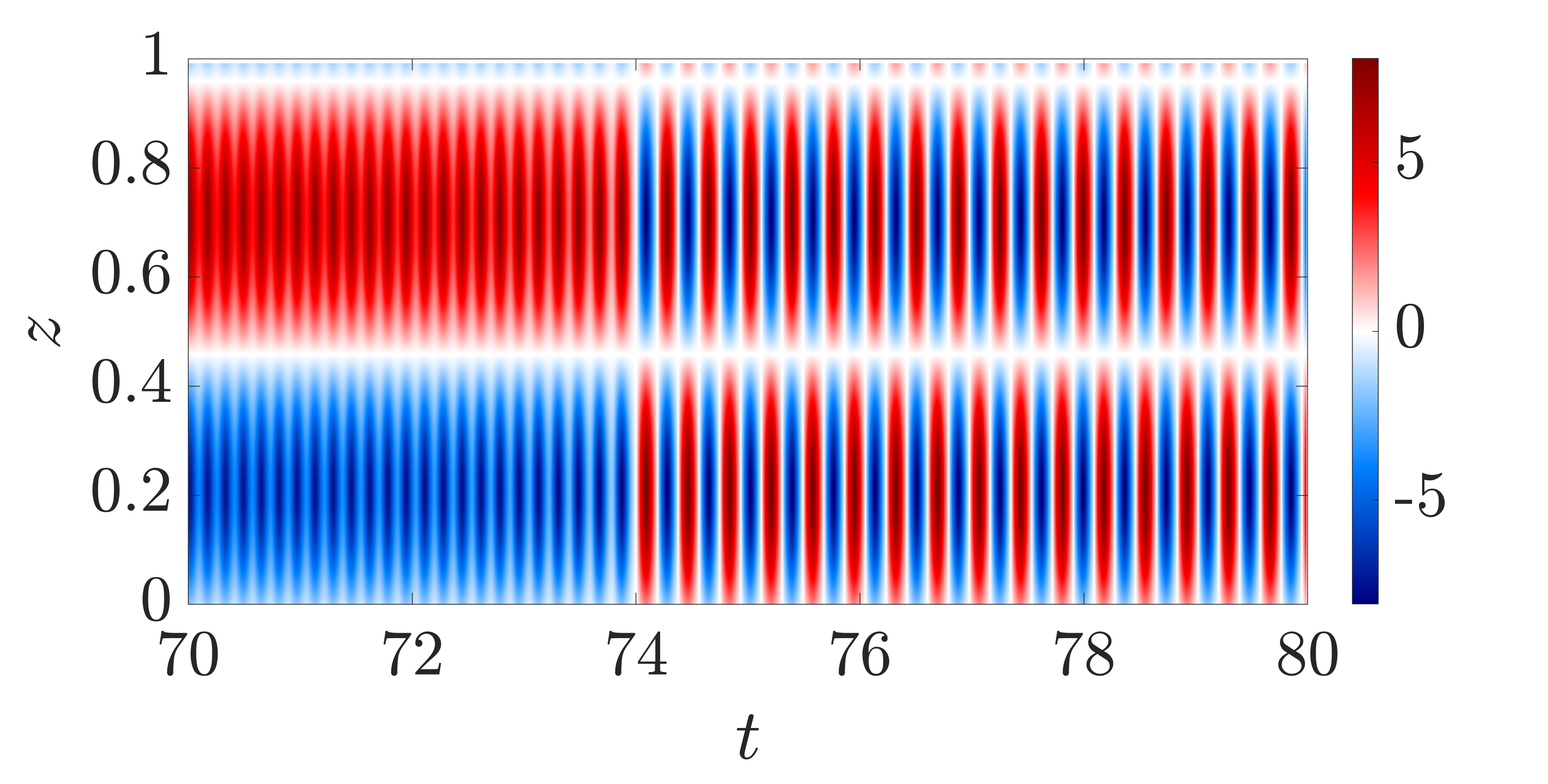}
\includegraphics[width=0.49\textwidth]{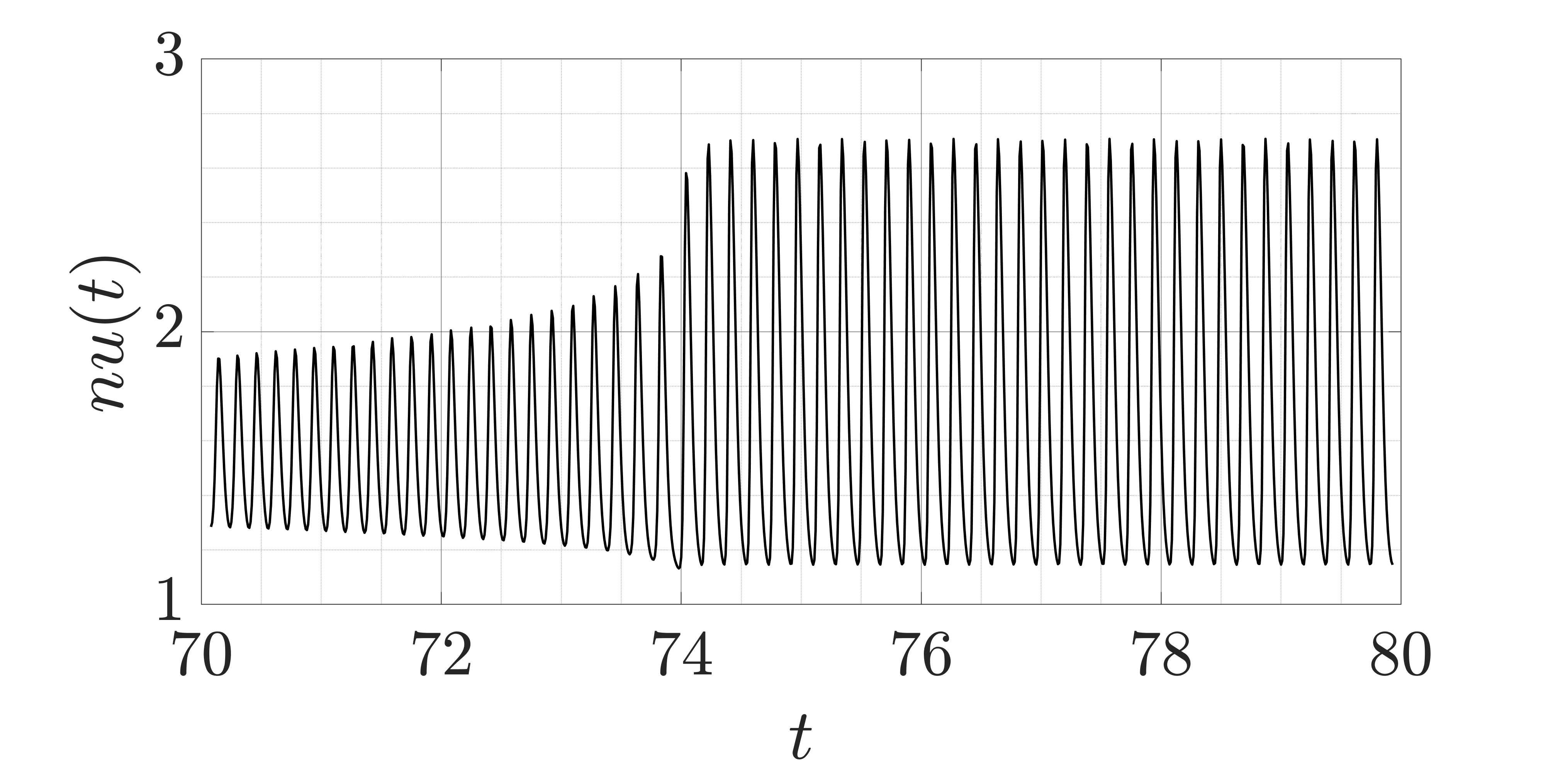}
    \caption{(a) Large-scale shear $\langle u\rangle_{h}(z,t)$ and (b) instantaneous Nusselt number $nu(t)$ at $Ra_{T,q}=3.21\times 10^4$, $Pr=1$ and $L_x=0.2\pi$ with a TEM initial condition (Movie 1 shows the corresponding temperature deviation $T(x,z,t)$).}
    \label{fig:DNS_Ra_Tq_3p21e4_Pr_1_transient}
\end{figure}

The steady tilted elevator mode loses stability at a Hopf bifurcation at $Ra_{T,q}^{(h)}=32085.1$ leading to oscillations about the TEM state with frequency $\omega_h=43.1$. This Hopf bifurcation is subcritical, however, implying that the tilted oscillations are unstable. Computations indicate that the system instead evolves into a symmetric direction-reversing state (DRS) with associated hysteresis near $Ra_{T,q}^{(h)}$ as shown in figure \ref{fig:bif_diag_full}(b). Here, we use $\langle nu(t)\rangle_t$ to distinguish the TEM state from the DRS state, reached from TEM initial conditions upon increasing $Ra_{T,q}$ ({\color{blue}$*$}, blue) or from DRS initial conditions upon decreasing $Ra_{T,q}$ ({\color{magenta}$\square$}, magenta). When $Ra_{T,q}<Ra_{T,q}^{(h)}$, the DNS with TEM initial conditions show excellent agreement with numerical continuation results (the blue thick line), while DNS with TEM initial conditions at $Ra_{T,q}>Ra_{T,q}^{(h)}$ evolve into DRS. Figure \ref{fig:DNS_Ra_Tq_3p21e4_Pr_1_transient} shows $\langle u \rangle_h(z,t)$ and $nu(t)$ at $Ra_{T,q}=32100$, a value slightly larger than $Ra_{T,q}^{(h)}$, with a TEM initial condition from a lower $Ra_{T,q}$. The large scale flow oscillates with frequency $\omega\approx 43.0$ in $t\in [0,10]$ (not shown in figure \ref{fig:DNS_Ra_Tq_3p21e4_Pr_1_transient}) that is close to the Hopf frequency $\omega_h=43.1$ but does not reverse. At $t\approx 74$, the flow abruptly transitions to a direction-reversing state as shown in figure \ref{fig:DNS_Ra_Tq_3p21e4_Pr_1_transient}(a). Figure \ref{fig:DNS_Ra_Tq_3p21e4_Pr_1_transient}(b) shows the corresponding instantaneous Nusselt number $nu(t)$. Similar direction-reversing states are also observed in RBC \citep{sugiyama2010flow,chandra2013flow,winchester2021zonal}, magnetoconvection \citep{matthews1993pulsating,proctor1994nonlinear} as well as in salt-finger convection \citep{liu2022staircase}.

\begin{figure}
    \centering
    (a) $t=2.21$ \hspace{0.19\textwidth} (b) $t=2.29$ \hspace{0.19\textwidth} (c) $t=2.37$
    \includegraphics[width=0.32\textwidth]{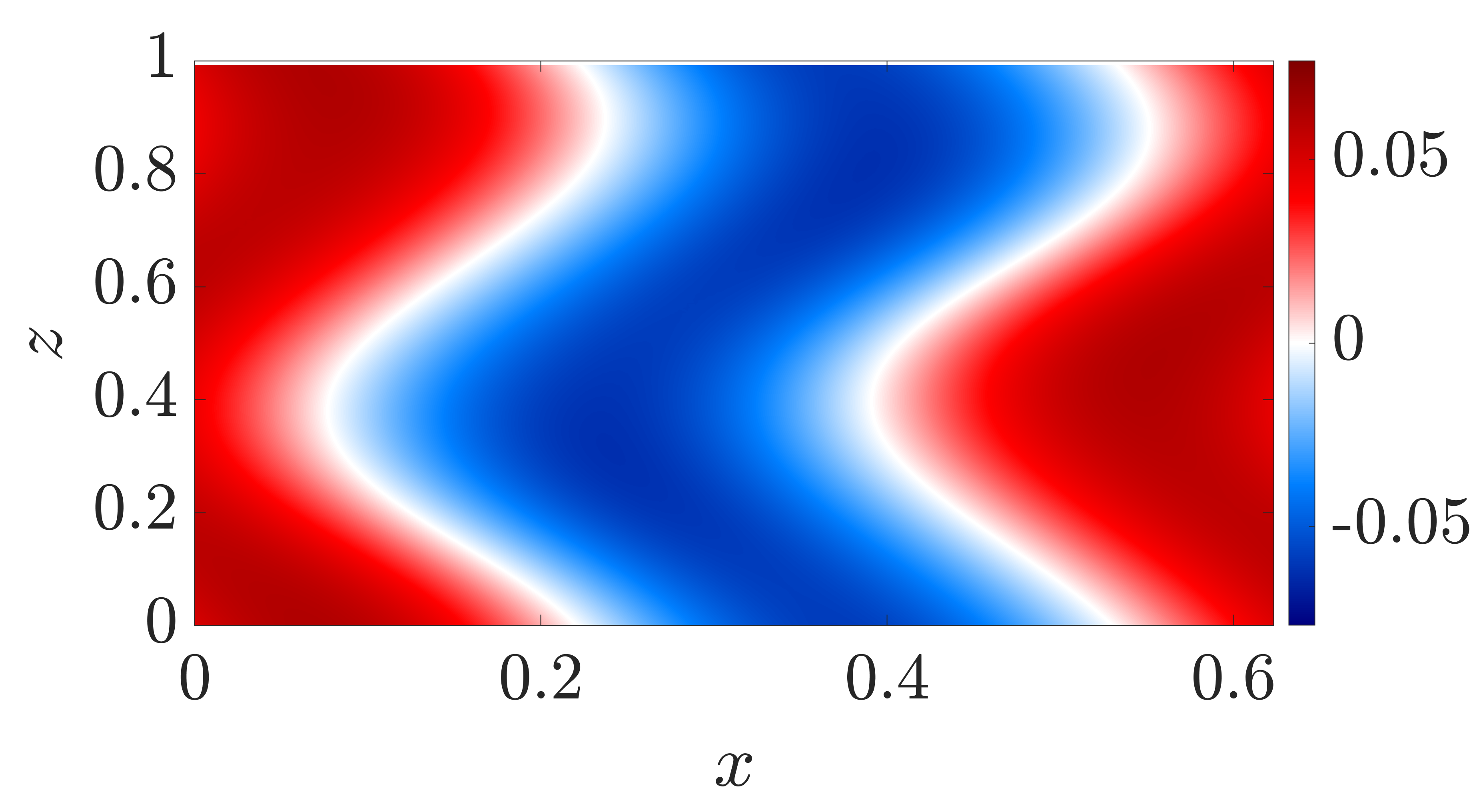}
    \includegraphics[width=0.32\textwidth]{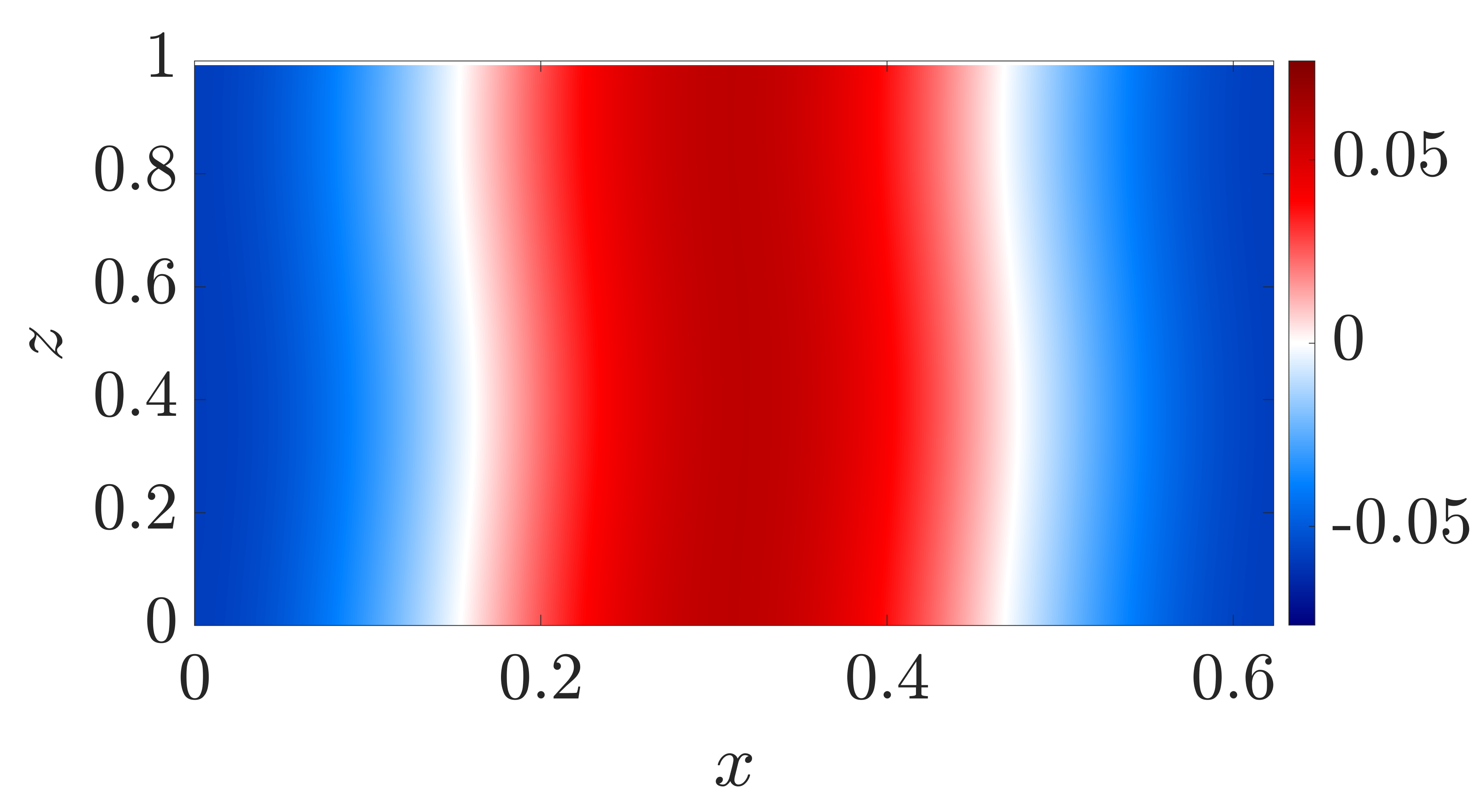}
    \includegraphics[width=0.32\textwidth]{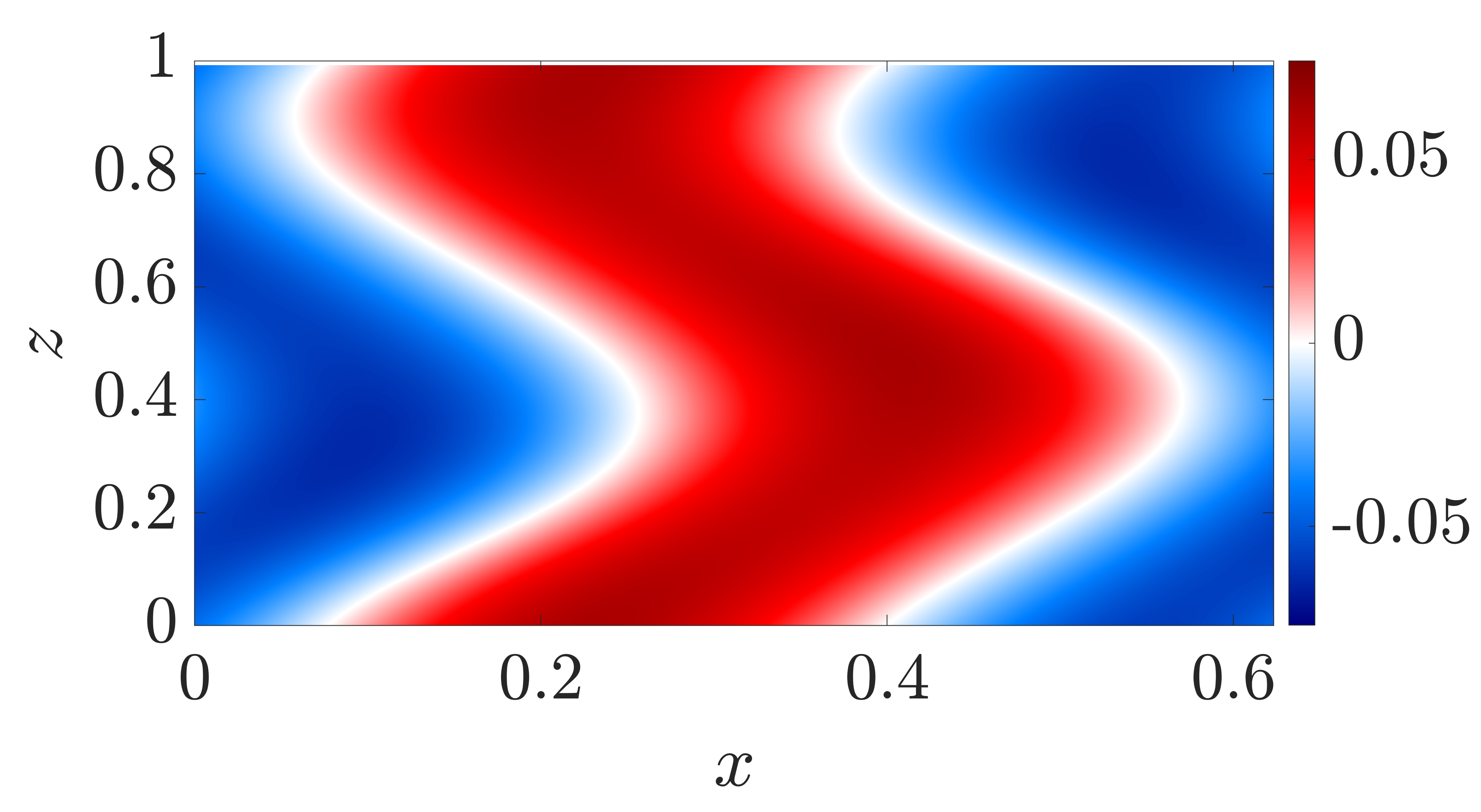}
    \caption{Temperature deviation $T(x,z,t)$ associated with the direction-reversing state at three different times when $Ra_{T,q}=4.6\times 10^4$, $Pr=1$ and $L_x=0.2\pi$.}
\label{fig:DNS_Ra_Tq_4p6e4_Pr_1_snapshot}
\end{figure}

At higher Rayleigh numbers, this direction-reversing state spends more time displaying flow structures close to an elevator mode. Figure \ref{fig:DNS_Ra_Tq_4p6e4_Pr_1_snapshot} shows three snapshots of the temperature deviation $T(x,z,t)$ at $Ra_{T,q}=4.6\times 10^4$. At $t=2.21$ and $t=2.37$, the temperature deviation tilts in opposite directions. At $t=2.29$, $T(x,z,t)$ displays flow structures close to an elevator mode, followed at $t=2.37$ by a restored and approximately reflected tilted state. At yet higher Rayleigh numbers, the direction-reversing state collides with the unstable steady elevator mode leading to a global bifurcation at $Ra_{T,q}^{(g)}\approx46892.03$ as indicated in figure \ref{fig:bif_diag_full}(a). At this global bifurcation, the direction-reversing state transitions to modulated traveling waves (MTW, green) that do not reverse direction. Figure \ref{fig:DNS_Ra_Tq_6e4_Pr_1}(a) shows the corresponding large-scale shear $\langle u\rangle_h(z,t)$ at $Ra_{T,q}=6\times 10^4$. Figure \ref{fig:DNS_Ra_Tq_6e4_Pr_1}(b) displays the corresponding temperature deviation $T(x,z,t)$ at $z=0.1$, including the modulated traveling wave that sets in at $t\approx0.8$. This global bifurcation is illustrated in the phase diagram shown in figure \ref{fig:phase_diagram_46892} near $Ra_{T,q}^{(g)}$, revealing an abrupt change in topology before and after this global bifurcation. Note that both states pass through $\langle T\rangle_h (z_p, t)=\langle u\rangle_h(z_p, t)=0$ corresponding to the elevator mode. A similar global bifurcation must take place on the subcritical DRS branch in order to generate the DRS from the oscillating TEM state but is inaccessible to DNS. Such gluing bifurcations are also seen in RBC, where oscillatory tilted convection rolls originating from a Hopf bifurcation of steady tilted convection rolls may collide with steady convection rolls and glue together in a global bifurcation; see, e.g., \citet{rucklidge1996analysis}. 

\begin{figure}
(a) $\langle u\rangle_{h}(z,t)$ \hspace{0.4\textwidth} (b) $T(x,z,t)$ at $z=0.1$

    \centering
    \includegraphics[width=0.49\textwidth]{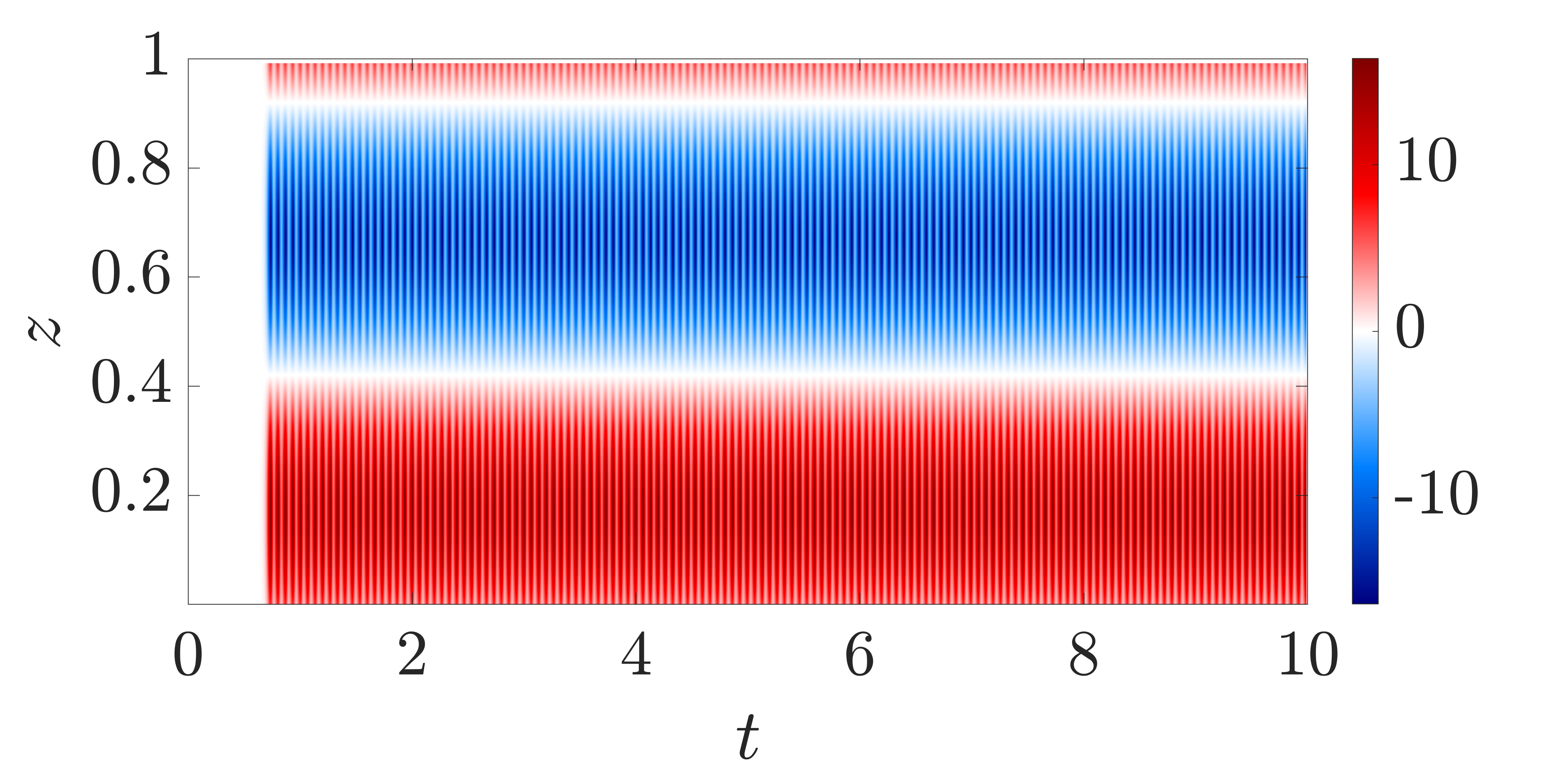}
    \includegraphics[width=0.49\textwidth]{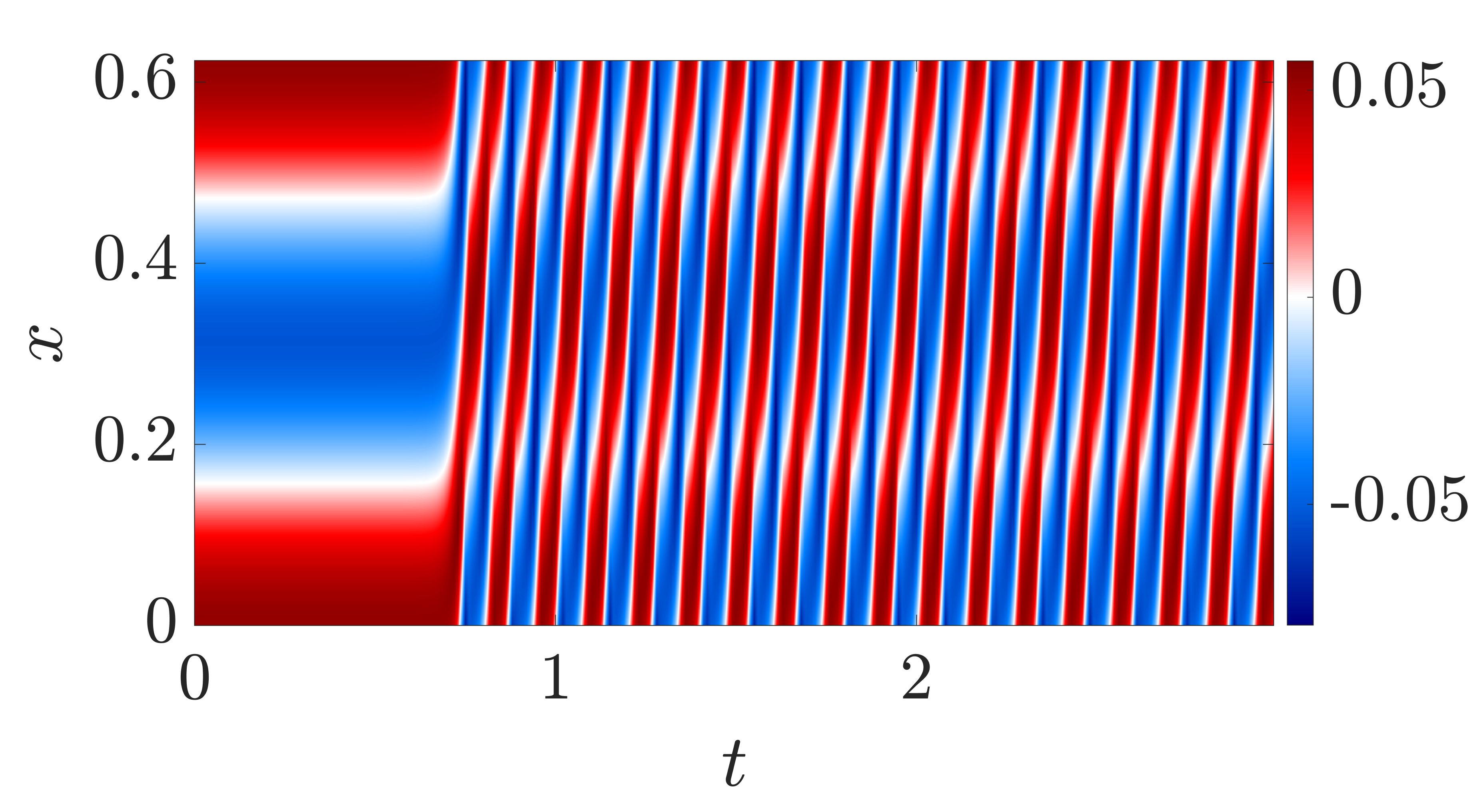}    
    
    \caption{(a) Large-scale shear $\langle u\rangle_h(z,t)$ and (b) temperature deviation $T(x,z,t)$ at $z=0.1$ for modulated traveling waves at $Ra_{T,q}=6\times 10^4$, $Pr=1$, and $L_x=0.2\pi$.}
    \label{fig:DNS_Ra_Tq_6e4_Pr_1}
\end{figure}

\begin{figure}
(a) $Ra_{T,q}=46892.0$ \hspace{0.28\textwidth} (b) $Ra_{T,q}=46892.1$

    \centering
    \includegraphics[width=0.49\textwidth]{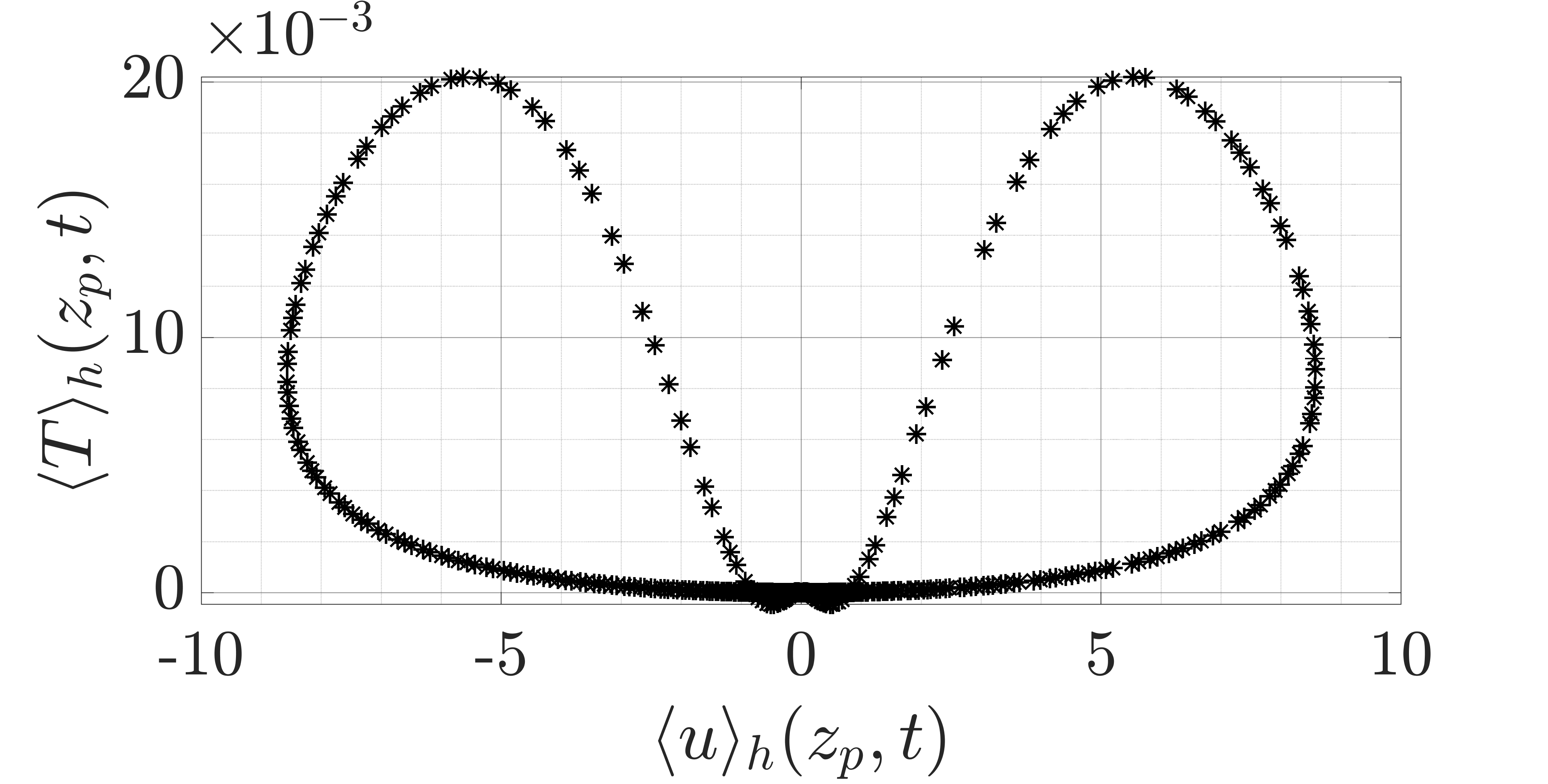}
    \includegraphics[width=0.49\textwidth]{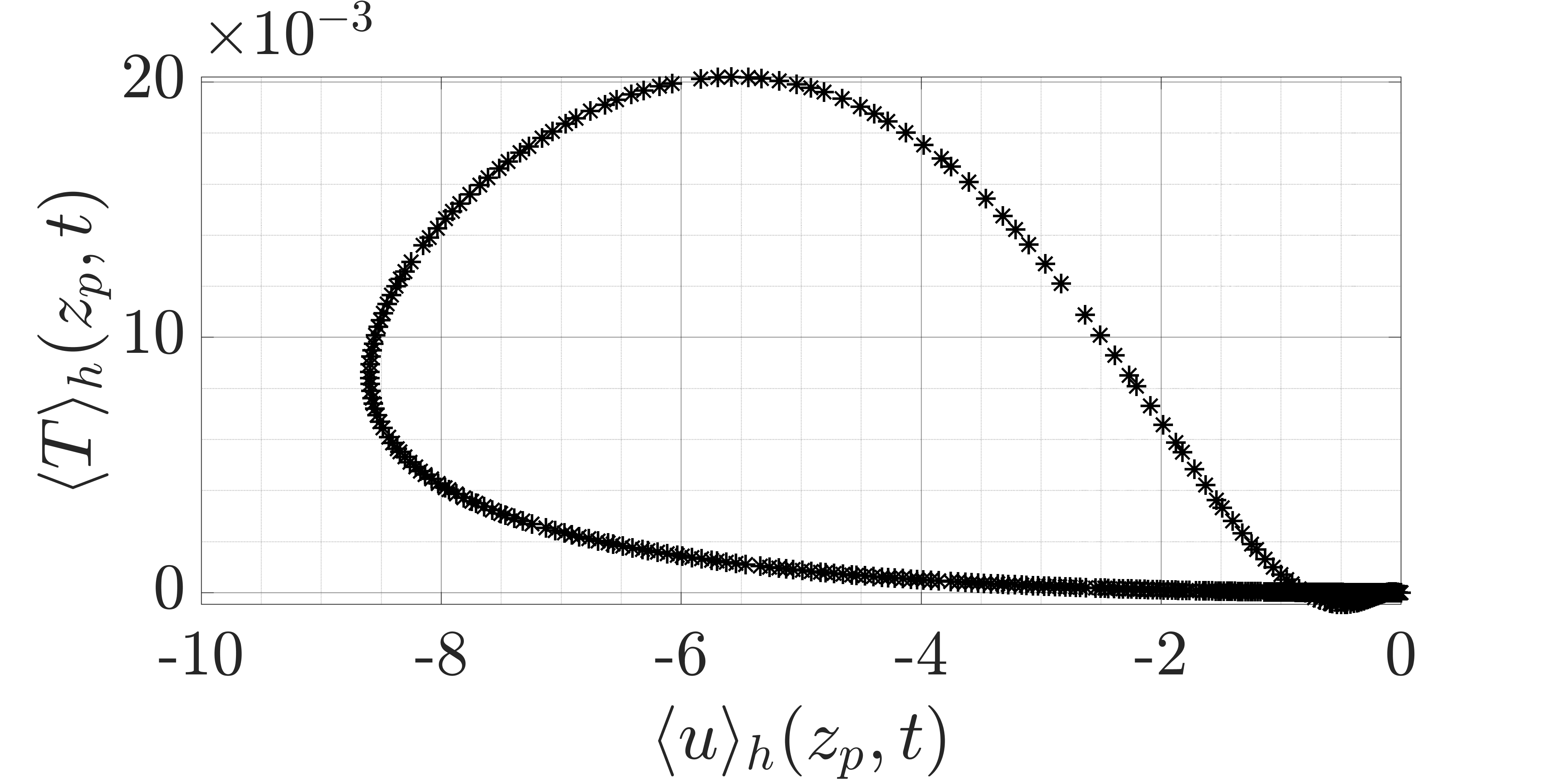}
    \caption{Phase diagram showing $\langle T\rangle_h(z_p,t)$ as a function of $\langle u\rangle_h(z_p,t)$ at (a) $Ra_{T,q}=46892.0$ and (b) $Ra_{T,q}=46892.1$ with $z_p:=\underset{z}{\text{arg max}} \langle T\rangle_h(z,t=10)$. The global bifurcation takes place in between. }
    \label{fig:phase_diagram_46892}
\end{figure}

\subsection{Single-mode equations}

The previous results show that the flow in the horizontal direction is dominated by a domain-filling mode. Moreover, in fixed-flux Rayleigh-B\'enard convection any long box will eventually contain a single pair of rolls \citep{chapman1980nonlinear}, and domain-filling modes also organize the flow in the turbulent regime at high Rayleigh number \citep{vieweg2021supergranule,vieweg2022inverse,kaufer2023thermal}. This motivates us to derive single-mode equations that have been successfully used in a wide range of convection problems \citep{herring1963investigation,toomre1977numerical,gough1982single,paparella1999sheared}, especially for well-organized columnar structures in the presence of strong restraining body forces including rapid rotation and strong magnetic field \citep{julien2007reduced}, or large-scale damping in salt-finger convection \citep{liu2022staircase} or convection in a porous medium \citep{liu2022single}.

Single-mode equations are obtained from a severely truncated Fourier expansion in the horizontal, which reduces the governing equations from three spatial dimensions to equations for the vertical solutions profile associated with a prescribed horizontal planform. Here, we derive the single-mode equations by decomposing variables into a mean mode in the horizontal and horizontal harmonics:
\begin{subequations}
\label{eq:normal_mode}
\begin{align}
    T(x,y,z,t)=&\bar{T}_0(z,t)+\hat{T}(z,t)\,e^{\text{i}(k_x x+k_y y)}+c.c.,\label{eq:normal_mode_T}\\
    \boldsymbol{u}(x,y,z,t)=&\bar{U}_0(z,t)\boldsymbol{e}_x+\hat{\boldsymbol{u}}(z,t)\,e^{\text{i}(k_x x+k_y y)}+c.c.,\label{eq:normal_mode_u}\\
    p(x,y,z,t)=&\bar{P}_0(z,t)+\hat{p}(z,t)\,e^{\text{i}(k_x x+k_y y)}+c.c.\label{eq:normal_mode_p}
\end{align}
\end{subequations}
We truncate the resulting equations at these harmonics to obtain the single-mode equations:
\begin{subequations}
\label{eq:single_mode}
\begin{gather}
    \partial_t \hat{u}+\bar{U}_0\text{i}k_x \hat{u}+\hat{w}\partial_z \bar{U}_0=-\text{i}k_x \hat{p}+Pr\hat{\nabla}^2 \hat{u},\\
    \partial_t \hat{v}+\bar{U}_0\text{i}k_x\hat{v}=-\text{i}k_y \hat{p}+Pr\hat{\nabla}^2 \hat{v},\\
    \partial_t \hat{w}+\bar{U}_0\text{i}k_x\hat{w}=-\partial_z \hat{p}+Pr\hat{\nabla}^2\hat{w}+Pr Ra_{T,q}\hat{T},\\
    \text{i}k_x \hat{u}+\text{i}k_y\hat{v}+\partial_z \hat{w}=0,\\
    \partial_t\hat{T}+\bar{U}_0\text{i}k_x\hat{T}+\hat{w}\partial_z \bar{T}_0-\hat{w}+\hat{w}\int_0^1\left(\hat{w}^* \hat{T}+\hat{w}\hat{T}^*\right)dz=\hat{\nabla}^2\hat{T},\label{eq:single_mode_T_hat}\\
    \partial_t \bar{U}_0+\partial_z\left(\hat{w}^*\hat{u}+\hat{w}\hat{u}^*\right)=Pr\partial_z^2\bar{U}_0,\label{eq:single_mode_U_0}\\
    \partial_t \bar{T}_0+\partial_z\left(\hat{w}^*\hat{T}+\hat{w}\hat{T}^*\right)=\partial_z^2 \bar{T}_0.\label{eq:single_mode_T_0}
\end{gather}
\end{subequations}
Here, the integral term in \eqref{eq:single_mode_T_hat} represents the fixed-flux constraint originating from $w\langle wT\rangle_{h,v}$ in Eq. \eqref{eq:NS_T}. The horizontal wavenumber is chosen as $k_x=10$ and $k_y=0$ corresponding to a domain-filling mode within a 2D domain with $L_x=0.2\pi$. Numerical continuation of the single-mode equations \eqref{eq:single_mode} is performed using pde2path \citep{uecker2014pde2path,uecker2021numerical} with $N_z=128$ while DNS of \eqref{eq:single_mode} is conducted using Dedalus \citep{burns2020dedalus} with $N_z=128$. 

\begin{figure}
(a) \hspace{0.49\textwidth} (b)

    \centering
    \includegraphics[width=0.49\textwidth]{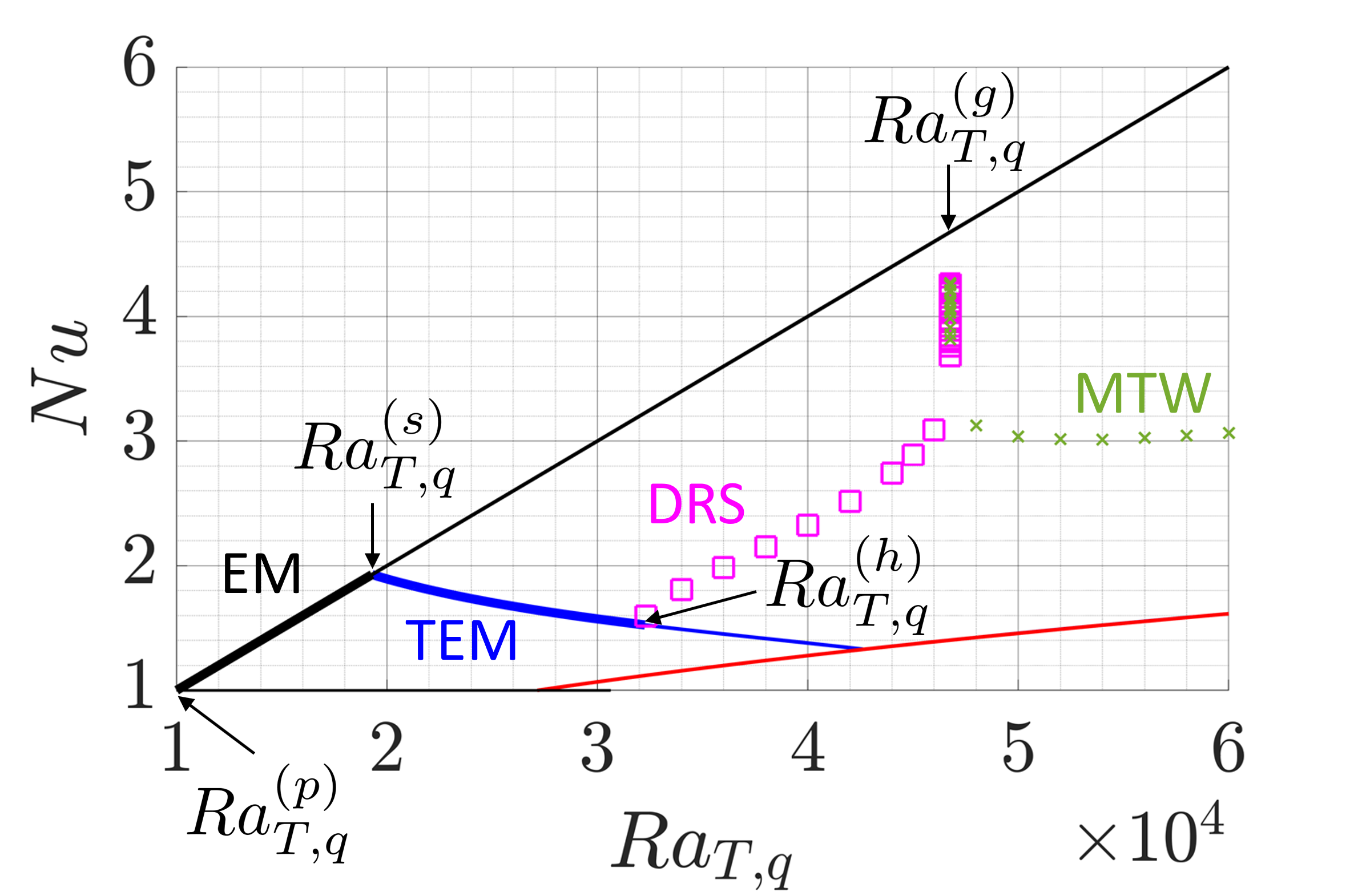}
    \includegraphics[width=0.49\textwidth]{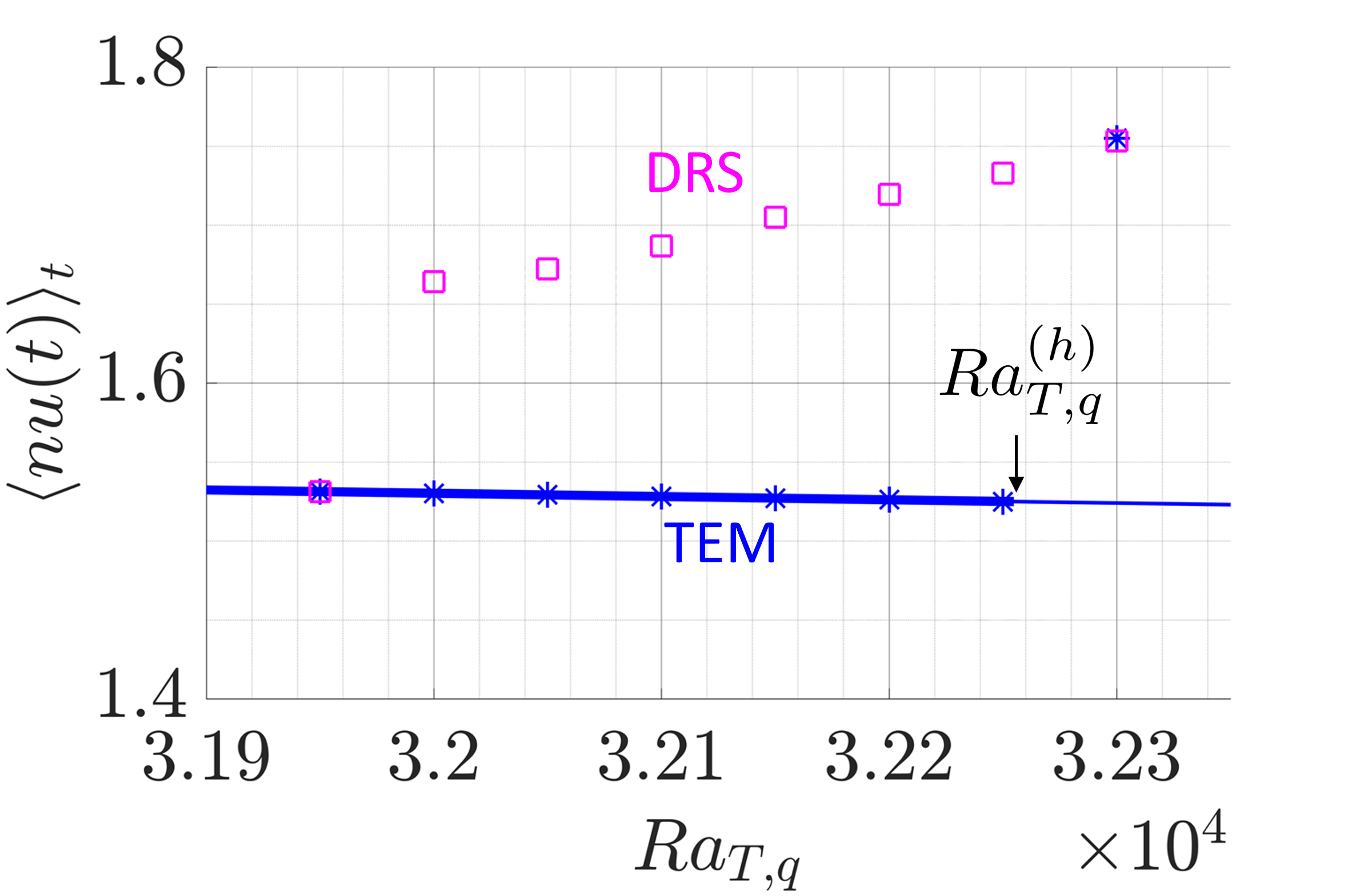}
    \caption{Same as figure \ref{fig:bif_diag_full} but obtained from the single-mode equations in Eqs. \eqref{eq:single_mode}.}
 \label{fig:bif_diag_single_mode}
\end{figure}

\begin{table}
    \centering
    \begin{tabular}{cccccc}
    \hline
        & $Ra_{T,q}^{(p)}$ & $Ra_{T,q}^{(s)}$ & $Ra_{T,q}^{(h)}$ & $\omega_h$ & $Ra_{T,q}^{(g)}$\\
        \hline
        2D full equations in \eqref{eq:NS} & $10^4$ &  19576.3 & 32085.1 & 43.1 & 46892.0 \\
        Single-mode equations in \eqref{eq:single_mode} & $10^4$ & 19291.3 & 32254.8 & 43.5 &  46761.1 \\
    \hline
    \end{tabular}
    \caption{Comparison of the bifurcation points between the full 2D equations in \eqref{eq:NS} with $L_x=0.2\pi$ and the single-mode equations in \eqref{eq:single_mode} with $k_x=10$, including the primary bifurcation $Ra_{T,q}^{(p)}$, the secondary bifurcation $Ra_{T,q}^{(s)}$, the Hopf bifurcation $Ra_{T,q}^{(h)}$ with Hopf frequency $\omega_h$, and the global bifurcation $Ra_{T,q}^{(g)}$, all at $Pr=1$.}
    \label{tab:bif_points_freq_compare}
\end{table}

\begin{figure}
(a) \hspace{0.49\textwidth} (b)

    \centering
    \includegraphics[width=0.49\textwidth]{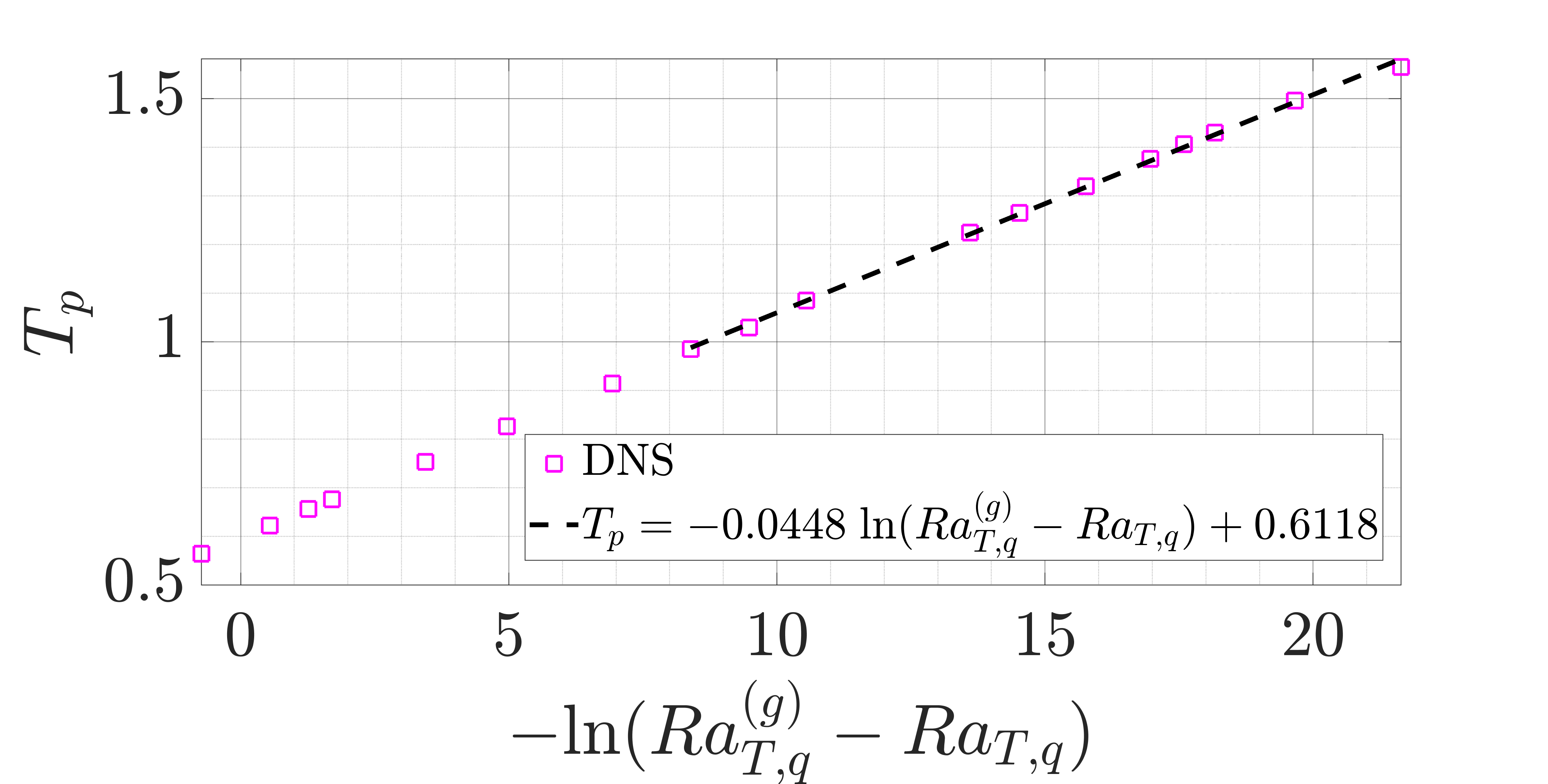}
    \includegraphics[width=0.49\textwidth]{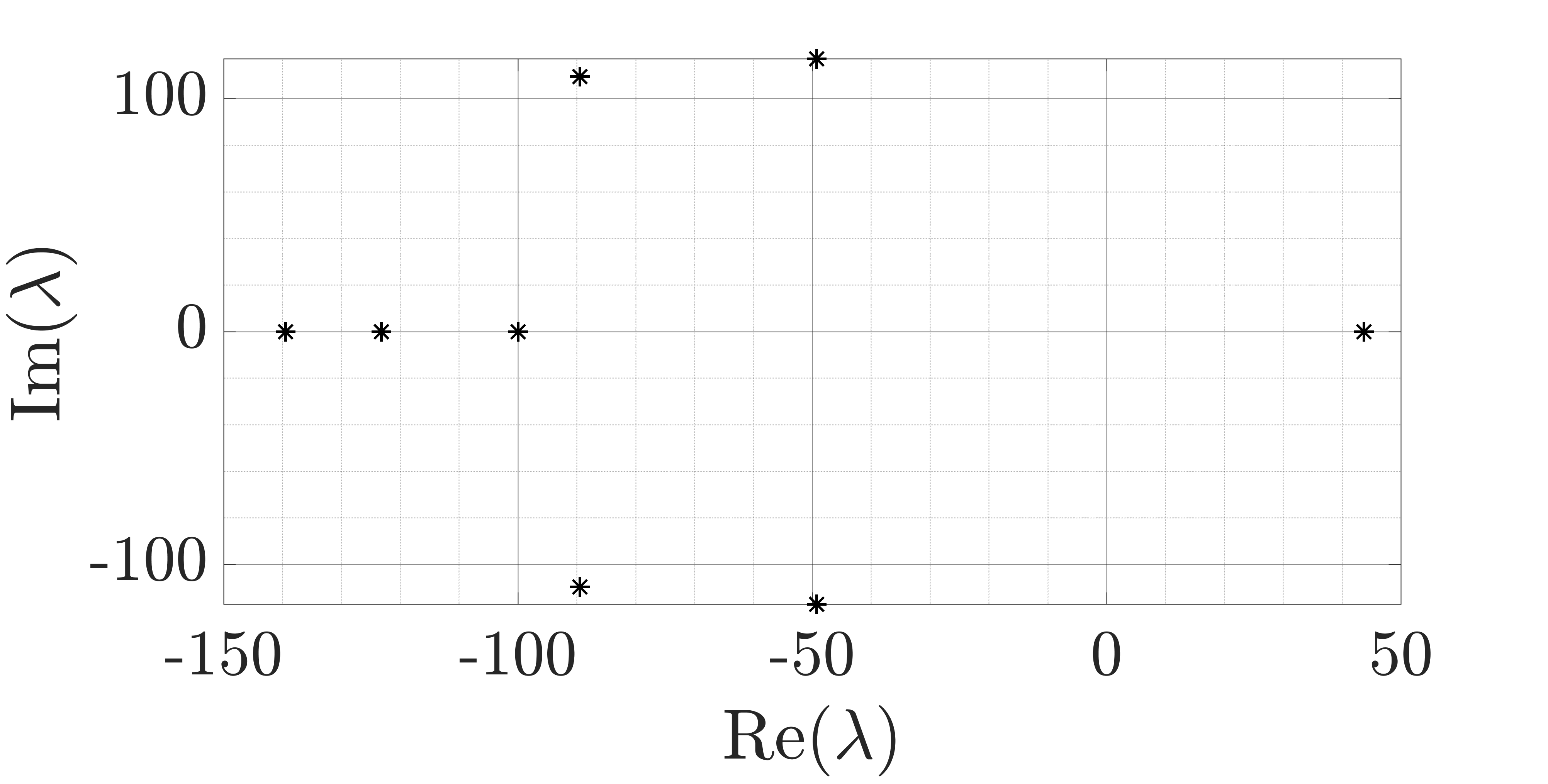}
    \caption{(a) The reversal period $T_p$ as a function of $Ra_{T,q}$ near the global bifurcation $Ra_{T,q}^{(g)}$ obtained from the single-mode equations in \eqref{eq:single_mode} at $Pr=1$ and $k_x=10$. The black dashed line is $T_p=-0.0448\;\text{ln}(Ra_{T,q}^{(g)}-Ra_{T,q})+0.6118$ and fits the DNS data with a relative residue as $0.4\%$, where $Ra_{T,q}^{(g)}=46761.0819762429$. (b) The eigenvalues of the (unstable) elevator mode at $Ra_{T,q}^{(g)}$ within the single-mode equations \eqref{eq:single_mode}.}
    \label{fig:freq_Ra_global}
\end{figure}

Figure \ref{fig:bif_diag_single_mode} shows (a) the bifurcation diagram and (b) the hysteresis diagram obtained from the single-mode equations in \eqref{eq:single_mode}. Here, we can see that the single-mode equations reproduce the bifurcation and hysteresis diagrams obtained from the full equations in 2D shown in figure \ref{fig:bif_diag_full}. The hysteresis behavior in figure \ref{fig:bif_diag_single_mode}(b) is present in a similar Rayleigh number range, $\Delta Ra_{T,q}\approx200$, as in figure \ref{fig:bif_diag_full}(b). Nevertheless, the bifurcation points are slightly shifted in the single-mode equations compared with the full equations as shown in table \ref{tab:bif_points_freq_compare}. The success of the single-mode equations in predicting the Hopf frequency is perhaps in the same spirit as the real zero imaginary frequency (RZIF) ansatz that has shown success in predicting the oscillation frequency in nonlinear thermosolutal convection and shear flow \citep{turton2015prediction,bengana2019bifurcation,bengana2021frequency}. Within the RZIF framework, the eigenvalues are computed based on dynamics linearized around a mean flow that can deviate from the laminar base flow, much as here the single-mode equations employ the large-scale modes $\bar{T}_0$ and $\bar{U}_0$ with superposed harmonics; see Eqs. \eqref{eq:single_mode_U_0} and \eqref{eq:single_mode_T_0}.  

We further leverage the computational efficiency of single-mode equations to analyze the frequency scaling near the global bifurcation. We perform a bisection over $Ra_{T,q}$ using DNS of the single-mode equations in \eqref{eq:single_mode} to identify the global bifurcation point with more significant digits, $Ra_{T,q}^{(g)}=46761.0819762429$, than possible from the full equations. Figure \ref{fig:freq_Ra_global} shows that the period $T_p$ of the direction reversals near $Ra_{T,q}^{(g)}$ diverges as $T_p=-0.0448\;\text{ln}(Ra_{T,q}^{(g)}-Ra_{T,q})+0.6118$, cf.~\citet{knobloch1981nonlinear,knobloch1986oscillatory}.

Since the direction-reversing state collides with an unstable elevator mode at the global bifurcation $Ra_{T,q}^{(g)}$ as indicated in figure \ref{fig:bif_diag_single_mode}(a), it is instructive to compute the eigenvalues of the elevator mode at this point within the single-mode equations \eqref{eq:single_mode}. Figure \ref{fig:freq_Ra_global}(b) shows that the unstable eigenvalue, $\lambda_1=43.69$, is real and that the least stable eigenvalues are complex, with $\lambda_{2,3}=-\rho \pm \text{i}\,\omega=-49.27 \pm \text{i}\,117.11$. Thus, this global bifurcation is associated with a saddle-focus equilibrium with $\delta\equiv\rho/\lambda_1=1.13>1$, i.e. the tame version of the Shil'nikov bifurcation \citep{shilnikov1965case,shilnikov2007shilnikov}. Here, we report these eigenvalues from the single-mode equations \eqref{eq:single_mode} to make a direct comparison with the logarithmic scaling law in figure \ref{fig:freq_Ra_global}(a) later. For the 2D full equations, the corresponding eigenvalues of the elevator mode at the corresponding $Ra_{T,q}^{(g)}$ are $\lambda_1=41.11$, $\lambda_{2,3}=-\rho\pm\text{i}\omega=-56.67\pm \text{i}\,114.67$, leading to a Shil'nikov bifurcation with $\delta=1.38$. 

The logarithmic scaling law and associated coefficient can be predicted by constructing a Poincar\'e map near the global bifurcation point $Ra_{T,q}^{(g)}$ and the saddle-focus equilibrium (here the steady elevator mode) by composing a local map near this saddle focus and a global map \citep{shilnikov2007shilnikov,champneyshomoclinic} as done by \citet{glendinning1984local}. For the local map, we consider the flow linearized around this elevator mode with $\mu:=Ra_{T,q}^{(g)}-Ra_{T,q}\ll 1$,
\begin{subequations}
\label{eq:global_linearized}
    \begin{align}
    \dot{\zeta}=&\lambda_1 \zeta +\text{h.o.t.}, \\
    \dot{\theta}=&\omega+\text{h.o.t.},\\
    \dot{r}=&-\rho r  +\text{h.o.t.},
\end{align}
\end{subequations}
where $\zeta$ is the coordinate corresponding to the unstable eigenvalue and $(r,\theta)$ are the polar coordinates associated with the least stable eigenvalues. Here, h.o.t. in \eqref{eq:global_linearized} refers to higher order terms. We consider the Poincar\'e section $\Sigma^{in}:=\{\theta=0\}$ and $\Sigma^{out}:=\{\zeta=H\}$ and construct the local map $\Pi_{loc}:\Sigma^{in}\rightarrow \Sigma^{out}$ according to the linearized dynamics in \eqref{eq:global_linearized}:
\begin{subequations}
\label{eq:local_all}
    \begin{align}
    H=\zeta(T_f)=&\zeta_0 e^{\lambda_1 T_f}, \label{eq:local_zeta} \\
    r(T_f)=&r_0e^{-\rho T_f},\\
    \theta(T_f)=&\omega T_f.
\end{align}
\end{subequations}
From \eqref{eq:local_zeta}, the time of flight $\Sigma^{in}:=\{\theta=0\}\to\Sigma^{out}:=\{\zeta=H\}$ is
\begin{align}
    T_f=-\frac{1}{\lambda_1}\ln \left( \frac{\zeta_0}{H}\right).
    \label{eq:time_of_flight}
\end{align}
Substituting \eqref{eq:time_of_flight} into \eqref{eq:local_all} we obtain the local map:
\begin{align}
    \Pi_{loc}:(r,\theta,\zeta)\rightarrow \left( r_0\left(\frac{\zeta_0}{H} \right)^{\delta},\frac{\omega}{\lambda_1}\ln\left(\frac{H}{\zeta_0}\right),H \right).
    \label{eq:local_map}
\end{align}
The global map $\Pi_{global}:\Sigma^{out}\rightarrow \Sigma^{in}$ is obtained from a Taylor series around the homoclinic orbit assumed to be present at $\mu=0$:
\begin{align}
  \Pi_{global}:  (r,\theta,h)\rightarrow (\bar{r}+a\mu+br\cos \theta +cr\sin\theta,0,d\mu+er\cos\theta+fr\sin\theta)+\text{h.o.t.},
  \label{eq:global_map}
\end{align}
where $a$, $b$, $c$, $d$, $e$, and $f$ are constants. By composing the local and global maps ($\Pi:\Sigma^{in}\rightarrow \Sigma^{in}=\Pi_{global}\circ \Pi_{loc}$), we obtain the Poincar\'e map:
\begin{align}
    \Pi: \begin{bmatrix}
        r\\
        \zeta
    \end{bmatrix}\rightarrow \begin{bmatrix}
        \bar{r}\\
        0
    \end{bmatrix}+\begin{bmatrix}
        a\\
        d
 \end{bmatrix}\mu+\begin{bmatrix}
        c_1 r \zeta^{\delta}\cos(k_1 \ln \zeta +\phi_1)\\
        c_2 r \zeta^{\delta} \cos(k_2 \ln \zeta +\phi_2)
    \end{bmatrix}+\text{h.o.t.},
    \label{eq:poincare_map}
\end{align}
where $c_i$, $k_i$ and $\phi_i$ ($i=1,2$) are constants. We may now search for a fixed point of the Poincar\'e map $\Pi$ that corresponds to the periodic orbit near $Ra_{T,q}^{(g)}$ in the original system. This point is approximated by the fixed point of the 1D map:
\begin{align}
    \zeta-d \mu=\Phi(\zeta)\equiv c_2 r \zeta^{\delta}\cos(k_2\ln \zeta +\phi_2)+\text{h.o.t.}
    \label{eq:fixed_point_poincare}
\end{align}
When $\delta>1$ there is a unique fixed point of \eqref{eq:fixed_point_poincare}, which scales as $\zeta\sim d\mu$ near the global bifurcation $\mu\rightarrow 0$. Thus, based on \eqref{eq:time_of_flight} and the assumption that the global return is much faster than the local passage past the fixed point, the period of the reversing orbit just before the global bifurcation scales as 
\begin{align}
    T_p=-\frac{2}{\lambda_1} \ln \mu +\text{const}=-\frac{2}{\lambda_1}\ln \left(Ra_{T,q}^{(g)}-Ra_{T,q}\right)+\text{const}.
\end{align}
Here the factor 2 arises because the orbit makes two passes near the fixed point in each reversal period. Using $\lambda_1$ from figure \ref{fig:freq_Ra_global}(b), this calculation predicts that $2/\lambda_1=0.04578$, a coefficient that is almost exactly that obtained from the fit to the simulation data in figure \ref{fig:freq_Ra_global}(a).

\section{Dynamics at high Rayleigh numbers}
\label{sec:high_Ra}

In this section, we study the dynamics at higher Rayleigh numbers, where chaotic behavior appears. We first analyze the secondary instability of the elevator mode, which continues to play an important role in the high Rayleigh number regime. We focus on the 2D elevator mode with a horizontal wavenumber $k_x=k_e$ in $x$ direction,
\begin{align}
    \bar{W}_e(x)=\hat{w}_e \exp(\text{i}k_e x)+c.c.,\quad
    \bar{T}_e(x)=\hat{T}_e\exp(\text{i}k_e x)+c.c.
\end{align}
and solution amplitude given by \eqref{eq:amplitude_elevator_mode}. The decomposition 
\begin{align}
    \boldsymbol{u}=\bar{W}_e\boldsymbol{e}_z+\boldsymbol{u}',\quad T=\bar{T}_e+T'
\end{align}
leads to the linearized equations:
\begin{subequations}
\label{eq:secondary}
    \begin{gather}
    \frac{\partial \boldsymbol{u}'}{\partial t}+u'\frac{d \bar{W}_e}{dx}\,\boldsymbol{e}_z+\bar{W}_e\partial_z \boldsymbol{u}'=-\boldsymbol{\nabla}p'+PrRa_{T,q}T'\boldsymbol{e}_z+Pr\nabla^2 \boldsymbol{u}',\label{eq:secondary_momentum}\\
    \boldsymbol{\nabla}{\cdot} \boldsymbol{u}'=0,\label{eq:secondary_mass}\\
    \frac{\partial T'}{\partial t}+u'\frac{d \bar{T}_e}{dx}+\bar{W}_e\partial_z T'-w'+w'\langle \bar{W}_e\bar{T}_e\rangle_{h,v}=\nabla^2 T'.\label{eq:secondary_T}
\end{gather}
\end{subequations}
The cubic flux-feedback nonlinearity $w\langle wT\rangle_{h,v}$ in \eqref{eq:NS_T} generates three linearized terms:
\begin{align}
    w'\langle \bar{W}_e\bar{T}_e\rangle_{h,v}+\bar{W}_e\langle w'\bar{T}_e\rangle_{h,v}+\bar{W}_e\langle \bar{W}_eT'\rangle_{h,v},
\label{eq:flux_linearization}
\end{align}
with the latter two terms in \eqref{eq:flux_linearization} vanishing for $k_z\neq 0$. As a result, only the term $ w'\langle \bar{W}_e\bar{T}_e\rangle_{h,v}$ originating from flux feedback appears in the linearized equation in \eqref{eq:secondary_T}. The normal mode assumption in general 3D form
\begin{subequations}
\begin{align}
\boldsymbol{u}'=\tilde{\boldsymbol{u}}(x) \exp[\text{i}(k_y y+k_z z)+\lambda t]+c.c.,\\
     T'=\tilde{T}(x)\exp[\text{i}(k_y y+k_z z)+\lambda t]+c.c.
\end{align}
\end{subequations}
contains the coefficients $\tilde{\boldsymbol{u}}(x)$ and $\tilde{T}(x)$ that depend on $x$ because the base flow (elevator mode) also depends on $x$. In terms of the horizontal vorticity $\tilde{\omega}_x:=\text{i}k_y \tilde{w}-\text{i}k_z\tilde{v}$ we have the linear eigenvalue problem 
\begin{align}
    &\lambda \begin{bmatrix}
        \tilde{u}\\
        \tilde{\omega}_x\\
        \tilde{T}
    \end{bmatrix}
    =
    \begin{bmatrix}
   \mathcal{A}_{11} & \mathcal{A}_{12} & \mathcal{A}_{13}\\
   \mathcal{A}_{21} & \mathcal{A}_{22} & \mathcal{A}_{23} \\
   \mathcal{A}_{31} & 
    \mathcal{A}_{32} & \mathcal{A}_{33}
    \end{bmatrix}
    \begin{bmatrix}
        \tilde{u}\\
    \tilde{\omega}_x\\
        \tilde{T}
    \end{bmatrix}=:\mathcal{A}\begin{bmatrix}
        \tilde{u}\\
    \tilde{\omega}_x\\
        \tilde{T}
    \end{bmatrix},
    \label{eq:secondary_linearization}
\end{align}
where
\begin{subequations}
\begin{align}
    \mathcal{A}_{11}&= \tilde{\nabla}^{-2}(-\text{i}k_z \bar{W}_e \tilde{\nabla}^2+\text{i}k_z\frac{d^2\bar{W}_e}{dx^2}+Pr \tilde{\nabla}^4),\\
    \mathcal{A}_{12}&=0,\\
    \mathcal{A}_{13}&= \tilde{\nabla}^{-2}PrRa_{T,q}(-\text{i}k_z\partial_x),\\
    \mathcal{A}_{21}&= -\text{i}k_y \frac{d\bar{W}_e}{dx},\\
    \mathcal{A}_{22}&=-\text{i}k_z \bar{W}_e+Pr\tilde{\nabla}^2,\\
    \mathcal{A}_{23}&=PrRa_{T,q}\text{i}k_y,\\
    \mathcal{A}_{31}&= -\frac{d\bar{T}_e}{dx}+[1-\langle \bar{W}_e\bar{T}_e\rangle_{h,v}]\frac{\text{i}k_z \partial_x }{k_y^2+k_z^2}, \label{eq:A_31}\\
    \mathcal{A}_{32}&=[1-\langle \bar{W}_e\bar{T}_e\rangle_{h,v}]\frac{-\text{i}k_y}{k_y^2+k_z^2},\label{eq:A_32}\\
    \mathcal{A}_{33}&=-\text{i}k_z \bar{W}_e+\tilde{\nabla}^2,
\end{align}
\end{subequations}
with $\tilde{\nabla}^2:=\partial_x^2-k_y^2-k_z^2$ and $\tilde{\nabla}^4:=\partial_x^4-2(k_y^2+k_z^2)\partial_x^2+(k_y^2+k_z^2)^2$. 

We compare the above formulation with that without the flux feedback, corresponding to setting the integral flux feedback terms $\langle \bar{W}_e\bar{T}_e\rangle_{h,v}$ in \eqref{eq:A_31}-\eqref{eq:A_32} to zero, leading to a modified eigenvalue problem with $\mathcal{A}$ in \eqref{eq:secondary_linearization} replaced by
\begin{align}
    \underline{\mathcal{A}}:=&\begin{bmatrix}
        \mathcal{A}_{11} & \mathcal{A}_{12} & \mathcal{A}_{13}\\
        \mathcal{A}_{21} & \mathcal{A}_{22} & \mathcal{A}_{23}\\
        \underline{\mathcal{A}}_{31} & \underline{\mathcal{A}}_{32} & \mathcal{A}_{33}
    \end{bmatrix},
    \label{eq:A_underline}
\end{align}
where
\begin{subequations}
    \begin{align}
     \underline{\mathcal{A}}_{31}:= -\frac{d\bar{T}_e}{dx}+\frac{\text{i}k_z \partial_x }{k_y^2+k_z^2},\quad\quad    \underline{\mathcal{A}}_{32}:=\frac{-\text{i}k_y}{k_y^2+k_z^2}.
\end{align}
\end{subequations}
The horizontal direction is discretized using a Fourier collocation method with the horizontal derivative computed using a Fourier differentiation matrix \citep{weideman2000matlab}. The numerical implementation is validated against Floquet-based linear stability analysis \citep{holyer1984stability,radko2016thermohaline,garaud2015stability,garaud2019interaction}. We choose the horizontal domain $L_x$ to contain one or more wavelengths of the elevator wavelength $2\pi/k_e$. For all the results reported here, we take $k_y=0$ corresponding to a 2D configuration. 

\begin{figure}
(a) \hspace{0.49\textwidth} (b)

    \centering
    \includegraphics[width=0.49\textwidth]{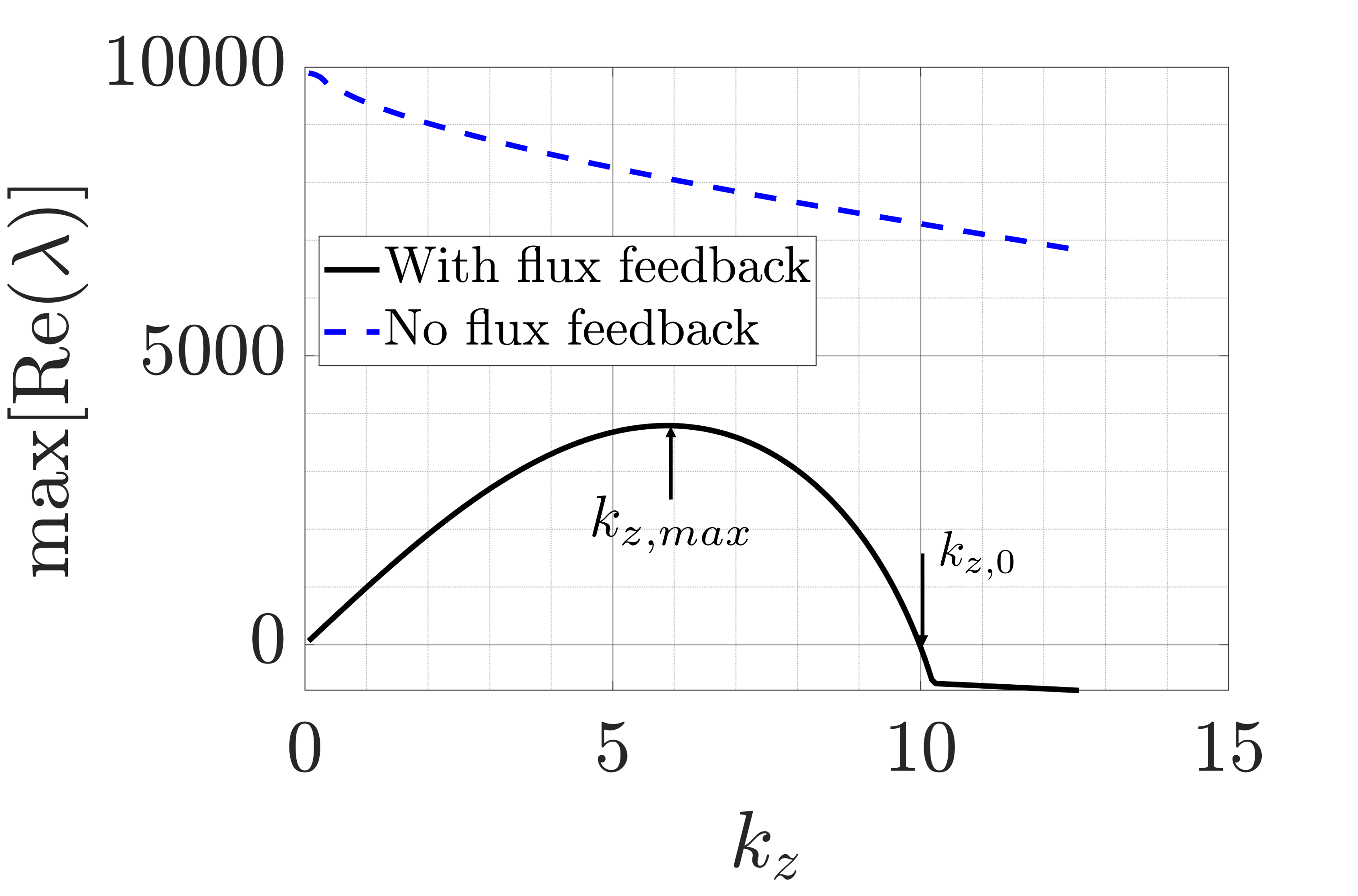}
    \includegraphics[width=0.49\textwidth]{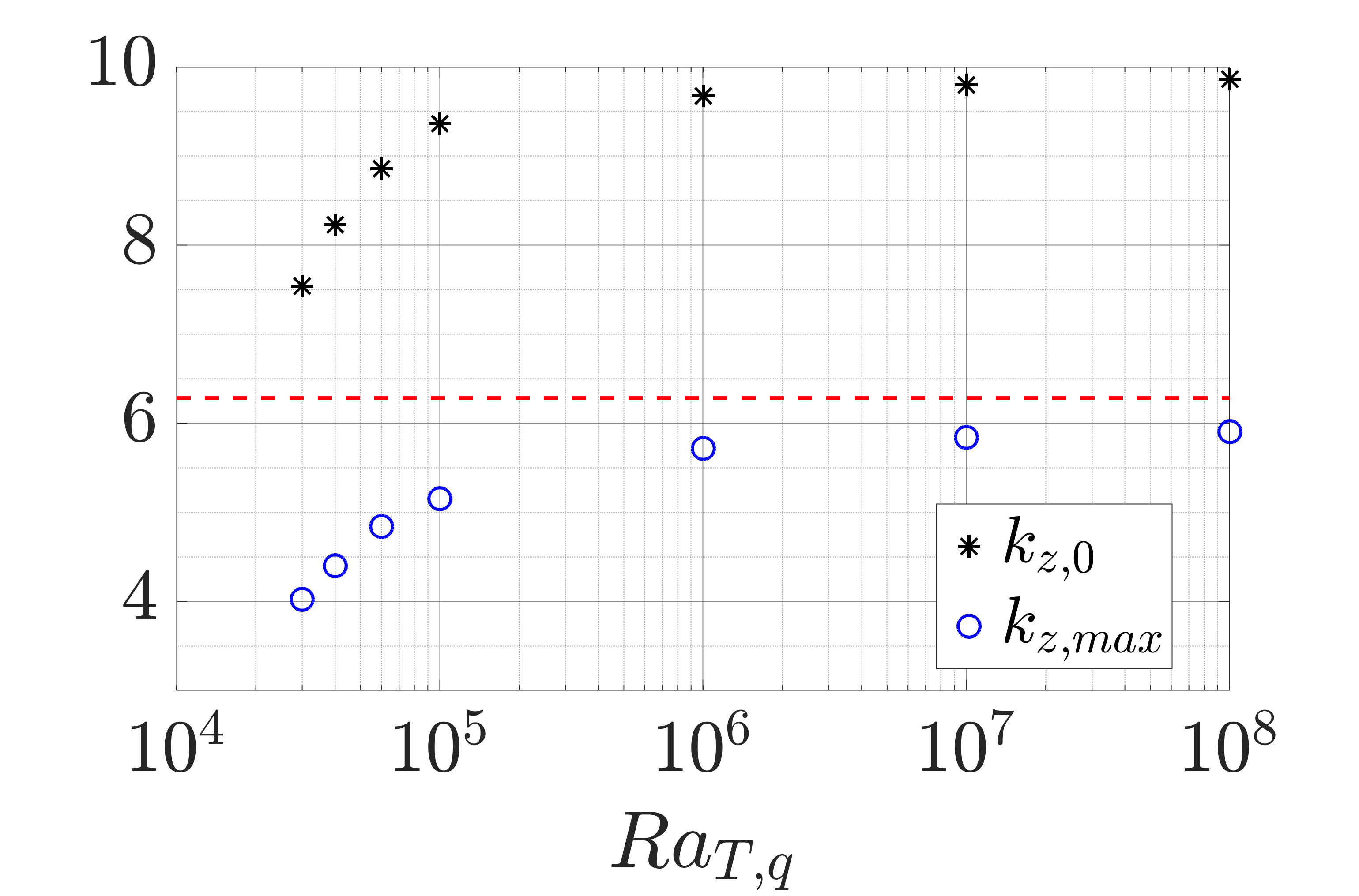}
    \caption{(a) The growth rate max[Re($\lambda$)] at $Ra_{T,q}=10^8$, $Pr=1$, $L_x=0.2\pi$ and $k_e=10$ with flux feedback computed from $\mathcal{A}$ in \eqref{eq:secondary_linearization} and without flux feedback, as computed from $\underline{\mathcal{A}}$ in \eqref{eq:A_underline}. (b) The wavenumbers $k_{z,0}$ and $k_{z,max}$ as a function of $Ra_{T,q}$ at $Pr=1$, $L_x=0.2\pi$, and $k_e=10$. The red dashed line ({\color{red}- -}, red) corresponds to $k_z=2\pi$.}
\label{fig:secondary_growth_rate_no_flux_com}
\end{figure}

Figure \ref{fig:secondary_growth_rate_no_flux_com}(a) shows the growth rate max[Re($\lambda$)] comparing the fixed-flux case computed from $\mathcal{A}$ in \eqref{eq:secondary_linearization} with the case without flux feedback computed from $\underline{\mathcal{A}}$ in \eqref{eq:A_underline}. The flux feedback leads to $\lambda=0$ at $k_z=0$ and the instability is limited to a small range of $k_z$, a feature widely observed in systems with a conservation law \citep{matthews2000pattern}. For the case without flux feedback, its growth rate is larger and decays to zero only at much higher wavenumbers $k_z$ (not shown in figure \ref{fig:secondary_growth_rate_no_flux_com}(a)). We further identify the wavenumbers $k_{z,0}$ and $k_{z,max}$ indicated in figure \ref{fig:secondary_growth_rate_no_flux_com}(a) corresponding, respectively, to zero growth rate and maximum growth rate:
\begin{subequations}
    \begin{align}
    k_{z,0}&:\; \lambda(k_{z,0})=0 \text{ with } k_{z,0}\neq 0,\\
    k_{z,max}&:=\underset{k_z}{\text{arg max}} \;\text{Re}(\lambda).
    \end{align}
\end{subequations}
Figure \ref{fig:secondary_growth_rate_no_flux_com}(b) displays $k_{z,0}$ and $k_{z,max}$, both of which increase as $Ra_{T,q}$ increases, with $k_{z,max}<k_{z,0}$. Here, we also plot $k_{z}=2\pi$ which is the smallest nontrivial vertical wavenumber that fits in the $L_z=1$ domain. The secondary instability wavelength is not required to lie within $L_z=1$ and thus a comparison of $k_{z,0}$ with $k_z=2\pi$ determines whether the secondary instability can occur within the domain, or whether it is suppressed by its finite size. Figure \ref{fig:secondary_growth_rate_no_flux_com}(b) shows that $k_{z,0}>2\pi$ when $L_x=0.2\pi$ and $Ra_{T,q}\geq 3\times 10^4$, a result consistent with the presence of a secondary instability in a $L_z=1$ domain identified in Section \ref{sec:moderate_Ra}. 

\begin{figure}
\hspace{0.05\textwidth} (a) $\langle u\rangle_{h}(z,t)$ \hspace{0.44\textwidth} (b)
\hspace{0.14\textwidth} (c)
    
    \centering
    \includegraphics[width=0.58\textwidth]{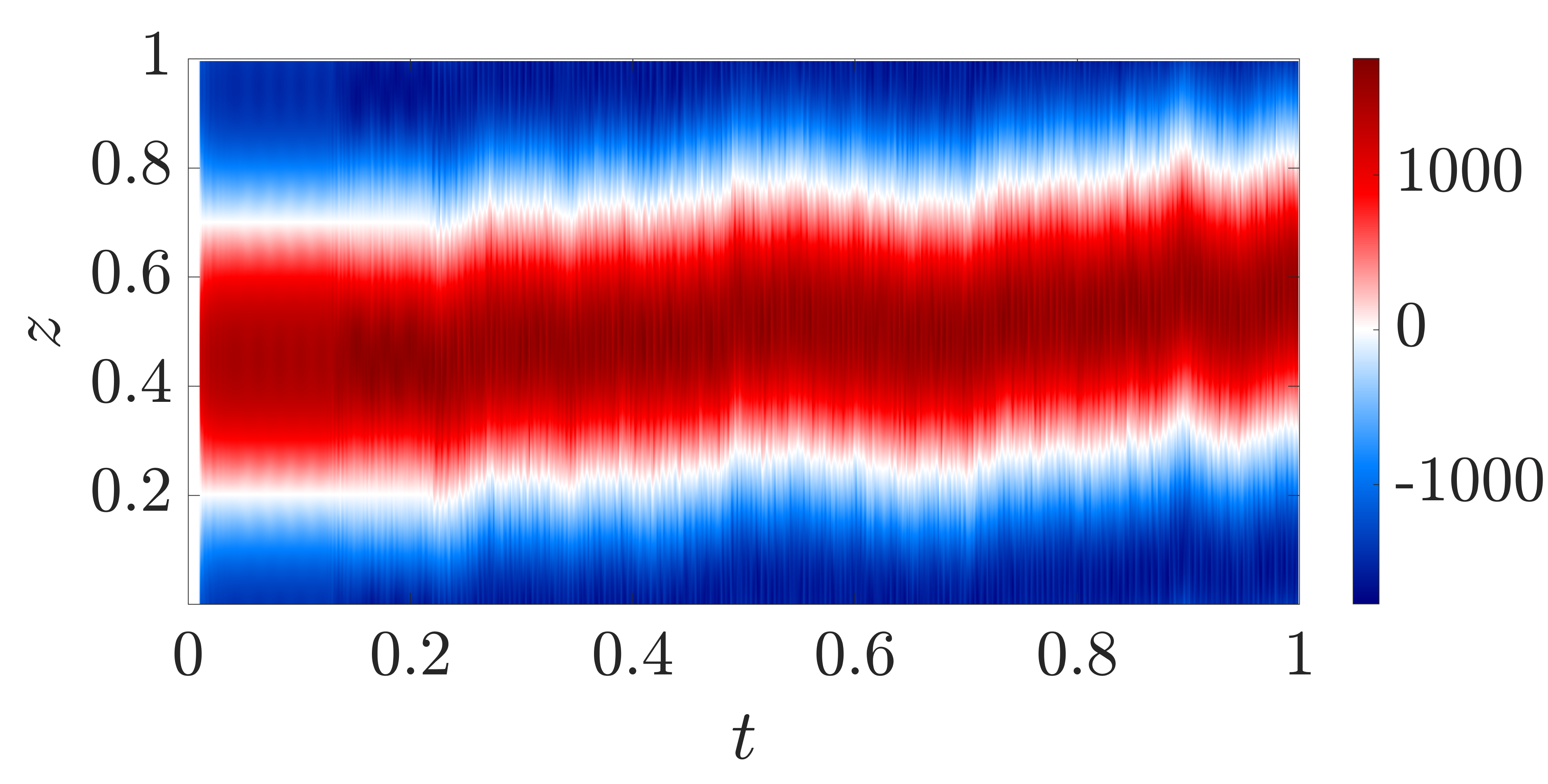}
    \includegraphics[width=0.33\textwidth,trim=-0 -0.75in 0 0]{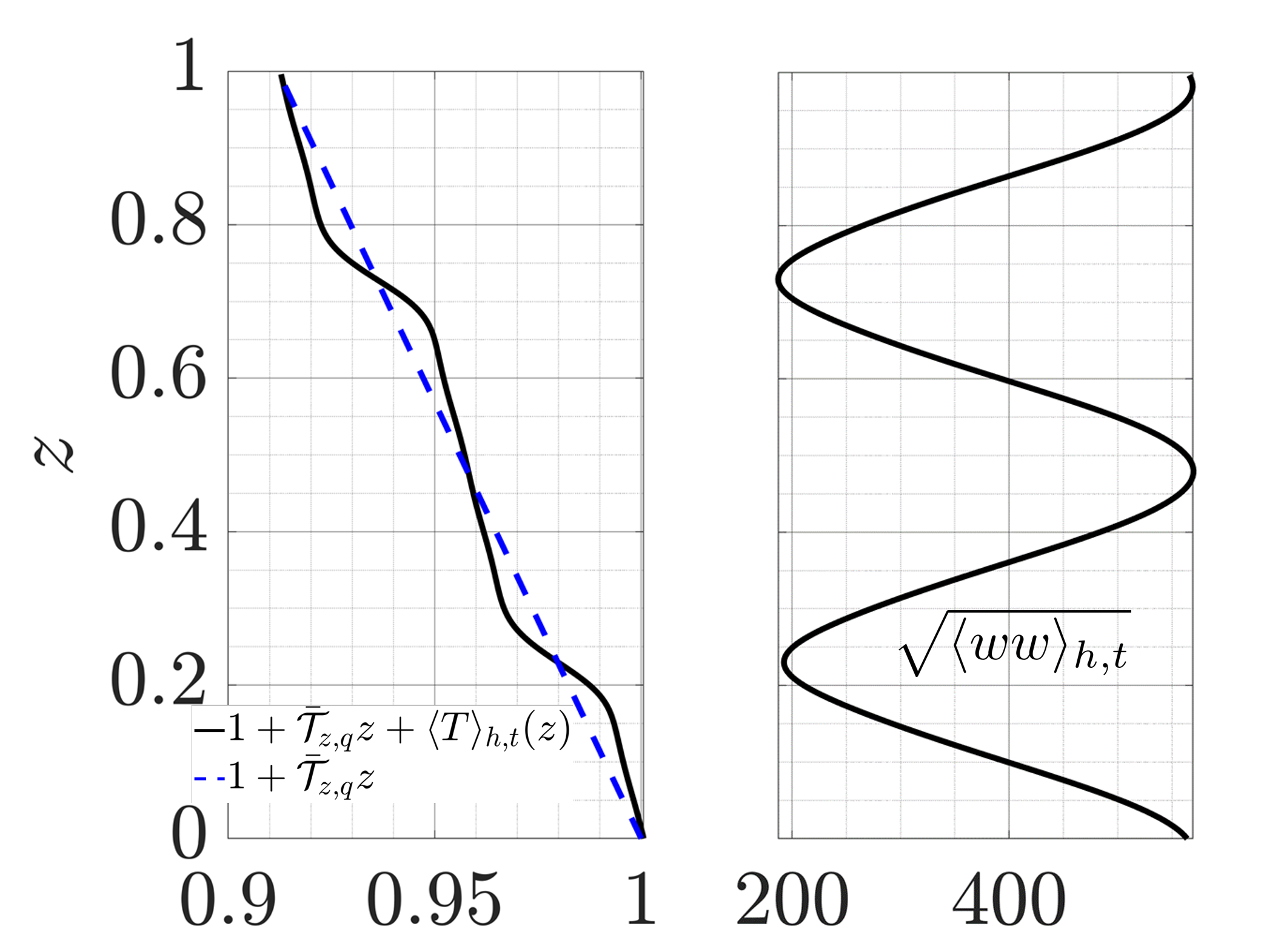}
    \caption{(a) Large-scale shear $\langle u\rangle_h(z,t)$, (b) total temperature $1{+} \bar{\mathcal{T}}_{z,q} z{+}\langle T\rangle_{h,t}(z)$ ($\mline$) compared with the linear profile $1{+} \bar{\mathcal{T}}_{z,q} z$ ({\color{blue}$\dashed$}) and (c) r.m.s. vertical velocity $\sqrt{\langle ww\rangle_{h,t}}$, the latter two averaged over $t\in [0.275,0.465]$, at $Ra_{T,q}=10^8$, $Pr=1$, and $L_x=0.2\pi$.   }
    \label{fig:DNS_Ra_Tq_1e8_Pr_1}
\end{figure}

The secondary instability of the elevator mode and the modulated traveling waves generated through the sequence of bifurcations examined in Section \ref{sec:moderate_Ra} continue to play an important role at larger Rayleigh numbers. Figure \ref{fig:DNS_Ra_Tq_1e8_Pr_1} displays DNS results at $Ra_{T,q}=10^8$ obtained with $N_x=N_z=256$ grid points ($N_x=N_z=512$ grid points generate the same behavior). The large-scale shear $\langle u\rangle_{h}(z,t)$ in figure \ref{fig:DNS_Ra_Tq_1e8_Pr_1}(a) is now dominated by modulated traveling waves, but displays chaotic behavior. In addition, it slowly migrates in the vertical direction, behavior that is permitted by the periodic boundary conditions in the vertical.

The mean total temperature averaged over $t\in [0.275,0.465]$ in figure \ref{fig:DNS_Ra_Tq_1e8_Pr_1}(b) exhibits a deviation from a linear profile similar to canonical Rayleigh-B\'enard convection. Moreover, in the region where the mean temperature deviation $\langle T\rangle_{h,t}(z)$ is close to zero, the corresponding large-scale shear $\langle u\rangle_{h}(z,t)$ also vanishes (white regions in figure \ref{fig:DNS_Ra_Tq_1e8_Pr_1}(a)) and the r.m.s. vertical velocity $\sqrt{\langle ww\rangle_{h,t}}$ displays local minima with a zero vertical derivative as shown in figure \ref{fig:DNS_Ra_Tq_1e8_Pr_1}(c). The local maxima of the r.m.s. vertical velocity $\sqrt{\langle ww\rangle_{h,t}}$ correspond instead to the mixed mean temperature regions in figure \ref{fig:DNS_Ra_Tq_1e8_Pr_1}(b). This behavior resembles that of RBC in a bounded domain with constant temperature boundaries and no-slip instead of stress-free velocity boundary conditions at the top and bottom \citep{van2014effect}. The figure shows that the fixed-flux constraint suppresses bursting behavior compared with homogeneous RBC driven by a constant temperature gradient \citep{borue1997turbulent,lohse2003ultimate,calzavarini2005rayleigh,calzavarini2006exponentially}. Figure \ref{fig:Ra_T_q_Nu_scaling} shows the Nusselt number scaling within $Ra_{T,q}\in [10^8,10^{10}]$ computed with $N_x=N_z=512$ grid points and $L_x=0.2\pi$, $Pr=1$. Here, the fitted scaling law $Nu=0.189\;Ra_{T,q}^{0.217}$ is associated with an exponent lower than $Nu\sim Ra_{T,q}^{1/3}$ corresponding to ultimate regime scaling $Nu\sim Ra_T^{1/2}$. However, the flow structures associated with figure \ref{fig:Ra_T_q_Nu_scaling} are dominated by large-scale shear with a $\langle u\rangle_h(z,t)$ profile similar to that in figure 10(a), potentially reducing heat transport.

\begin{figure}
    \centering
    \includegraphics[width=0.6\textwidth]{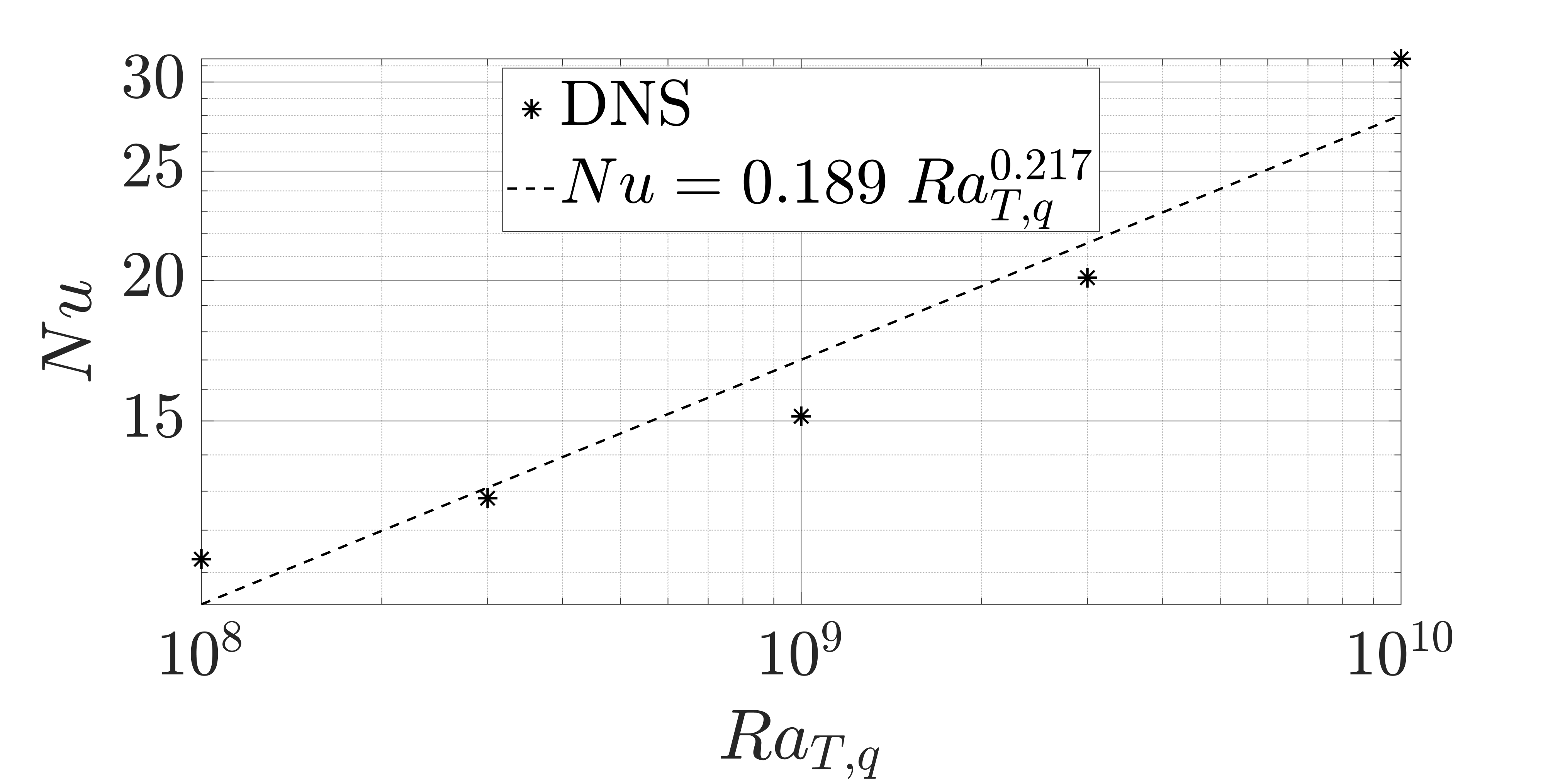}
    \caption{$Nu$ as a function of $Ra_{T,q}$ from DNS with $L_x=0.2\pi$ and $Pr=1$.}
    \label{fig:Ra_T_q_Nu_scaling}
\end{figure}

\begin{figure}
(a) \hspace{0.49\textwidth} (b)

    \centering
    \includegraphics[width=0.49\textwidth]{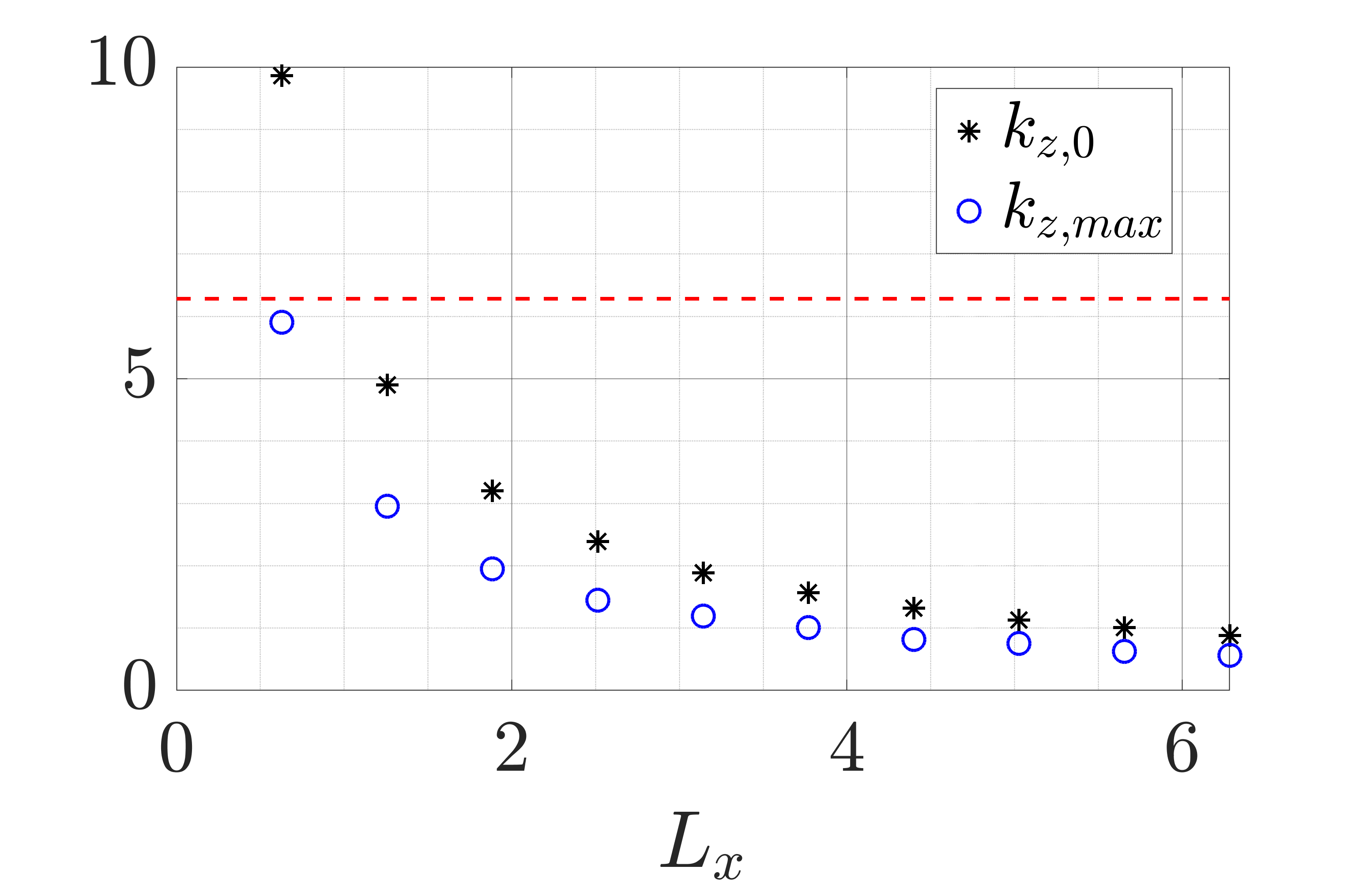}
\includegraphics[width=0.49\textwidth]{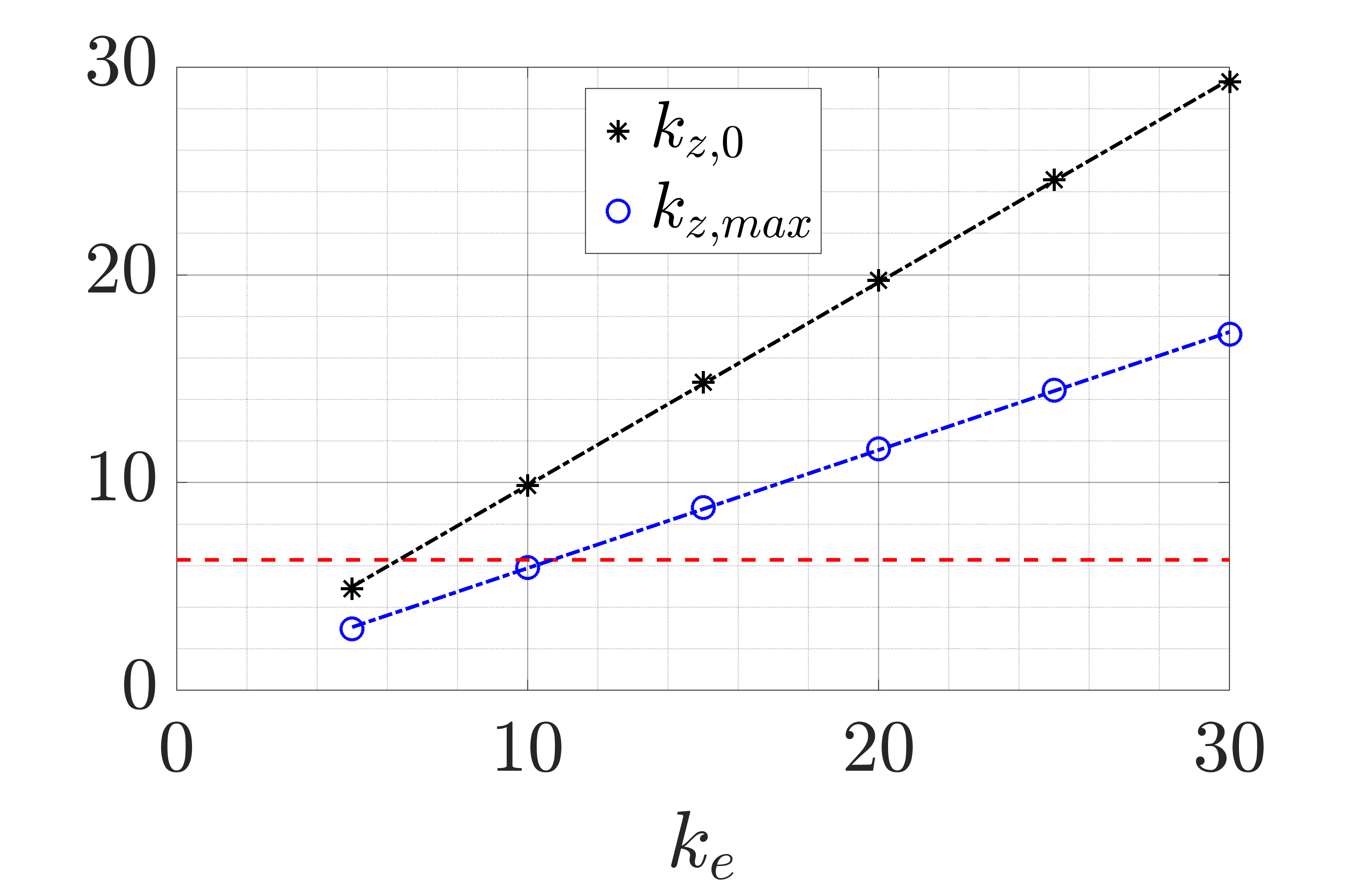}
    \caption{The wavenumbers $k_{z,0}$ and $k_{z,max}$ as a function of (a) $L_{x}$ when $k_e=2\pi/L_x$ and (b) $k_e$ with $L_x=0.4\pi$. The linear fit is $k_{z,0}=0.98k_e+0.10$ ({\color{black}-.-}, black) and $k_{z,max}=0.57k_e+0.20$ ({\color{blue}-.-}, blue). For both panels, ({\color{red}- -}, red) indicates $k_z=2\pi$. Instability requires that $k_{z,0}>2\pi$. Other parameters are $Ra_{T,q}=10^8$ and $Pr=1$. }
\label{fig:secondary_growth_rate_Lx_ke}
\end{figure}

\begin{figure}
(a) $\langle u\rangle_h(z,t)$ with $k_e=20$ \hspace{0.25\textwidth} (b) $\langle u\rangle_h(z,t)$ with $k_e=30$

    \centering
    \includegraphics[width=0.49\textwidth]{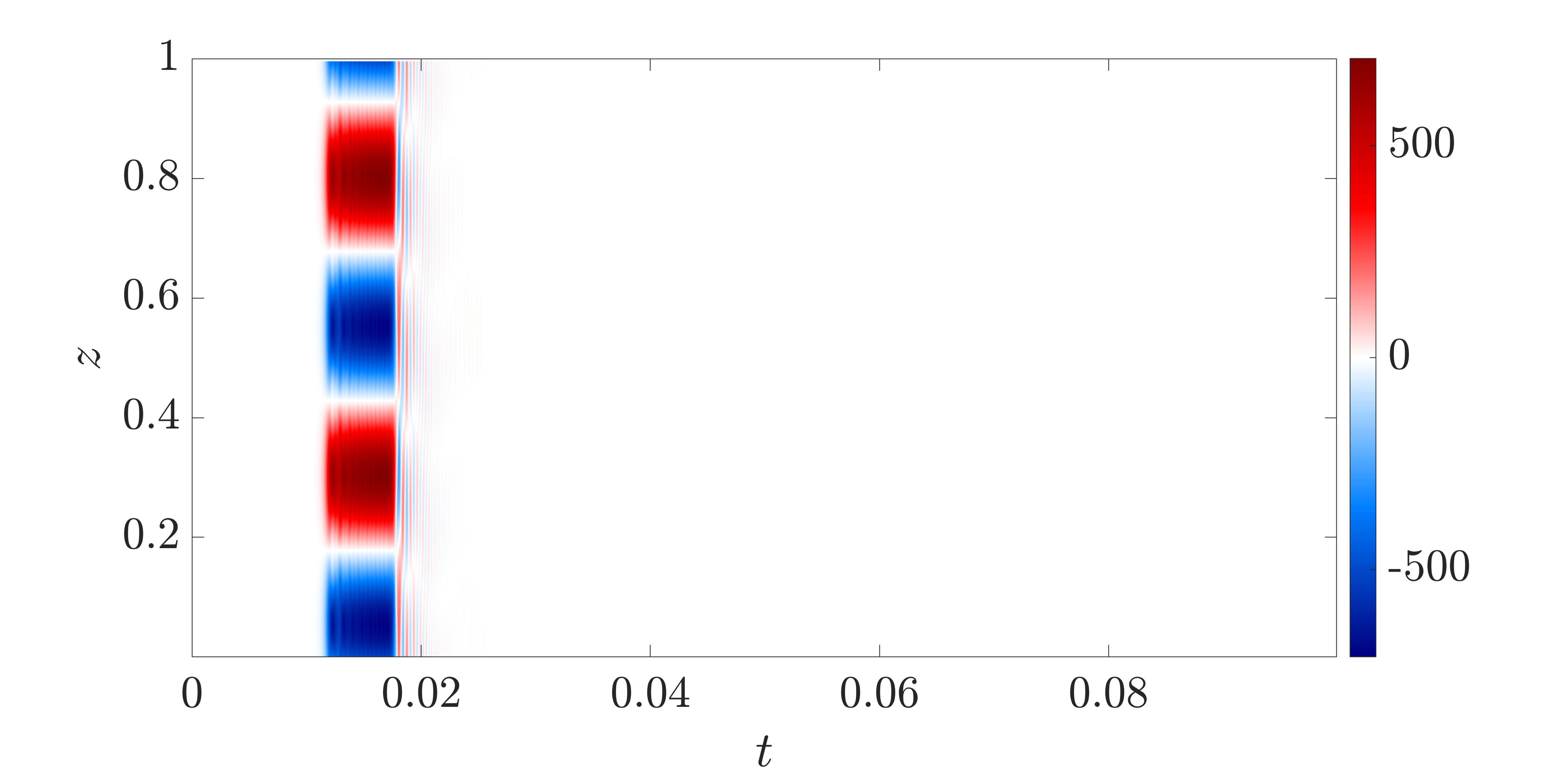}
    \includegraphics[width=0.49\textwidth]{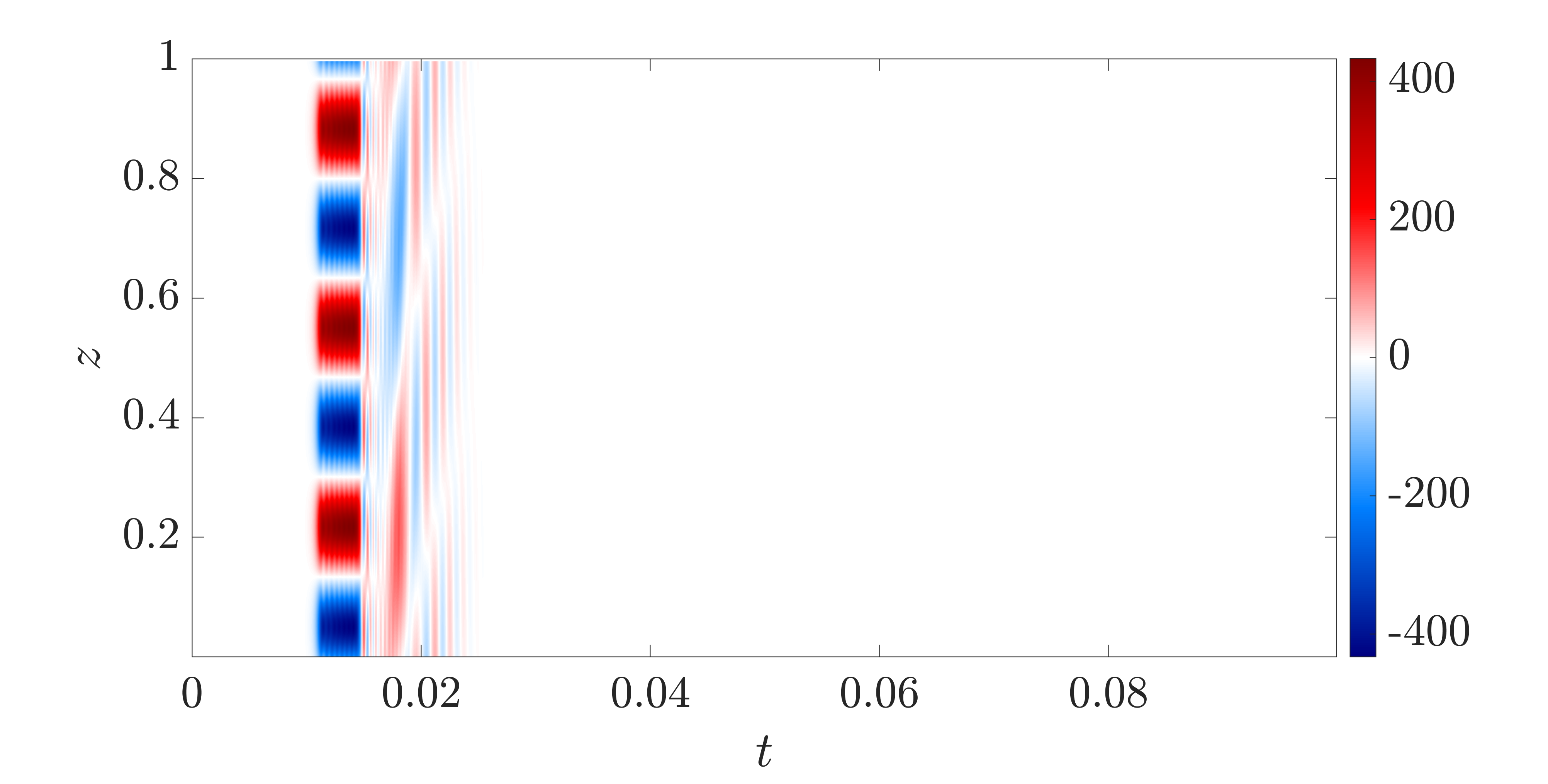}
    \caption{Large-scale shear $\langle u\rangle_h(z,t)$ with initial elevator mode wavenumber (a) $k_e=20$ and (b) $k_e=30$, both at $Ra_{T,q}=10^8$, $Pr=1$ and $L_x=0.4\pi$. }
    \label{fig:DNS_u_h_ke_20_ke_30}
\end{figure}

The impact of the domain size and of the horizontal wavenumber of the elevator mode within the domain is further analyzed in figure \ref{fig:secondary_growth_rate_Lx_ke}(a) which shows $k_{z,0}$ and $k_{z,max}$ as a function of $L_x$ when $k_e=2\pi/L_x$, $Ra_{T,q}=10^8$ and $Pr=1$. As the horizontal domain and wavelength increase both $k_{z,0}$ and $k_{z,max}$ decrease, suggesting a small vertical wavenumber of the secondary instability. With $L_z$ fixed at $L_z=1$ this requires that we choose a small enough horizontal domain such that $k_{z,0}\geq 2\pi$. A similar requirement applies for mean flow generation in canonical Rayleigh-B\'enard convection \citep{rucklidge1996analysis,fitzgerald2014mechanisms,wang2020zonal} and is the reason for choosing $L_x=0.2\pi$ for most of the results in this paper. Figure \ref{fig:secondary_growth_rate_Lx_ke}(b) then fixes $L_x$ at $L_x=0.4\pi$ but varies $k_e$ allowing multiple elevator modes within the domain. Evidently both $k_{z,0}$ and $k_{z,max}$ increase as $k_e$ increases and both are fitted well by a linear scaling law. This trend can be also confirmed in DNS. The associated large-scale shear $\langle u\rangle_h(z,t)$ in figure \ref{fig:DNS_u_h_ke_20_ke_30}(a) shows that $k_z=4\pi$ for an initial elevator mode wavenumber $k_e=20$, while $\langle u\rangle_h(z,t)$ in figure \ref{fig:DNS_u_h_ke_20_ke_30}(b) displays a $k_z=6\pi$ instability associated with $k_e=30$. This observation directly corresponds to the $k_{z,max}$ value at $k_e=20$ and $k_e=30$ predicted by the secondary instability analysis in figure \ref{fig:secondary_growth_rate_Lx_ke}(b). 

\begin{figure}
(a) \hspace{0.48\textwidth} (b)

    \centering
    \includegraphics[width=0.49\textwidth]{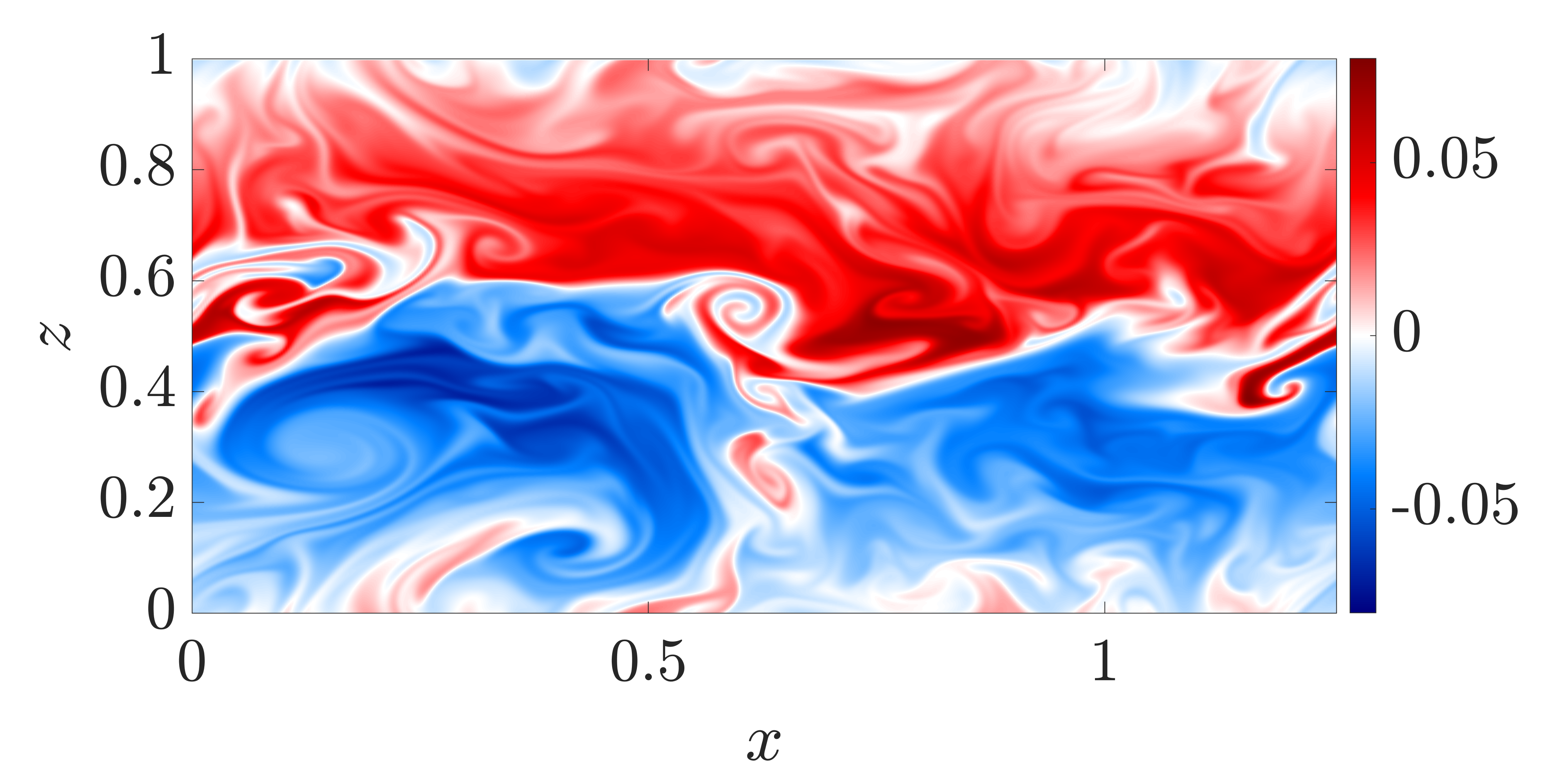}
    \includegraphics[width=0.49\textwidth]{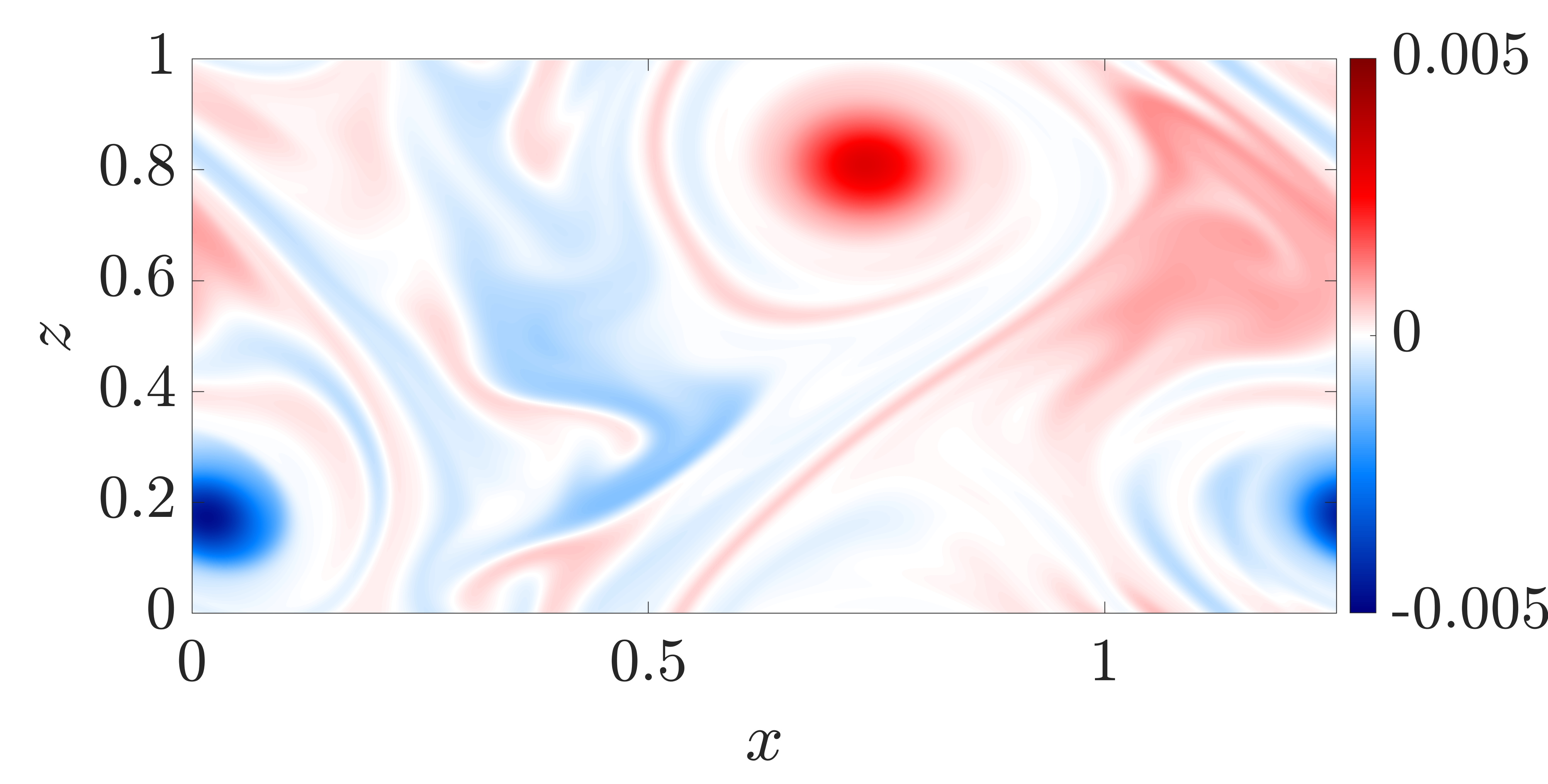}
    \caption{Snapshots of temperature deviation $T(x,z,t)$ at (a) $t=0.03$ and (b) $t=0.032$. Parameters are $Ra_{T,q}=10^{10}$, $L_x=0.4\pi$ and $Pr=1$. }
    \label{fig:snapshot_Ra_Tq_1e10_Lx_0p4_pi}
\end{figure}
In a domain of horizontal size $L_x=0.4\pi$, the final state is a stable domain-filling elevator mode with $k_e=5$ corresponding to the white region in figure \ref{fig:DNS_u_h_ke_20_ke_30} at $Ra_{T,q}=10^8$. A similar transition to a larger horizontal scale is also observed in fixed-flux RBC within both the weakly nonlinear regime \citep{chapman1980nonlinear} and the turbulent regime \citep{vieweg2021supergranule,vieweg2022inverse,kaufer2023thermal}. The stable elevator mode that results suggests that vertical jets (i.e. elevator modes) are favored in wide domains, while horizontal jets are found in narrow domains. This behavior resembles that in rapidly rotating convection where jets parallel to the short side of an anisotropic domain are found \citep{julien2018impact} or in 2D turbulence driven by stochastic forcing \citep{bouchet2009random}. 
At higher Rayleigh numbers the flow in wider domains does become turbulent despite the absence of a shear-generating instability. For example, when $L_x=0.4\pi$ this instability is absent since $k_{z,0}<2\pi$. However, when the Rayleigh number is increased to $Ra_{T,q}=10^{10}$ the flow nonetheless becomes turbulent or at least chaotic. Figure \ref{fig:snapshot_Ra_Tq_1e10_Lx_0p4_pi} shows two snapshots of a solution with $N_x=N_z=512$ grid points at these parameter values, displaying intermittent layering and vortex dipole generation. The mechanism responsible for destabilizing the stationary elevator state at these large Rayleigh numbers remains to be studied.

\section{Prandtl number effect}
\label{sec:Pr}

In this section, we investigate the effect of the Prandtl number using the full 2D equations. A low Prandtl number is of interest in astrophysical applications, where the heat transport is dominated by photon diffusion \citep{garaud2018double,garaud2021journey}. Prandtl number $Pr=7$ corresponds to thermal diffusivity in water appropriate to oceanographic applications. When the temperature field is exchanged for a concentration such as salinity, the corresponding thermal diffusivity is replaced by molecular diffusivity, leading to Prandtl numbers as large as $Pr=700$.

\begin{figure}
(a) $Pr=0.1$ \hspace{0.38\textwidth} (b) $Pr=7$

    \centering
    \includegraphics[width=0.49\textwidth]{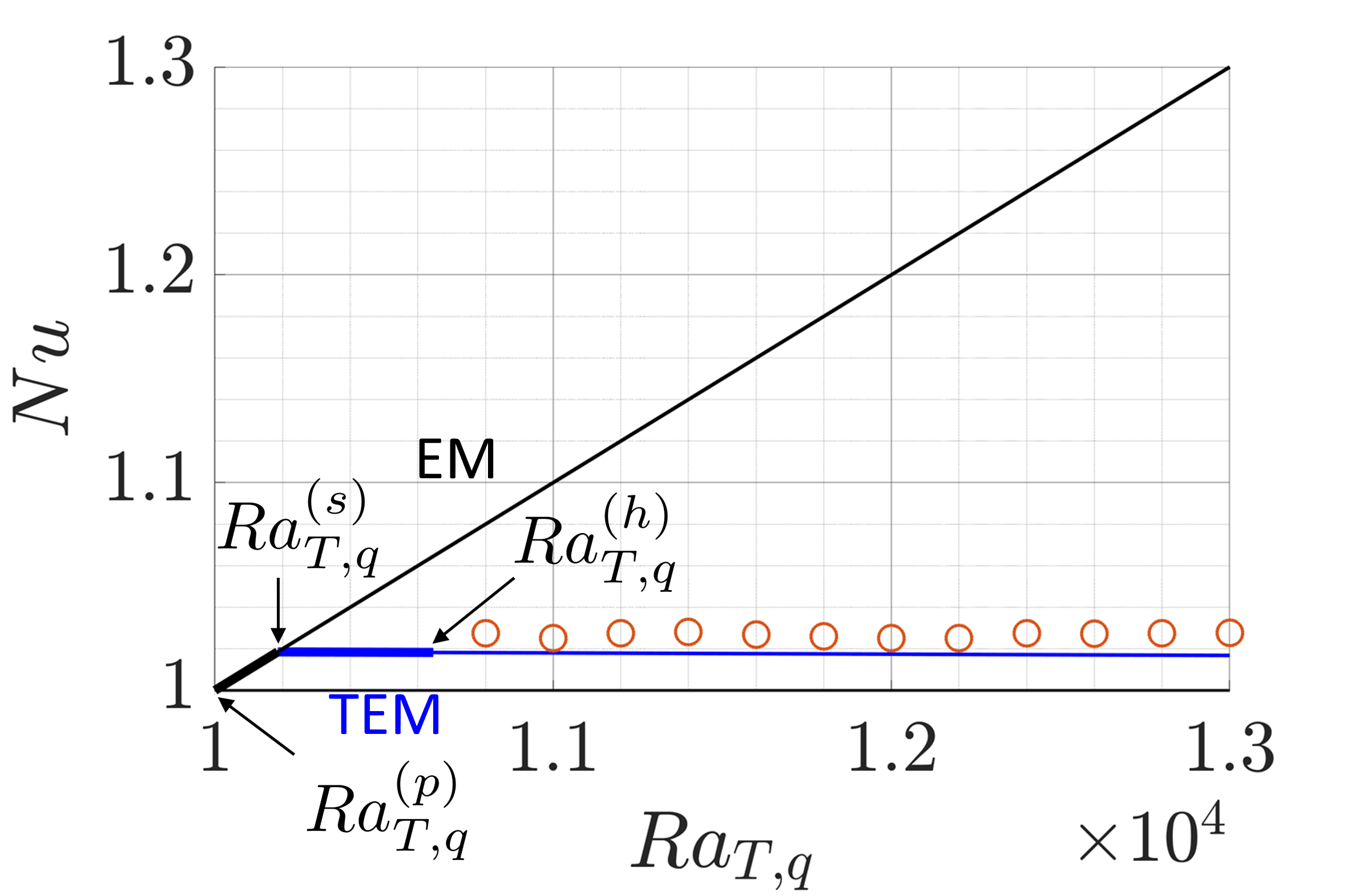}
    \includegraphics[width=0.49\textwidth]{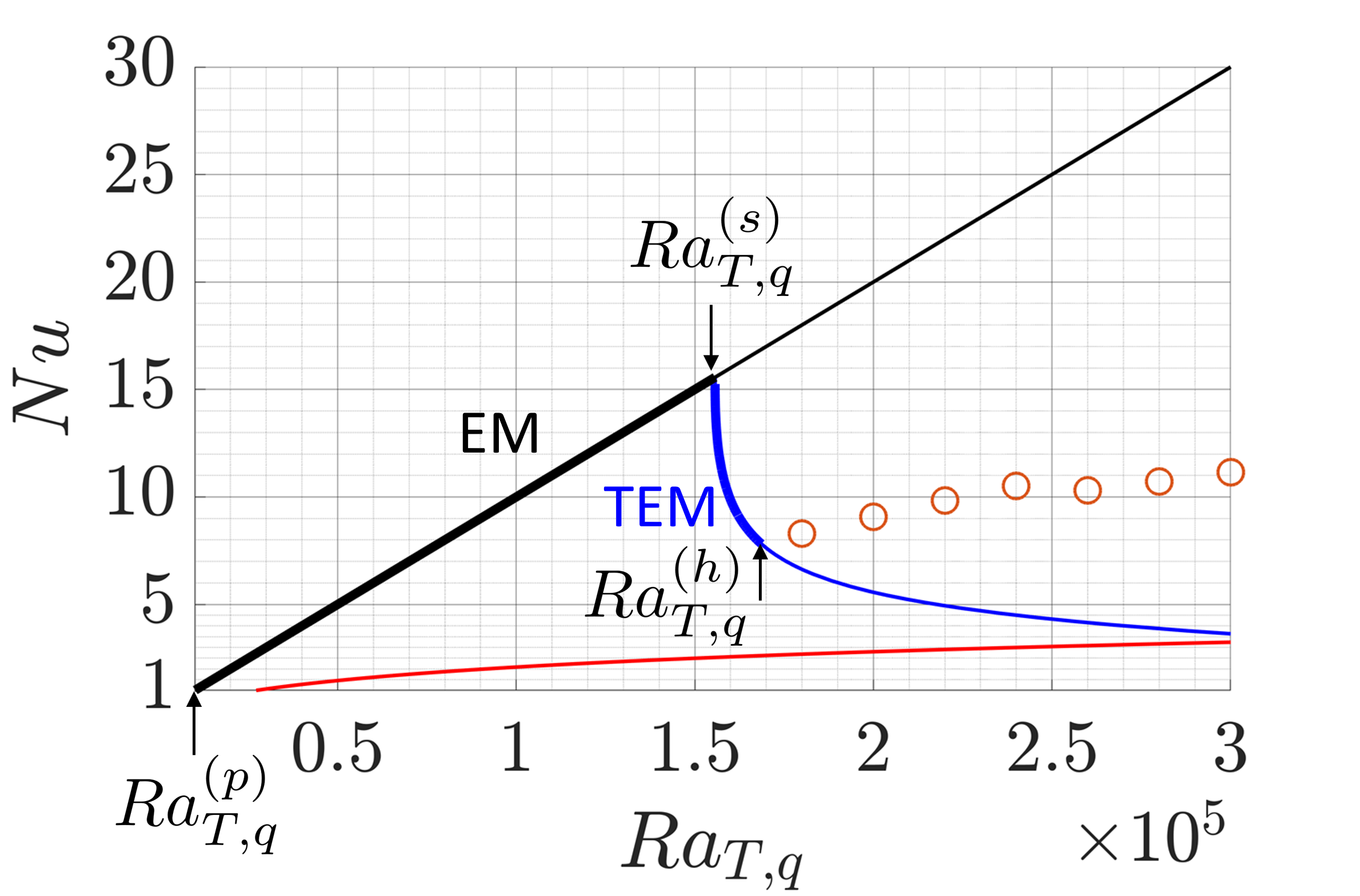}
    
    \caption{Bifurcation diagram with an elevator mode (EM, $\mline$, black), tilted elevator mode (TEM, {\color{blue}$\mline$}, blue) and time-dependent states ({\color{darkorange}$\circ$}, orange) at (a) $Pr=0.1$ and (b) $Pr=7$. The bifurcation points include the primary bifurcation $Ra_{T,q}^{(p)}$, the secondary bifurcation $Ra_{T,q}^{(s)}$ and a Hopf bifurcation $Ra_{T,q}^{(h)}$. The horizontal domain size is $L_x=0.2\pi$. }
    \label{fig:bif_diag_Pr_0p1_7}
\end{figure}

Figure \ref{fig:bif_diag_Pr_0p1_7} shows a bifurcation diagram similar to figure \ref{fig:bif_diag_full}(a) but with $Pr=0.1$ and $Pr=7$ in panels (a) and (b), respectively. The onset of the primary instability $Ra_{T,q}^{(p)}$ and the amplitude of the resulting elevator mode branch are not affected by $Pr$, consistent with the analytical results in Section \ref{subsec:primary_instability_elevator_mode}. At low Prandtl numbers, the secondary bifurcation point $Ra_{T,q}^{(s)}$ and the Hopf bifurcation point $Ra_{T,q}^{(h)}$ are both shifted to much lower Rayleigh numbers, while increasing $Pr$ shifts these bifurcation points to a higher Rayleigh number. These bifurcation points and the associated Hopf frequency $\omega_h$ are listed in table \ref{tab:bif_points_freq_compare_Pr}. The Hopf frequency in table \ref{tab:bif_points_freq_compare_Pr} obtained from numerical continuation is also very close to the oscillation frequency obtained from DNS at a parameter close to $Ra_{T,q}^{(h)}$ (not shown). Table \ref{tab:bif_points_freq_compare_Pr} compares the bifurcation points computed from the full 2D equations in \eqref{eq:NS} with the corresponding results from the single-mode equations in \eqref{eq:single_mode}. The latter are in closer agreement with the full 2D equations at $Pr=0.1$ than at $Pr=7$ because a low Prandtl number shifts these bifurcation points closer to the onset of primary instability at $Ra_{T,q}^{(p)}$. A related phenomenon is found in Rayleigh–B\'enard convection, where at low $Pr$ a steady convection roll becomes immediately unstable to a large-scale (zonal) mode \citep{winchester2022onset}. The markers ({\color{darkorange}$\circ$}) in figure \ref{fig:bif_diag_Pr_0p1_7} show the Nusselt number of the time-dependent states obtained from DNS, where the low Prandtl number case closely follows that of the TEM branch. 

\begin{table}
    \centering
    \begin{tabular}{clcccc}
        \hline
         & & $Ra_{T,q}^{(p)}$ & $Ra_{T,q}^{(s)}$ & $Ra_{T,q}^{(h)}$ & $\omega_h$ \\
         \hline
        \multirow{2}{*}{$Pr=0.1$} & full 2D equations in \eqref{eq:NS} & $10^4$ & 10185.4 & 10607.6 & 2.0 \\
         & Single-mode equations in \eqref{eq:single_mode}  & $10^4$ & 10181.5 & 10608.8 & 2.0 \\
       \hline 
       \multirow{2}{*}{$Pr=7$} &  full 2D equations in \eqref{eq:NS} & $10^4$ & 155671.7 & 168463.5 & 247.5 \\
        & Single-mode equations in \eqref{eq:single_mode} & $10^4$ & 117033.7 & 157801.8 & 250.1 
        \\
        \hline
    \end{tabular}
    \caption{Comparisons of bifurcation points between the full 2D equations in \eqref{eq:NS} with $L_x=0.2\pi$ and the single-mode equations in \eqref{eq:single_mode} with $k_x=10$ including the primary bifurcation $Ra_{T,q}^{(p)}$, the secondary bifurcation $Ra_{T,q}^{(s)}$, the Hopf bifurcation $Ra_{T,q}^{(h)}$ and the Hopf frequency $\omega_h$ at $Pr=0.1$ and $Pr=7$. }    \label{tab:bif_points_freq_compare_Pr}
\end{table}

\begin{figure}
(a) $\langle u\rangle_h(z,t)$ \hspace{0.4\textwidth} (b) 

    \centering
    \includegraphics[width=0.49\textwidth]{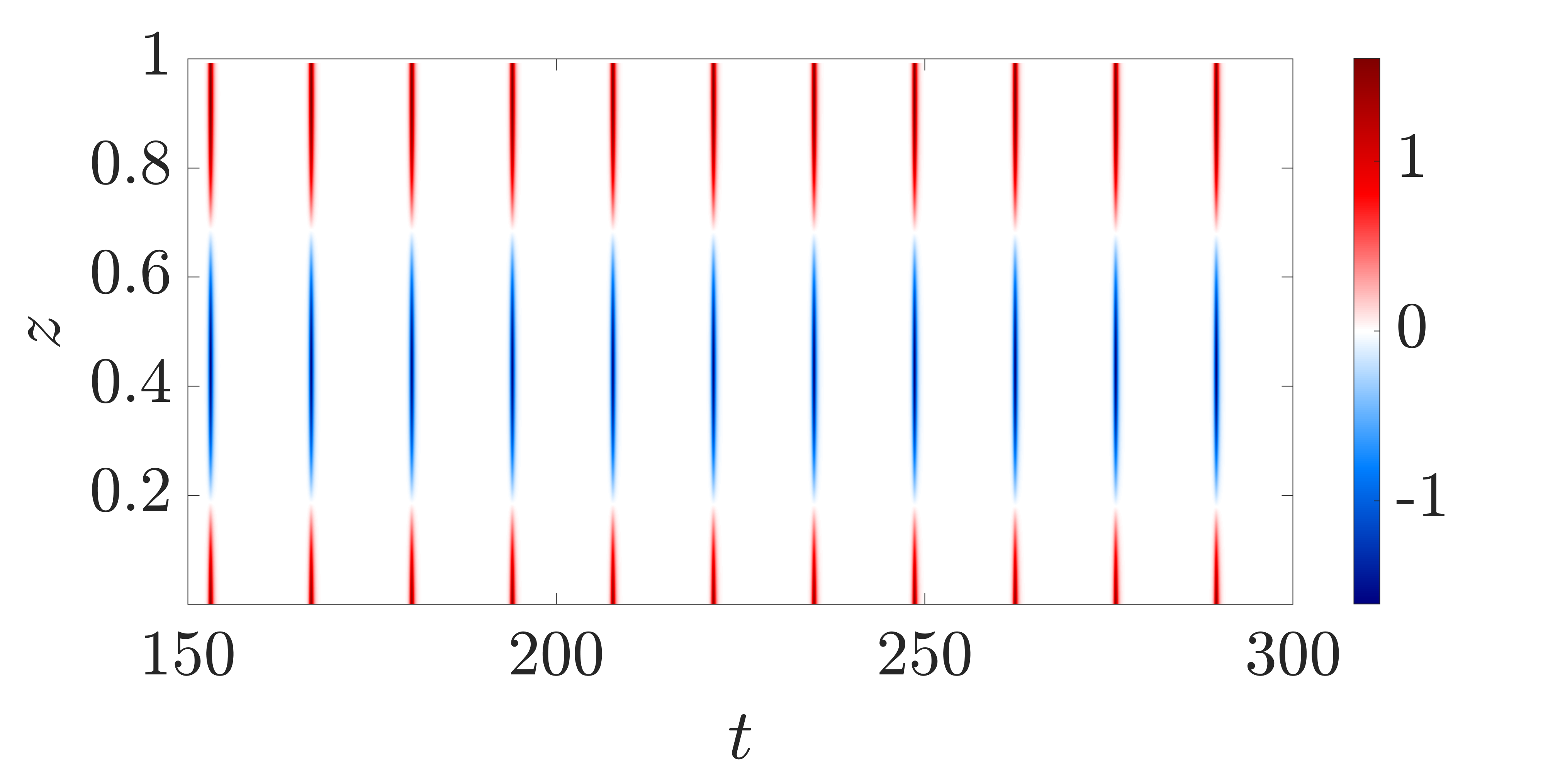}
    \includegraphics[width=0.49\textwidth]{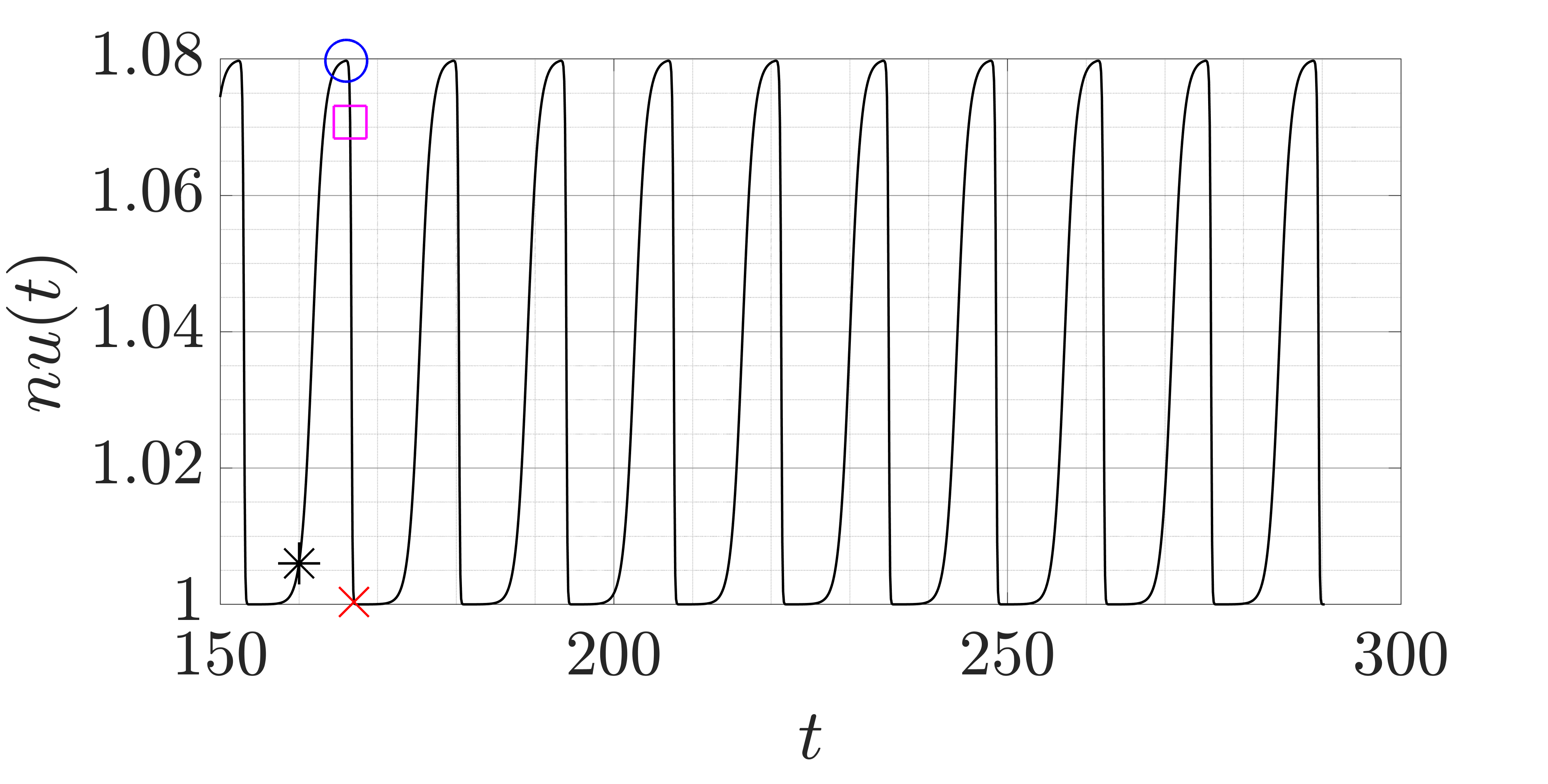}
    \caption{(a) Large-scale shear $\langle u\rangle_{h}(z,t)$ and (b) instantaneous Nusselt number $nu(t)$ displaying relaxation oscillations at $Ra_{T,q}=10800$, $Pr=0.1$ and $L_x=0.2\pi$. }
    \label{fig:DNS_Ra_Tq_10800_Pr_0p1}
\end{figure}

\begin{figure}
(a) $t=160$ ($*$) \hspace{0.10\textwidth} (b) $t=166$ ({\color{blue}$\circ$})  \hspace{0.1\textwidth} (c) $t=166.5$ ({\color{magenta}$\square$})  \hspace{0.08\textwidth} (d) $t=167$ ({\color{red}$\times$})

    \centering
    
        \includegraphics[width=\textwidth]{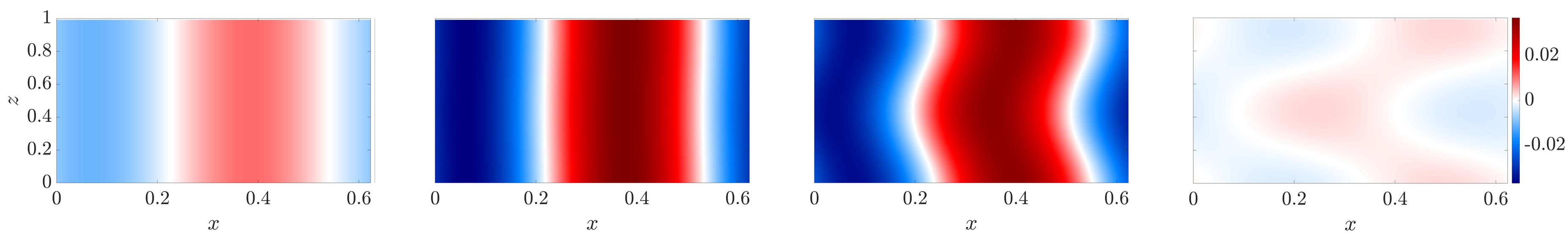}
    \caption{Evolution of the temperature deviation $T(x,z,t)$ during a burst corresponding to the four times indicated in figure \ref{fig:DNS_Ra_Tq_10800_Pr_0p1}(b). Parameters are $Ra_{T,q}=10800$, $Pr=0.1$ and $L_x=0.2\pi$ (see Movie 2). }
    \label{fig:snapshots_DNS_Ra_Tq_10800_Pr_0p1}
\end{figure}

Figure \ref{fig:DNS_Ra_Tq_10800_Pr_0p1}(a) displays the large-scale shear $\langle u\rangle_h(z,t)$ found at $Ra_{T,q}=10800$ and $Pr=0.1$. The instantaneous Nusselt number grows exponentially from $Nu\approx1$ corresponding to the conduction state and saturates at $Nu\approx1.08$ associated with the elevator mode based on \eqref{eq:Nu_elevator_mode}. The Nusselt number then rapidly decreases to $Nu\approx1$ as a result of shear generation. Four different snapshots corresponding to the four different markers in figure \ref{fig:DNS_Ra_Tq_10800_Pr_0p1}(b) are shown in figure \ref{fig:snapshots_DNS_Ra_Tq_10800_Pr_0p1}. Here, the elevator mode grows from $t=160$ to $t=166$ and then tilts at $t=166.5$, followed by rapid decay back towards the conduction state at $t=167$. This behavior resembles a relaxation oscillation between the conduction base state and the steady elevator mode associated with different time scales between the growth and decay of the elevator mode. A similar oscillation cycle between a sheared state and convection rolls is also observed in RBC \citep{matthews1996three,goluskin2014convectively} and magnetoconvection \citep{matthews1995compressible}. More generally, relaxation oscillations associated with disparate time scales are also widely observed in binary fluid convection \citep{batiste2006spatially}, double-diffusive convection \citep{beaume2018three} and convection in a rapidly rotating sphere \citep{busse2002convective}. Burst behavior similar to the relaxation oscillation cycle observed here can be described by a two-dimensional ODE model, where a burst is triggered by a symmetry-breaking instability of a growing symmetric state when it reaches certain amplitude \citep{batiste2001simulations,bergeon2002natural}. The prevalence of large-scale shear at low $Pr$ is known to lead to suppression of convection, an effect widely observed in the low $Pr$ regime of RBC; see, e.g., \citet[figure 9]{schumacher2020colloquium}. The suppression of convection by large-scale shear resembles the suppression of turbulence by self-generated mean flows \citep{shats2007suppression}.

\begin{figure}
(a) \hspace{0.49\textwidth} (b)

    \centering
    \includegraphics[width=0.49\textwidth]{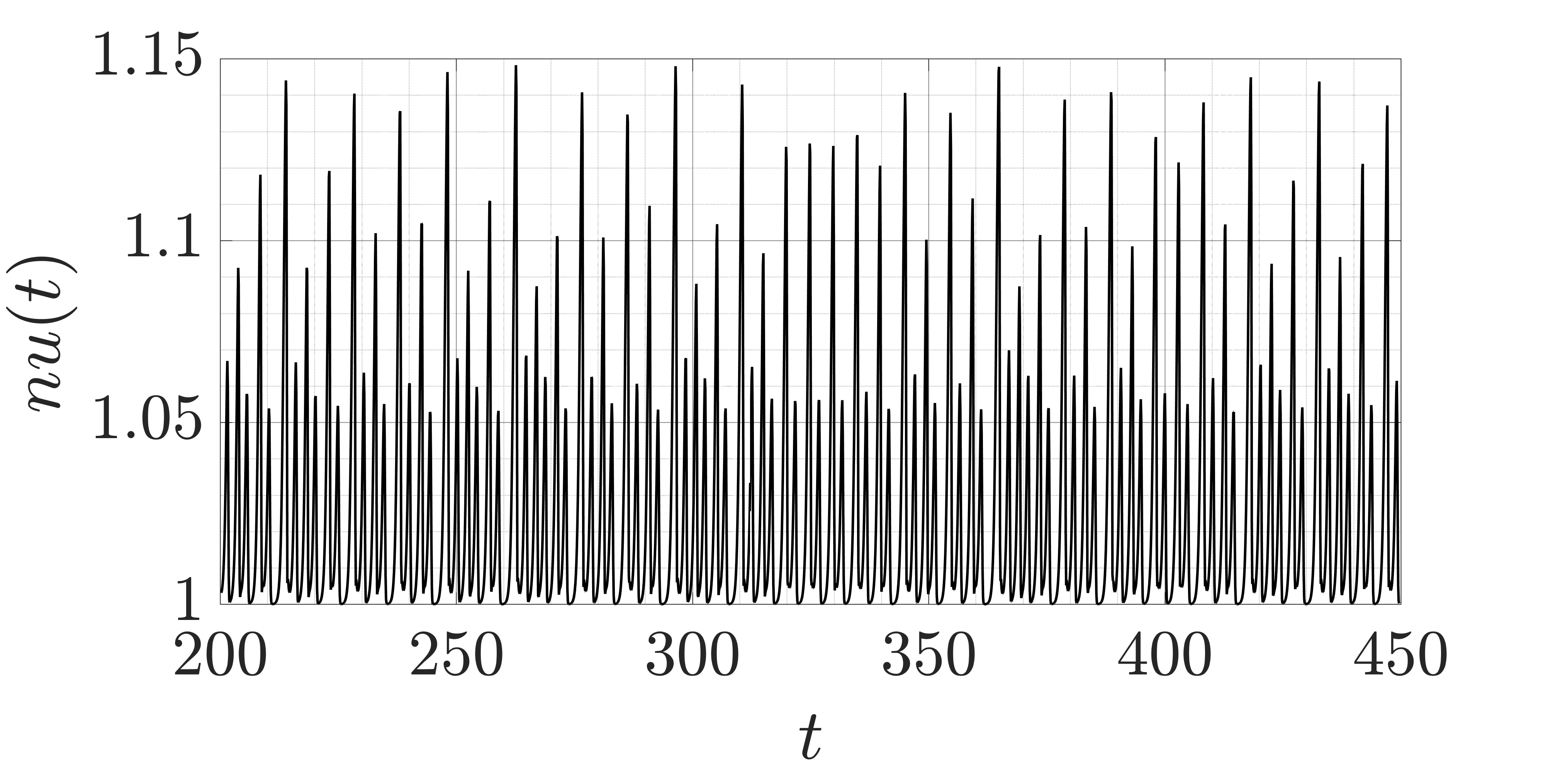}
    \includegraphics[width=0.49\textwidth]{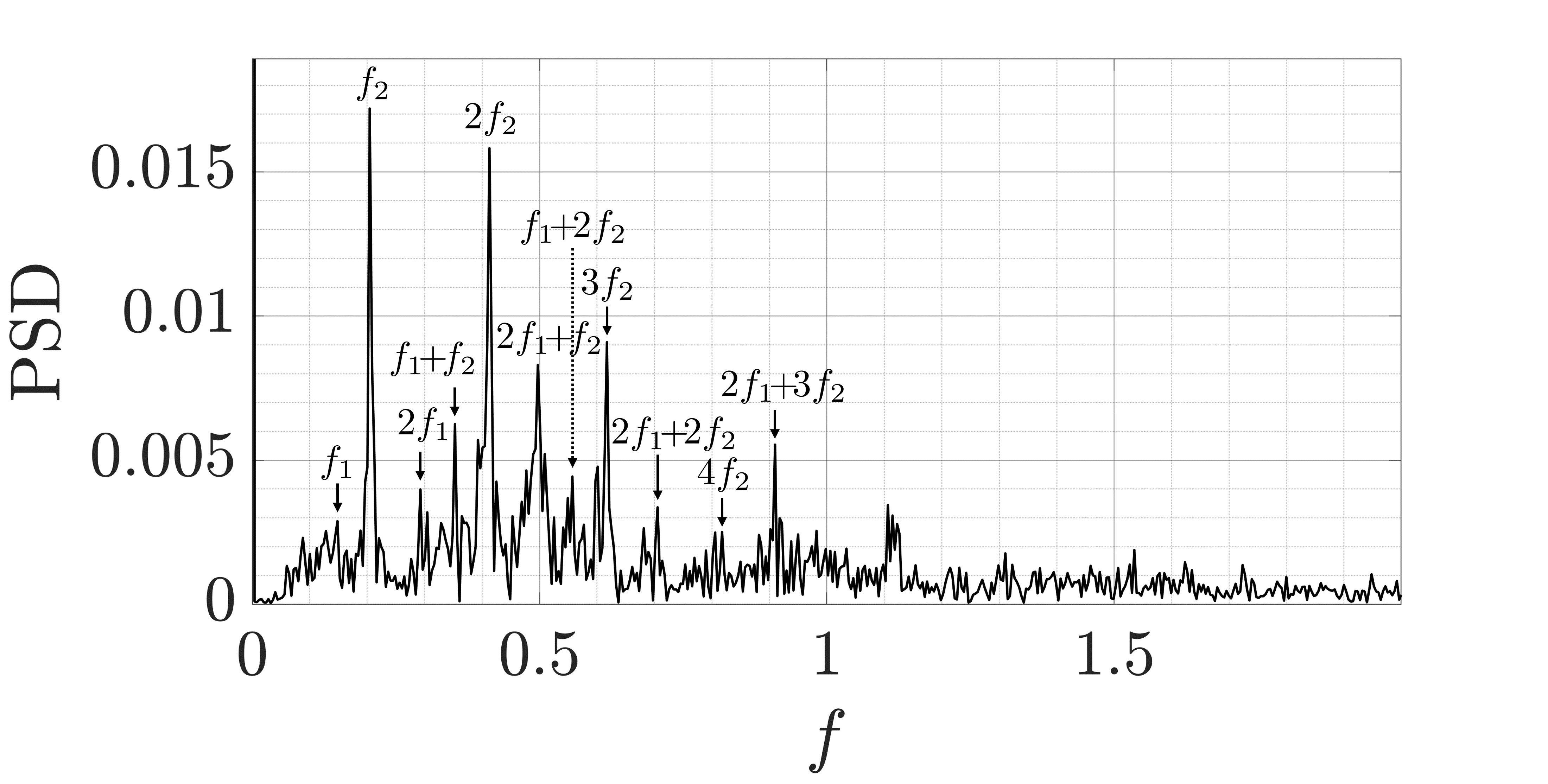}
    \caption{(a) The instantaneous Nusselt number $nu(t)$ and (b) its  PSD within $t\in [200,450]$ displaying spectral peaks at frequencies $mf_1+nf_2$, where $f_1=0.1483$, $f_2=0.2044$ and $m,n\in \mathbb{Z}^+$, indicating quasiperiodic dynamics at $Ra_{T,q}=11900$, $Pr=0.1$ and $L_x=0.2\pi$.}
    \label{fig:DNS_Ra_Tq_11900_Pr_0p1}
\end{figure}

\begin{figure}
(a)  \hspace{0.49\textwidth} (b)

    \centering

\includegraphics[width=0.49\textwidth]{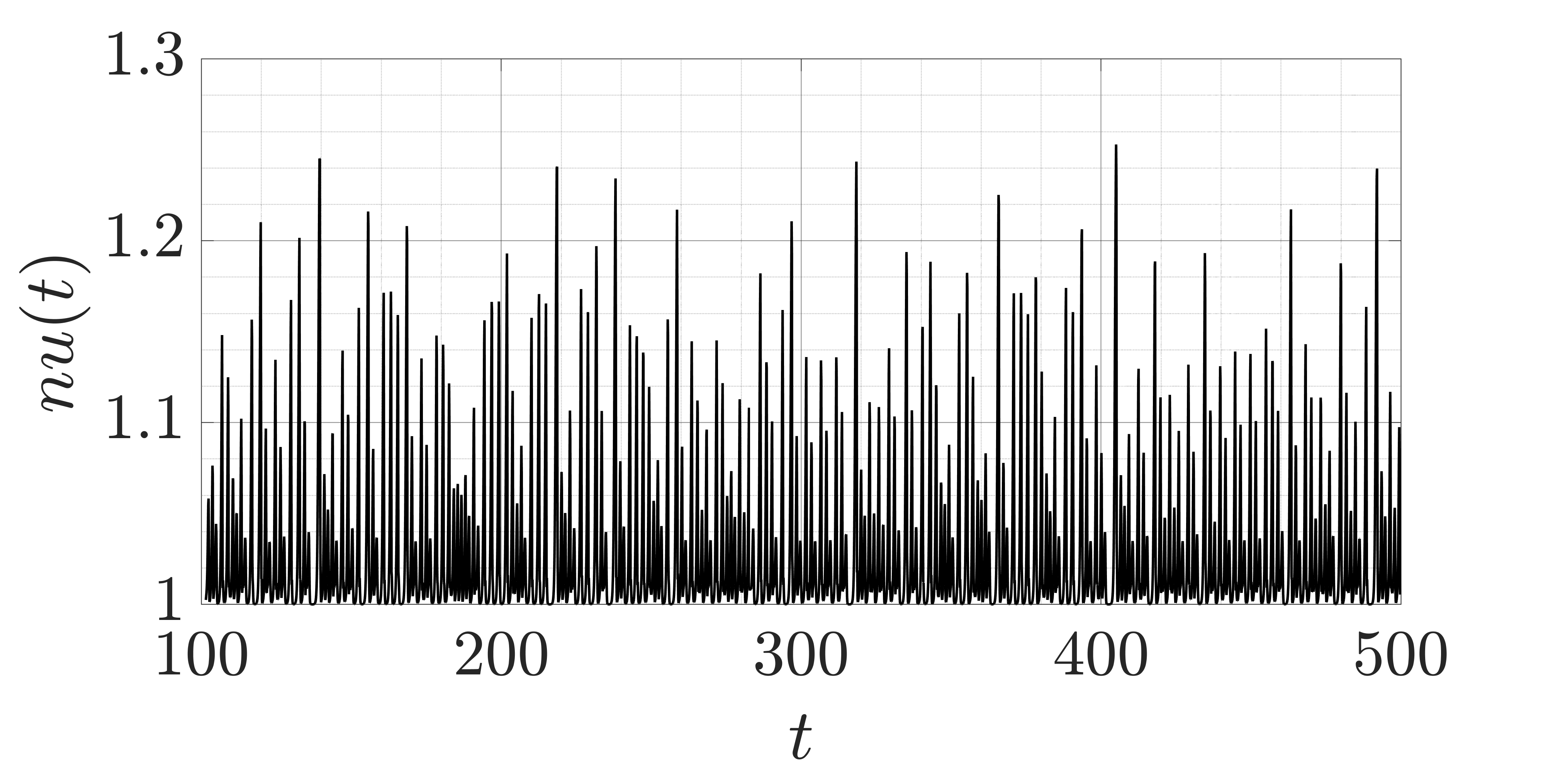}
\includegraphics[width=0.49\textwidth]{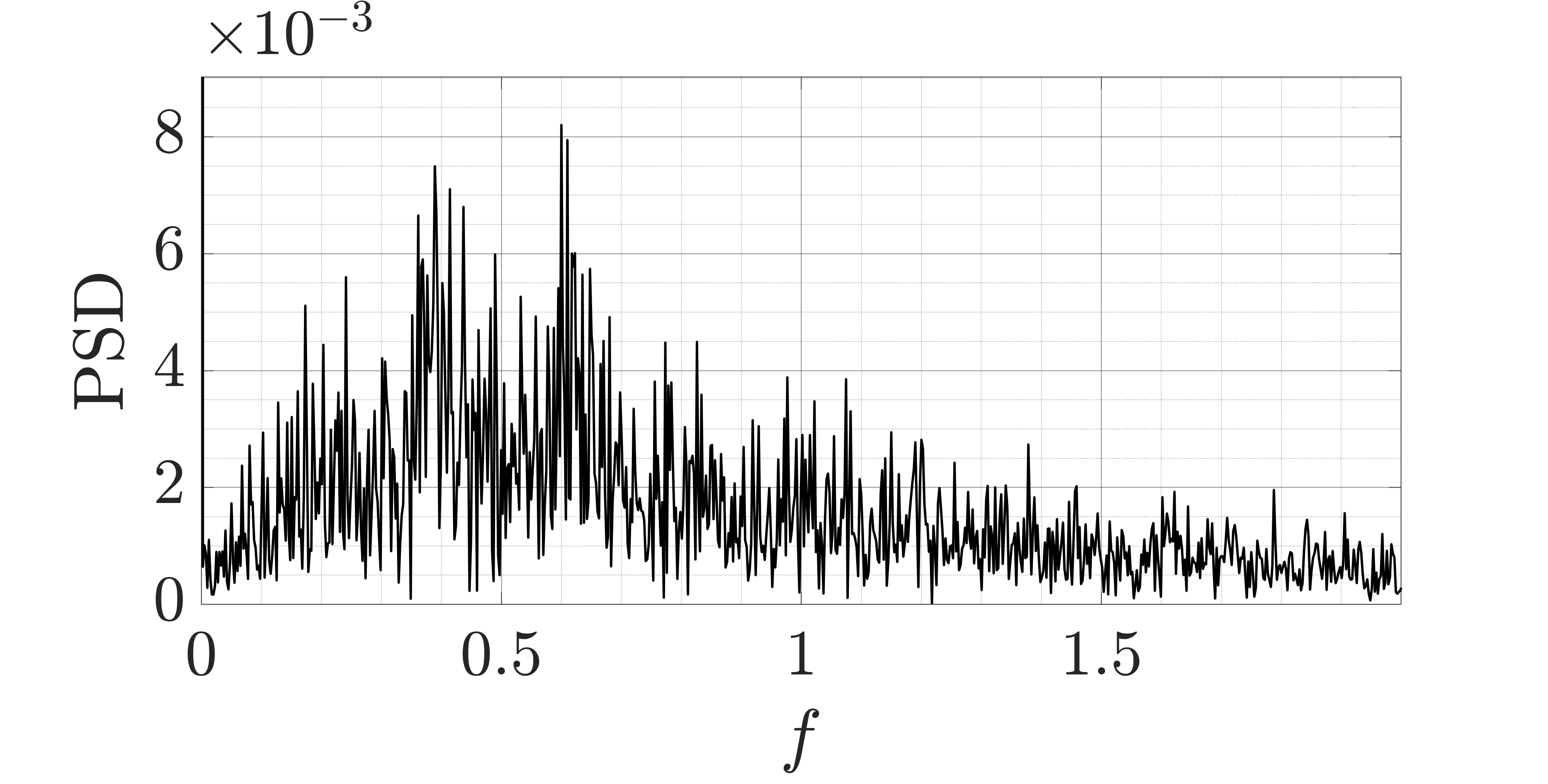}
    
    \caption{ (a) The instantaneous Nusselt number $nu(t)$ and (b) its PSD at parameters $Ra_{T,q}=13000$, $Pr=0.1$ and $L_x=0.2\pi$.}
    \label{fig:DNS_Ra_Tq_13000_Pr_0p1}
\end{figure}

Figure \ref{fig:DNS_Ra_Tq_11900_Pr_0p1}(a) shows $nu(t)$ at a higher Rayleigh number, $Ra_{T,q}=11900$. This time series no longer displays visible periodic behavior but its power spectrum density in figure \ref{fig:DNS_Ra_Tq_11900_Pr_0p1}(b) shows multiple peaks corresponding to frequencies $m\,f_1+n\,f_2$ with $f_1=0.1483$, $f_2=0.2044$ and $m,n\in \mathbb{Z}^+$. We conclude that the state in figure \ref{fig:DNS_Ra_Tq_11900_Pr_0p1}(a) is quasiperiodic, and likely generated by a torus bifurcation (equivalently, a Hopf bifurcation of a periodic orbit). Quasiperiodic orbits are also observed prior to the onset of nonperiodic motion in Rayleigh-B\'enard convection \citep{ahlers1978evolution,yahata1982transition}. At yet higher Rayleigh numbers chaotic behavior appears as shown in figure \ref{fig:DNS_Ra_Tq_13000_Pr_0p1} for $Ra_{T,q}=13000$. The broad power spectrum density (PSD) is indicative of a chaotic state. The instantaneous Nusselt number does not reach the amplitude corresponding to the saturated steady elevator mode with $Nu=1.3$ as shown in figure \ref{fig:DNS_Ra_Tq_13000_Pr_0p1}(a). This is because the growing EM are disrupted before they are able to saturate.

\begin{figure}
(a) $\langle u\rangle_h(z,t)$ \hspace{0.38\textwidth} (b)

    \centering
    \includegraphics[width=0.49\textwidth]{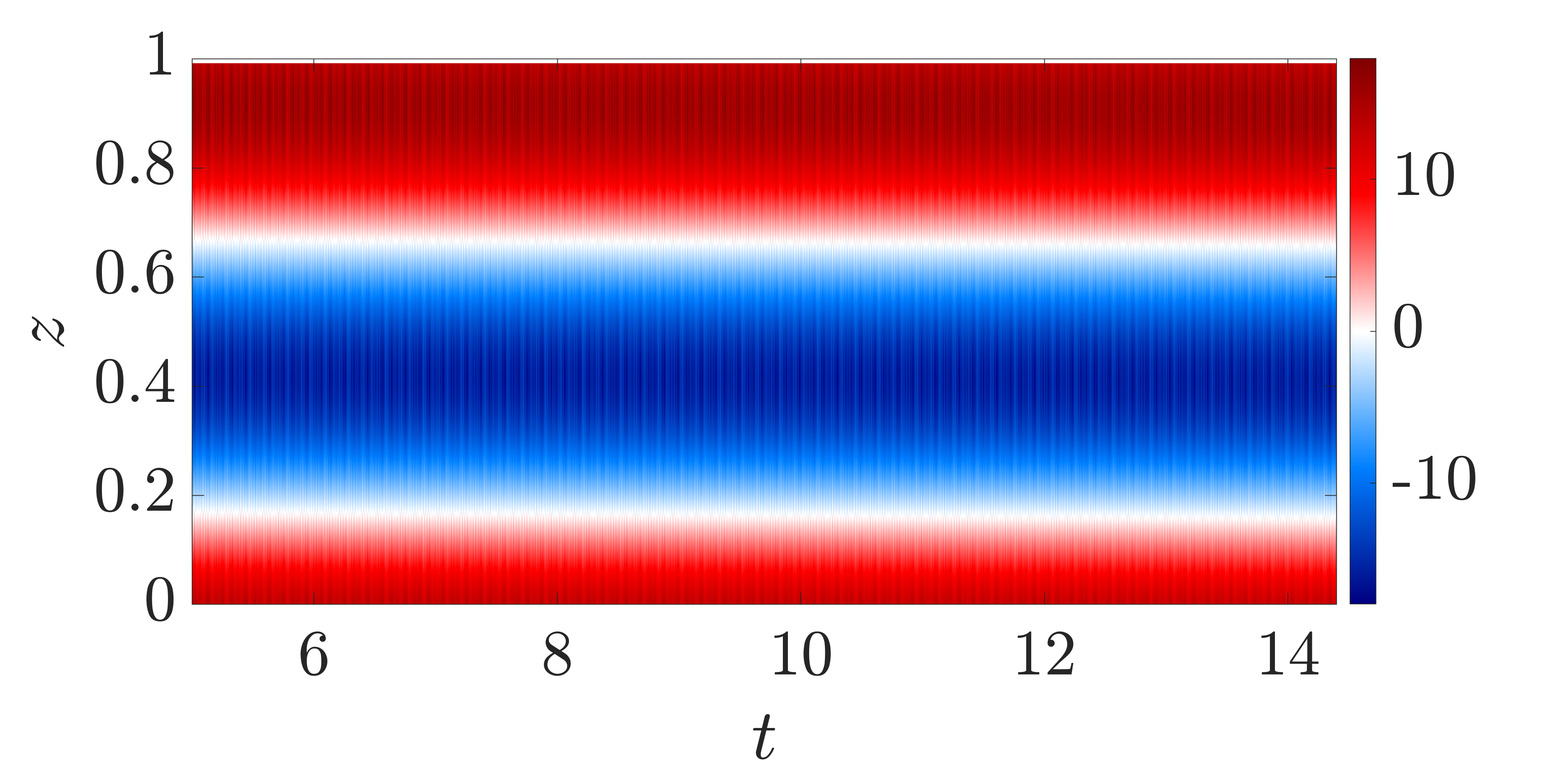}
    \includegraphics[width=0.49\textwidth]{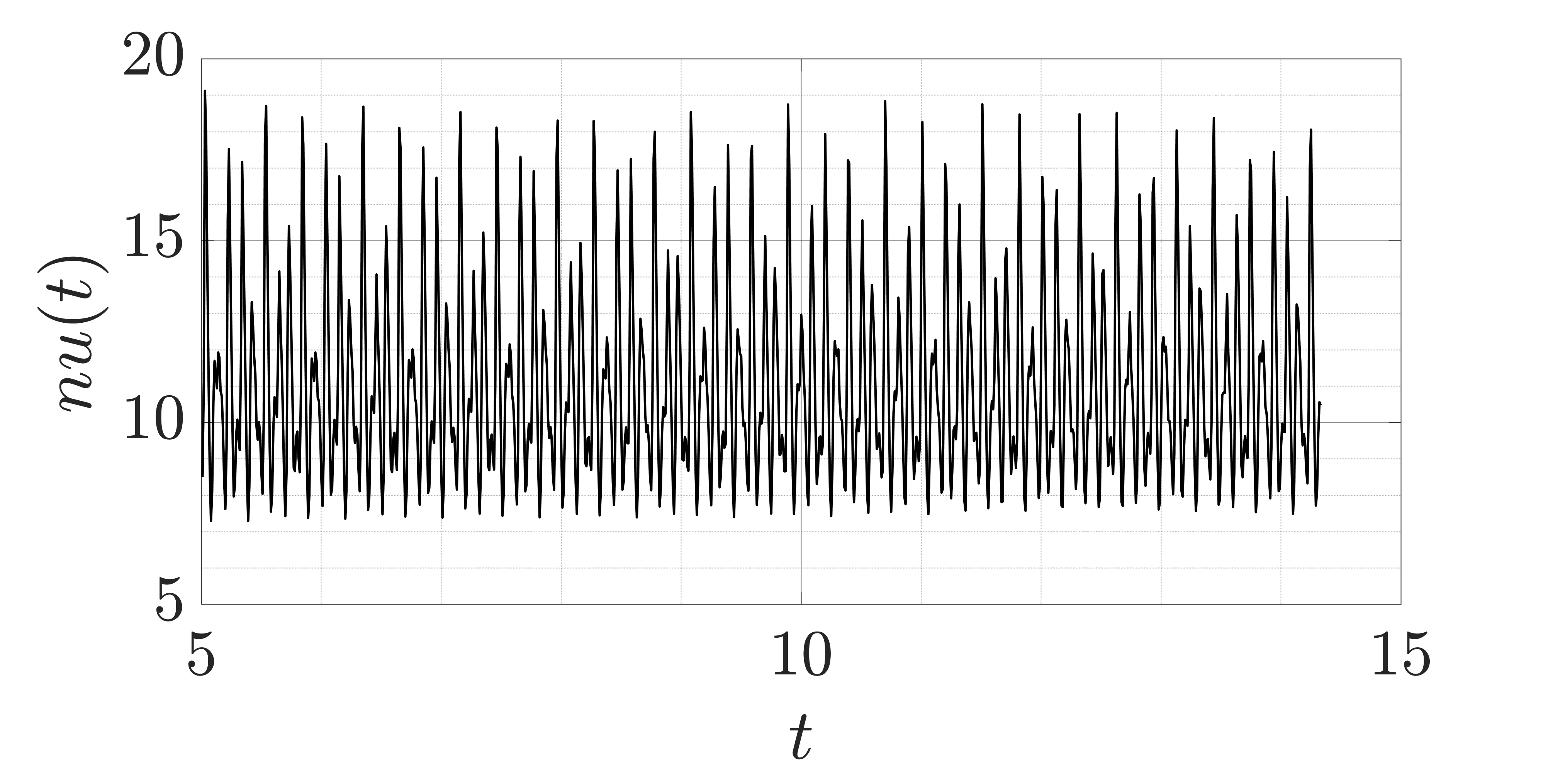}
    \caption{(a) The mean horizontal flow $\langle u \rangle_h(z,t)$ and (b) instantaneous Nusselt number $nu(t)$ at $Ra_{T,q}=2.4\times 10^5$, $Pr=7$ and $L_x=0.2\pi$.  }
    \label{fig:DNS_Ra_Tq_2p4e5_Pr_7}
\end{figure}

\begin{figure}
(a) $\langle u \rangle_h(z,t)$ \hspace{0.38\textwidth} (b) 

    \centering
    \includegraphics[width=0.49\textwidth]{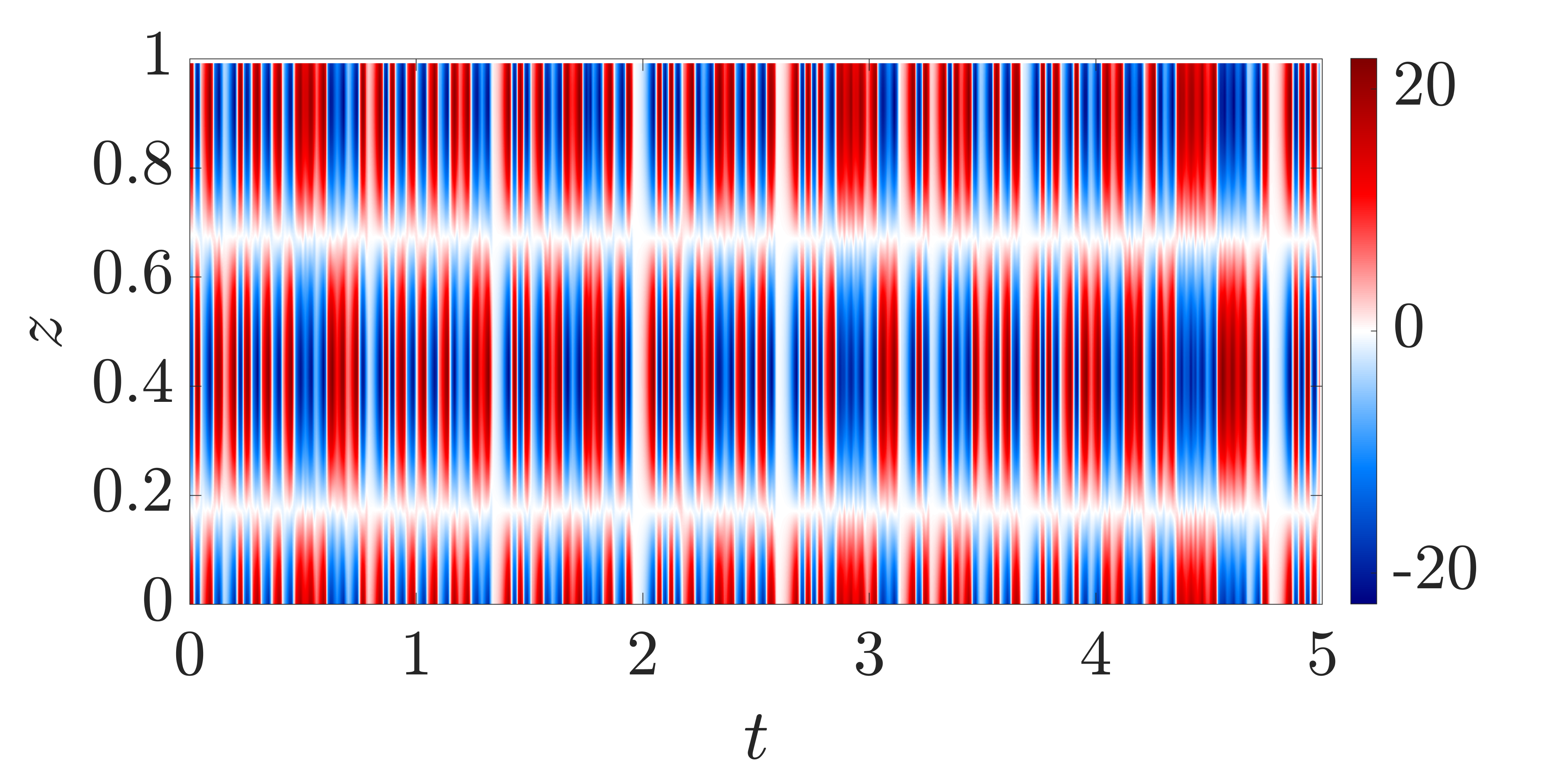}
    \includegraphics[width=0.49\textwidth]{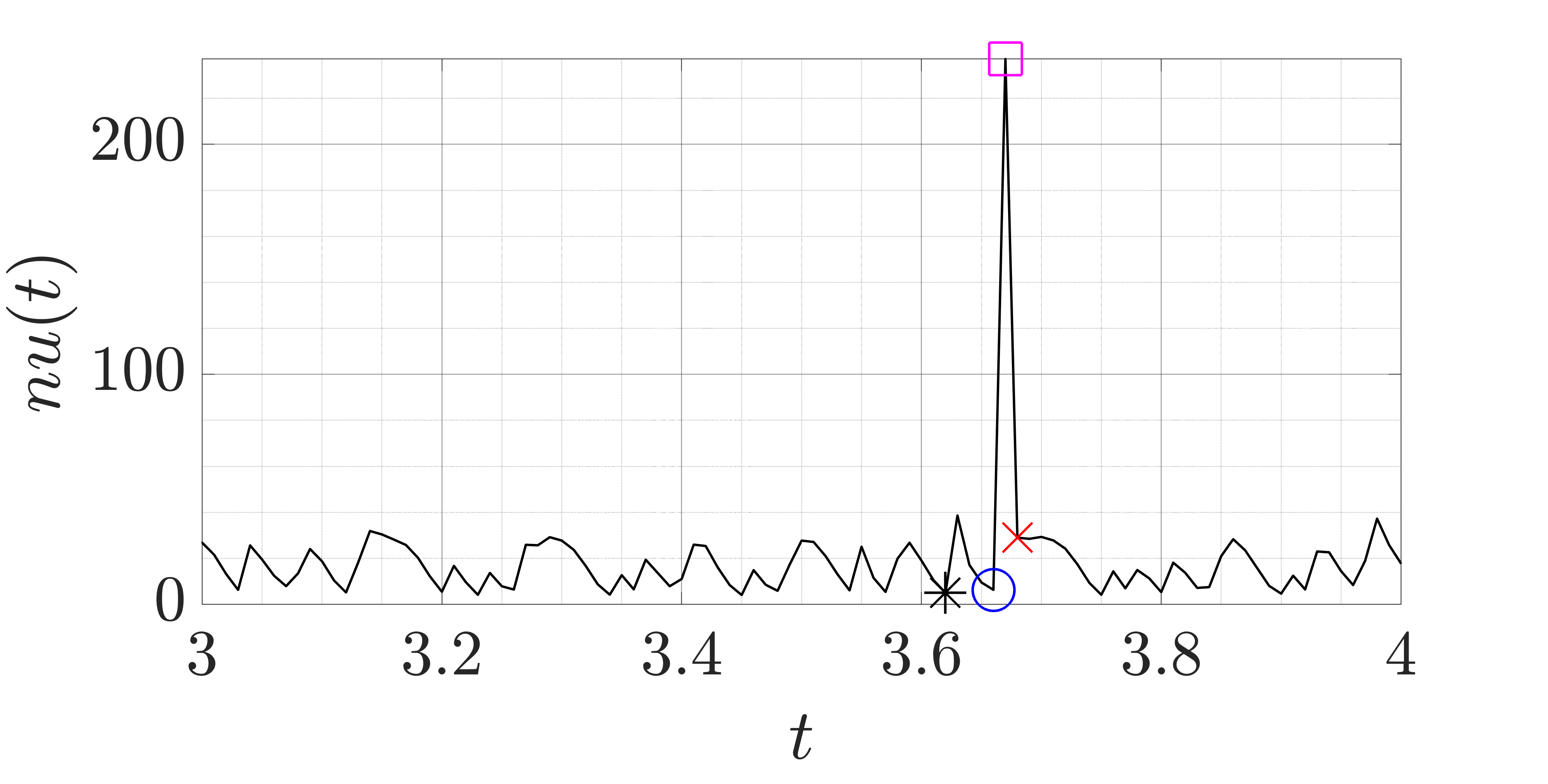}
    \caption{(a) The mean horizontal flow $\langle u \rangle_h(z,t)$ and (b) instantaneous Nusselt number $nu(t)$ at $Ra_{T,q}=3\times 10^5$, $Pr=7$ and $L_x=0.2\pi$.}
    \label{fig:DNS_Ra_Tq_3e5_Pr_7}
\end{figure}

\begin{figure}
(a) $t=3.62$ ($*$) \hspace{0.10\textwidth} (b) $t=3.67$ ({\color{blue}$\circ$}) \hspace{0.09\textwidth} (c) $t=3.68$ ({\color{magenta}$\square$})  \hspace{0.08\textwidth} (d) $t=3.69$ ({\color{red}$\times$})

    \centering
    \includegraphics[width=\textwidth]{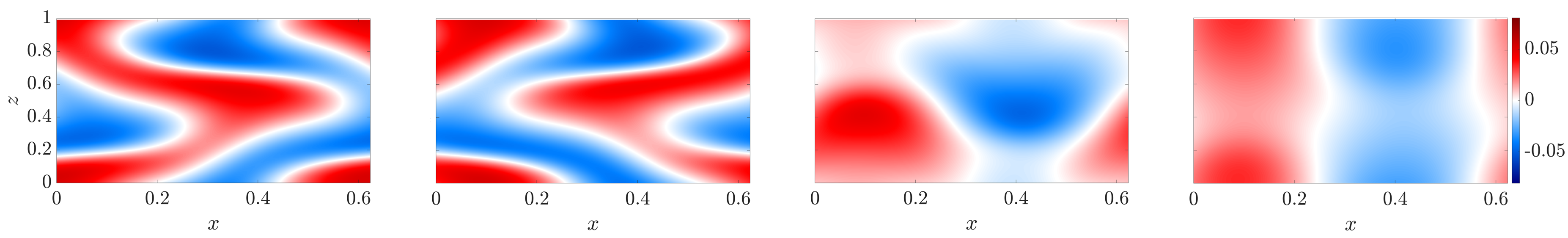}
    \caption{Evolution of the temperature deviation $T(x,z,t)$ during a burst corresponding to the four times indicated in figure \ref{fig:DNS_Ra_Tq_3e5_Pr_7}(b). Parameters are $Ra_{T,q}=3\times 10^5$, $Pr=7$ and $L_x=0.2\pi$.}
    \label{fig:snapshot_DNS_Ra_Tq_3e5_Pr_7}
\end{figure}

We next move on to the case $Pr=7$. Figure \ref{fig:DNS_Ra_Tq_2p4e5_Pr_7} displays $\langle u \rangle_h(z,t)$ and $nu(t)$ at $Ra_{T,q}=2.4\times 10^5$ and $Pr=7$. Here, modulated traveling waves appear similar to those present at $Pr=1$ in figure \ref{fig:DNS_Ra_Tq_6e4_Pr_1}(a). The instantaneous Nusselt number shows irregular behavior. At the higher Rayleigh number of $Ra_{T,q}=3\times 10^5$, the large-scale shear can reverse its direction as shown in figure \ref{fig:DNS_Ra_Tq_3e5_Pr_7}(a). At this Rayleigh number, the instantaneous Nusselt number displays intermittent behavior that can be instantaneously much larger than that of the elevator mode ($Nu\approx30$) as shown in \ref{fig:DNS_Ra_Tq_3e5_Pr_7}(b). Similar intermittency in the heat transport is also observed in homogeneous RBC driven by a constant temperature gradient \citep{borue1997turbulent,calzavarini2005rayleigh,calzavarini2006exponentially} owing to the intermittent appearance of an elevator mode. Here, a larger Prandtl number suppresses the large-scale shear that disrupts elevator modes, leading to the observed intermittent bursting behavior resembling that in homogeneous RBC driven by a constant temperature gradient. 

Figure \ref{fig:snapshot_DNS_Ra_Tq_3e5_Pr_7} shows four snapshots of the temperature deviation $T(x,z,t)$ close to the burst event in figure \ref{fig:DNS_Ra_Tq_3e5_Pr_7}(b) at the four times indicated in the figure. Here the local minima in the Nusselt number correspond to tilted states that can be associated with direction reversals as shown in figure \ref{fig:snapshot_DNS_Ra_Tq_3e5_Pr_7}(a) at $t=3.62$ and figure \ref{fig:snapshot_DNS_Ra_Tq_3e5_Pr_7}(b) at $t=3.67$. The local maximum of the Nusselt number at $t=3.68$ (figure \ref{fig:snapshot_DNS_Ra_Tq_3e5_Pr_7}(c)) corresponds to a burst state with hot fluid moving upwards and cold fluid moving downwards without impediment. This burst significantly reduces the absolute value of the temperature gradient, which leads to a large Nusselt number based on their reciprocal relation in \eqref{eq:nusselt_time}. This behavior is similar to homogeneous RBC in the high Prandtl number regime, where the temperature field often leads to vertical jet formation associated with a strong influence on the vertical temperature gradient \citep{calzavarini2005rayleigh}. 

\begin{figure}
    \centering
    \includegraphics[width=0.49\textwidth]{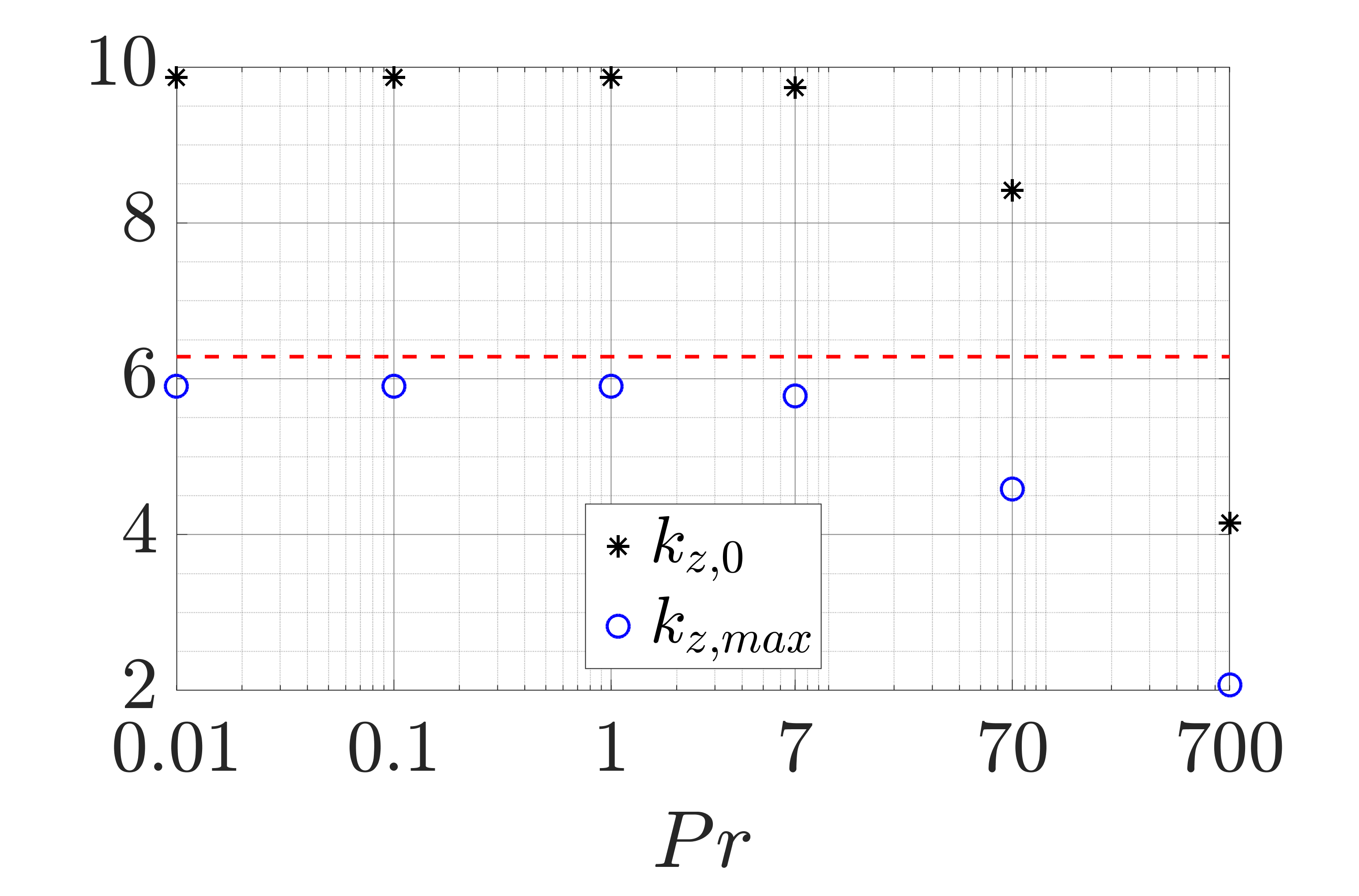}
\caption{The wavenumbers $k_{z,0}$ and $k_{z,max}$ as a function of $Pr$ at $Ra_{T,q}=10^8$, $L_x=0.2\pi$, and $k_e=10$. The red dashed line ({\color{red}$\dashed$}, red) corresponds to $k_z=2\pi$. }
    \label{fig:secondary_growth_rate_Pr}
\end{figure}

\begin{figure}
\hspace{0.1\textwidth} (a) $\langle u\rangle_h(z,t)$ \hspace{0.35\textwidth} (b) \hspace{0.15\textwidth} (c)

    \centering
    \includegraphics[width=0.49\textwidth]{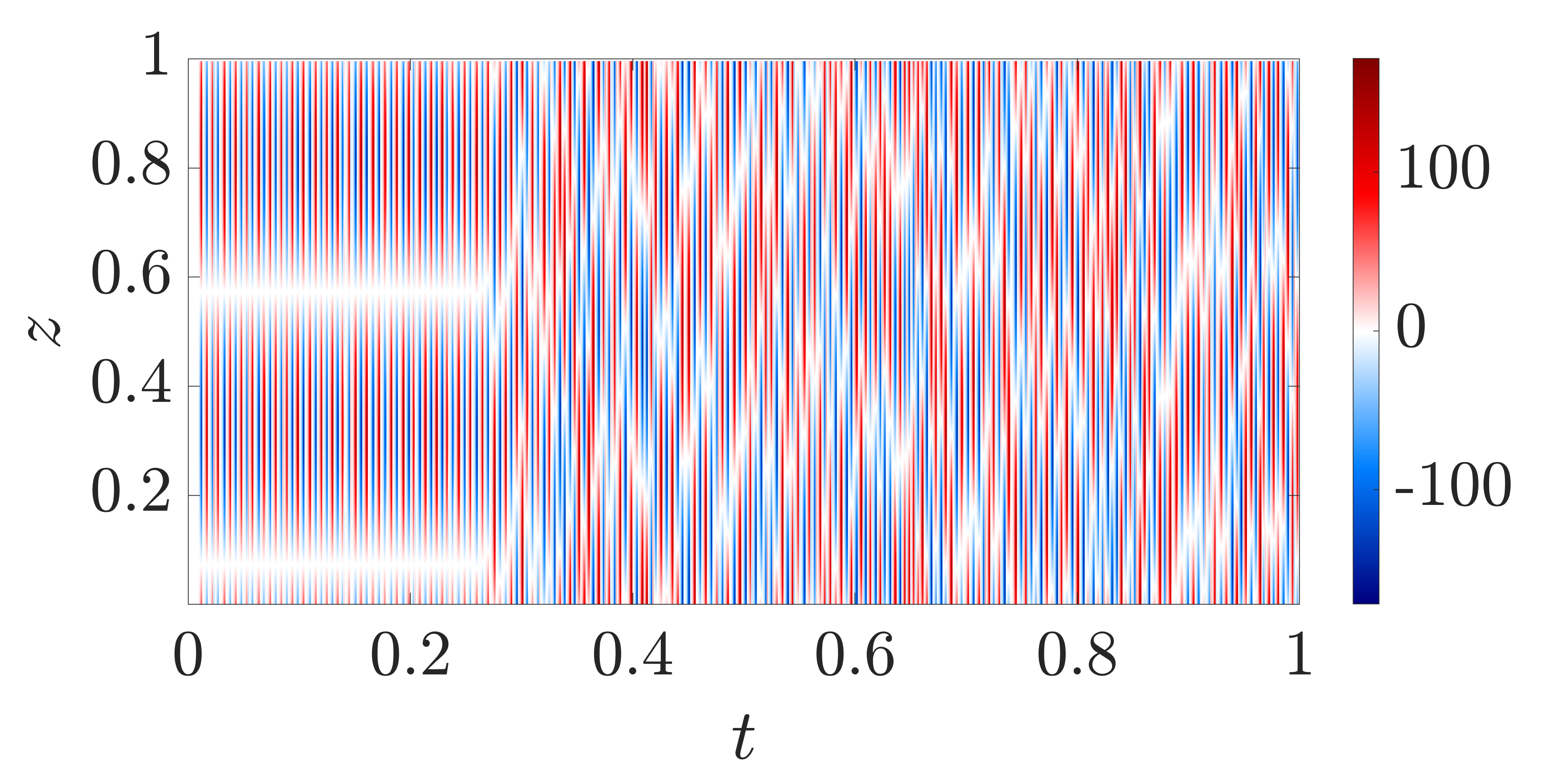}
\includegraphics[width=0.33\textwidth,trim=-0 -0.05in 0 0]{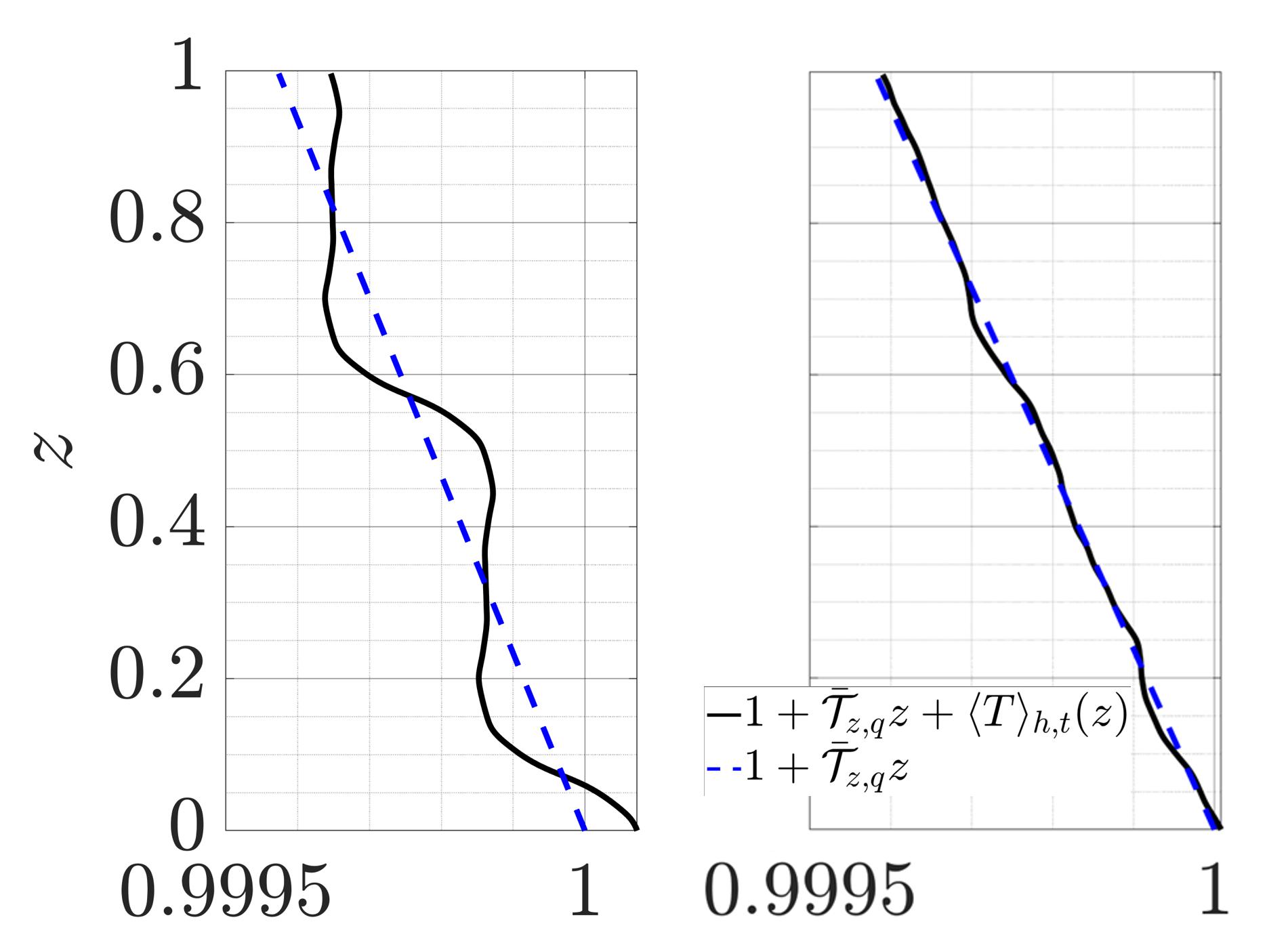}
    \caption{(a) The mean horizontal flow $\langle u\rangle_h(z,t)$ and (b)-(c) the mean temperature $1{+} \bar{\mathcal{T}}_{z,q} z{+}\langle T\rangle_{h,t}(z)$ ($\mline$) compared with the linear profile $1{+} \bar{\mathcal{T}}_{z,q} z$ ({\color{blue}$\dashed$}) averaged, respectively, over (b) $t\in [0.02,0.26]$ and (c) $t\in [0.5,1]$. The parameters are $Ra_{T,q}=8\times 10^8$, $Pr=700$ and $L_x=0.1\pi$.}
    \label{fig:DNS_Ra_Tq_8e8_Pr_700}
\end{figure}

We next move on to $Pr=700$ corresponding to the molecular diffusivity of salt in water. At this parameter, the elevator mode remains `stable' within DNS with $L_x=0.2\pi$ up to $Ra_{T,q}=10^8$. This can be understood from the secondary instability of elevator mode as shown in figure \ref{fig:secondary_growth_rate_Pr}, where the onset vertical wavenumber $k_{z,0}$ at $Pr=700$ is lower than $k_z=2\pi$ which is the minimum wavenumber for the presence of a secondary instability in a domain with $L_z=1$. In order to incorporate the secondary instability of the elevator mode, we change the horizontal domain size to $L_x=0.1\pi$ associated with domain-filling wavenumber $k_x=2\pi/L_x=20$ and increase the Rayleigh number to $Ra_{T,q}=8\times 10^8$. This is equivalent to changing the vertical domain to $L_z=2$ and keeping $L_x=0.2\pi$ and $Ra_{T,q}=10^8$.

\begin{figure}
    (a) $t=0.456$ \hspace{0.12\textwidth} (b) $t=0.457$ \hspace{0.12\textwidth} (c) $t=0.459$ \hspace{0.1\textwidth} (d) $t=0.461$

    \centering
    
\includegraphics[width=\textwidth]{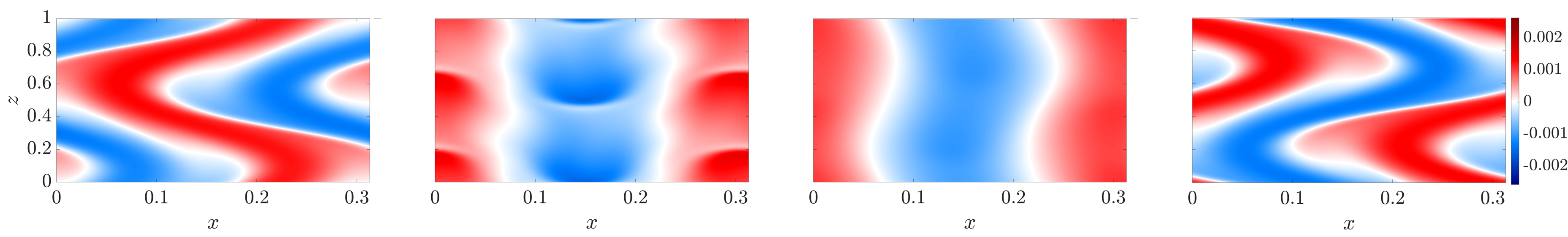}
    
    \caption{Temperature deviation $T(x,z,t)$ at (a) $t=0.456$ with $nu(t)=1165.5$, (b) $t=0.457$ with $\bar{T}_{z,q}(t)=1.75\times 10^{-5}$, (c) $t=0.459$ with $nu(t)=4784.7$, and (d) $t=0.461$ with $nu(t)=961.5$ at $Ra_{T,q}=8\times 10^8$, $Pr=700$ and $L_x=0.1\pi$ showing the generation of momentary stable stratification in case (b) (see Movie 3). }
    \label{fig:snapshot_DNS_Ra_Tq_1e8_Pr_700}
\end{figure}

Figure \ref{fig:DNS_Ra_Tq_8e8_Pr_700}(a) displays the mean horizontal flow $\langle u \rangle_h(z,t)$ at $Ra_{T,q}=8\times 10^8$, $L_x=0.1\pi$ and $Pr=700$. This large-scale shear begins to drift vertically at around $t\approx 0.3$ although its magnitude is in fact much reduced from that at $Pr=1$ (figure \ref{fig:DNS_Ra_Tq_1e8_Pr_1}(a)). Moreover, as shown in figure \ref{fig:DNS_Ra_Tq_8e8_Pr_700}(b) the mean temperature profile averaged over $t\in [0.02,0.26]$ exhibits apparent layer formation and several narrow regions displaying a stably stratified temperature profile. Such stably stratified regions are also present in extended domain DNS of RBC with $Pr=7$ but disappear when the Prandtl number decreases \citep[figure 9(g)]{schumacher2020colloquium}. Related locally stably stratified regions are also observed in potential vorticity staircases \citep[figure 5]{read2006mapping} and rotating double-diffusive convection \citep[figures 6-8]{moll2016effect}. Note that due to the vertical drift the mean temperature profile is close to a linear profile if averaged over longer times as shown in figure \ref{fig:DNS_Ra_Tq_8e8_Pr_700}(c).

Figure \ref{fig:snapshot_DNS_Ra_Tq_1e8_Pr_700} shows four snapshots of the temperature deviation $T(x,z,t)$ at $Pr=700$ and $Ra_{T,q}=8\times 10^8$. The temperature deviation in figures \ref{fig:snapshot_DNS_Ra_Tq_1e8_Pr_700}(a) and \ref{fig:snapshot_DNS_Ra_Tq_1e8_Pr_700}(d) displays tilted states associated with different tilting directions, with Nusselt numbers that are smaller than that of the elevator mode. Figure \ref{fig:snapshot_DNS_Ra_Tq_1e8_Pr_700}(b) shows a burst associated with the temporary creation of a positive or stable instantaneous mean temperature gradient $\bar{T}_{z,q}(t):=\langle wT\rangle_{h,v}-1$. At a slightly later instant shown in figure \ref{fig:snapshot_DNS_Ra_Tq_1e8_Pr_700}(c), the temperature deviation is slightly tilted but still close to an elevator mode with an instantaneous Nusselt number close to that of an elevator mode ($Nu=Ra_{T,q}/k_x^4=5000$). In homogeneous RBC driven by a constant temperature gradient, elevator modes appear more frequently at high Prandtl numbers \citep{calzavarini2005rayleigh}, which is also the case here in the fixed-flux setup, potentially due to a weaker large-scale shear that plays such an important role in the secondary instability of elevator modes.

\section{Conclusions and future work}
\label{sec:conclusion}

This work formulated a fixed-flux problem for homogeneous Rayleigh-B\'enard convection and analyzed its underlying dynamics in detail using numerical continuation, secondary instability analysis, and direct numerical simulations. The fixed-flux formulation leads to steady elevator mode solutions with a well-defined amplitude, something that is not the case in homogeneous RBC driven by a constant temperature gradient. The secondary instability of this elevator mode leads to tilted elevator modes accompanied by the generation of horizontal shear flow or jet, provided the elevator mode is sufficiently slender. We have chosen to admit this secondary instability by increasing the wavenumber $k_e$ of the elevator modes while keeping the vertical extent of the domain fixed. This procedure is equivalent to keeping $k_e$ fixed and increasing the domain height, provided the Rayleigh number is adjusted accordingly. Thus narrow domains favor generation of horizontal jets while wider domains favor stable elevator modes or equivalently vertical jets.

At $Pr=1$ a further increase in the Rayleigh number destabilizes the TEM state via a subcritical Hopf bifurcation leading to an interval of coexistence between the steady TEM state and a time-dependent state we have called a direction-reversing state. With increasing Rayleigh number this direction-reversing state encounters a global bifurcation of Shil'nikov type \citep{shilnikov2007shilnikov} leading to a modulated traveling wave state without flow reversal. Single-mode equations that severely truncate the horizontal structure reproduce this moderate Rayleigh number behavior well and confirm the tame (nonchaotic) nature of the Shil'nikov bifurcation for the parameter values used. At higher Rayleigh numbers, chaotic behavior appears but is dominated by modulated traveling waves in narrow domains while in wider domains simulations with $Ra_{T,q}=10^{10}$ display intermittent layering and vortex dipole generation. The correspondence between mean temperature and velocity profiles resembles behavior encountered with no-slip instead of stress-free boundary conditions in RBC with fixed temperature.  In contrast, the low Prandtl number regime displays relaxation oscillations between the conduction state and the elevator mode and exhibits quasiperiodic and then chaotic behavior as the Rayleigh number increases. At high Prandtl numbers, the large-scale shear generated by the secondary instability is much weaker, and the flow exhibits bursting behavior resembling that in homogeneous RBC driven by a constant temperature gradient \citep{borue1997turbulent,calzavarini2005rayleigh,calzavarini2006exponentially}. These bursts are associated with a significant increase in heat transport or even intermittent stable temperature stratification. Secondary bifurcations points are shifted closer to the primary instability at lower Prandtl numbers, leading to greater fidelity of our single-mode description. The relaxation rate $\beta$ of the heat flux does not influence the late-time flow behavior of the system.

This work opens up several directions for future work. In particular, it is crucial to analyze the corresponding dynamics in three dimensions, where large-scale shears may form in an arbitrary horizontal direction or result from the excitation of the vertical vorticity mode. High Rayleigh number convection inevitably generates multiple scales and the resulting interaction between different scales in the turbulent regime represents a continuing challenge to theory. Of particular interest is the recent discovery that temperature boundary conditions play a significant role in the multiscale structure of convection even in the turbulent regime \citep{{vieweg2021supergranule,vieweg2022inverse}} through a mechanism that remains elusive. The details of such high Rayleigh number flows in our configuration are beyond the scope of the present investigation but a systematic study of the Nusselt number scaling with $Ra_{T,q}$ and $Pr$ for this case is clearly desirable. Moreover, the fixed-flux formulation can be naturally extended to double-diffusive convection or rotating convection setups, and to reduced equations valid in the low Prandtl number limit \citep{spiegel1962thermal,thual1992zero,lignieres1999small,garaud2021journey} or in the high Prandtl number limit \citep{constantin1999infinite,wang2004infinite}. The usefulness of these approaches for studying finite Prandtl number dynamics in fixed-flux systems merits detailed investigation. Moreover, a systematic comparison between single-mode equations and the full 2D equations at different Prandtl numbers will provide further quantification of the validity of the single-mode description, and potentially provide further reduction in the low Prandtl number limit.

\section*{Acknowledgment}
This work was supported by the National Science Foundation under Grant Nos. OCE-2023541 (C.L. and E.K.) and OCE-2023499 (M.S. and K.J.). This work utilized the Blanca condo computing resource at the University of Colorado Boulder. Blanca is jointly funded by computing users and the University of Colorado, Boulder. C.L. also acknowledges support from the Connecticut Sea Grant PD-23-07 and NASA CT Space Grant P-2104 during the completion of this work. Part of the computational work performed on this project was done with the help from the Storrs High Performance Computing cluster. C.L. would like to thank the UConn Storrs HPC and HPC team for providing the resources and support that contributed to these results.

\section*{Declaration of Interests}
The authors report no conflict of interest.

\appendix

\section{Effect of finite $\beta$ in \eqref{eq:Q}}
\label{sec:finite_beta}

\begin{figure}
(a) $\langle u\rangle_{h}(z,t)$ \hspace{0.4 \textwidth} (b) $T(x,z,t=10)$

    \centering
    \includegraphics[width=0.49\textwidth]{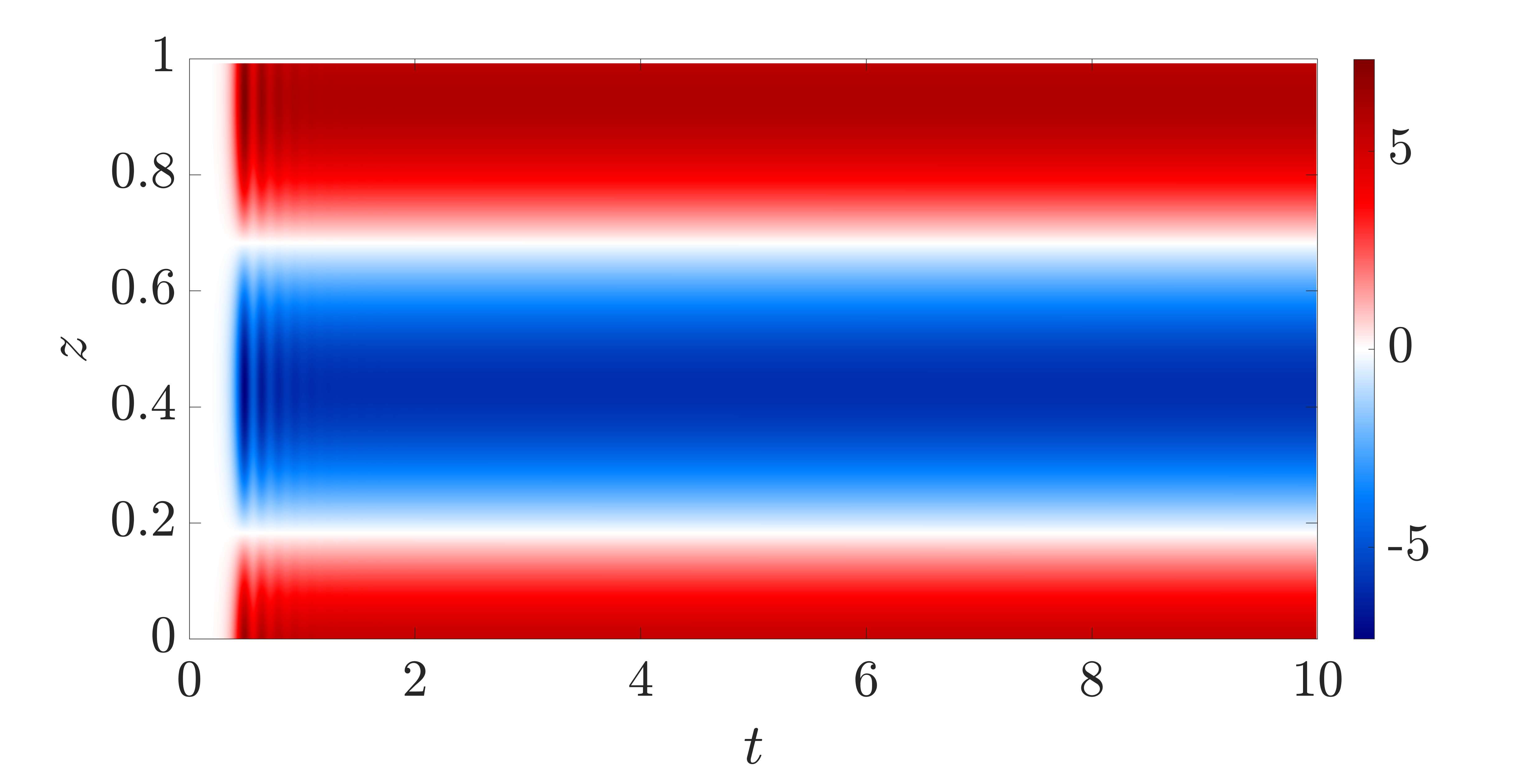}
    \includegraphics[width=0.49\textwidth]{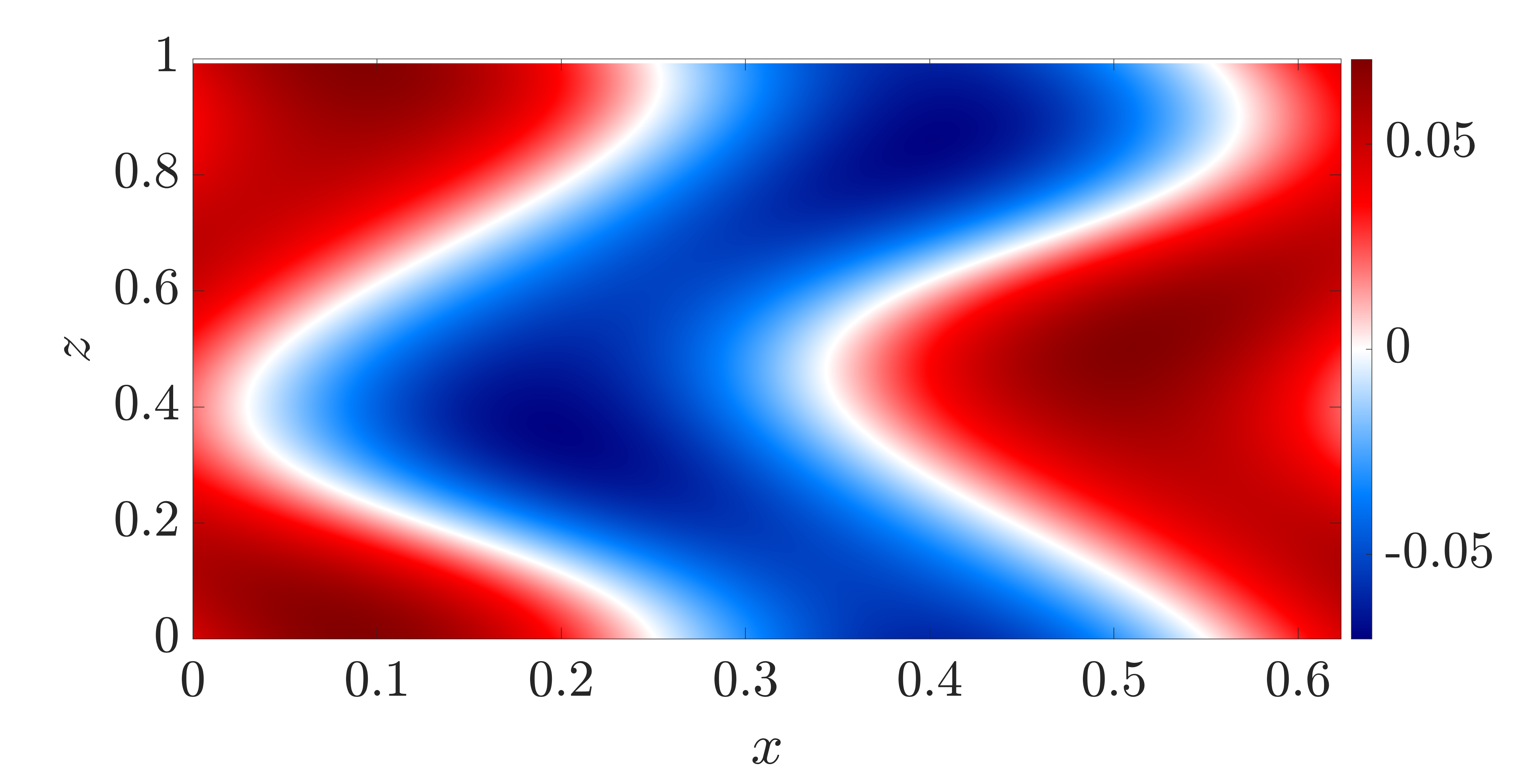}
    \caption{Same as figure \ref{fig:DNS_Ra_Tq_3e4_Pr_1} but obtained with  $\beta=10^4$ and a random $Q(t=0)$. }
    \label{fig:DNS_Ra_Tq_3e4_Pr_1_beta_1e4}
\end{figure}

\begin{figure}
(a) $\langle u\rangle_{h}(z,t)$ \hspace{0.4\textwidth} (b) $T(x,z,t)$ at $z=0.1$

    \centering
    \includegraphics[width=0.49\textwidth]{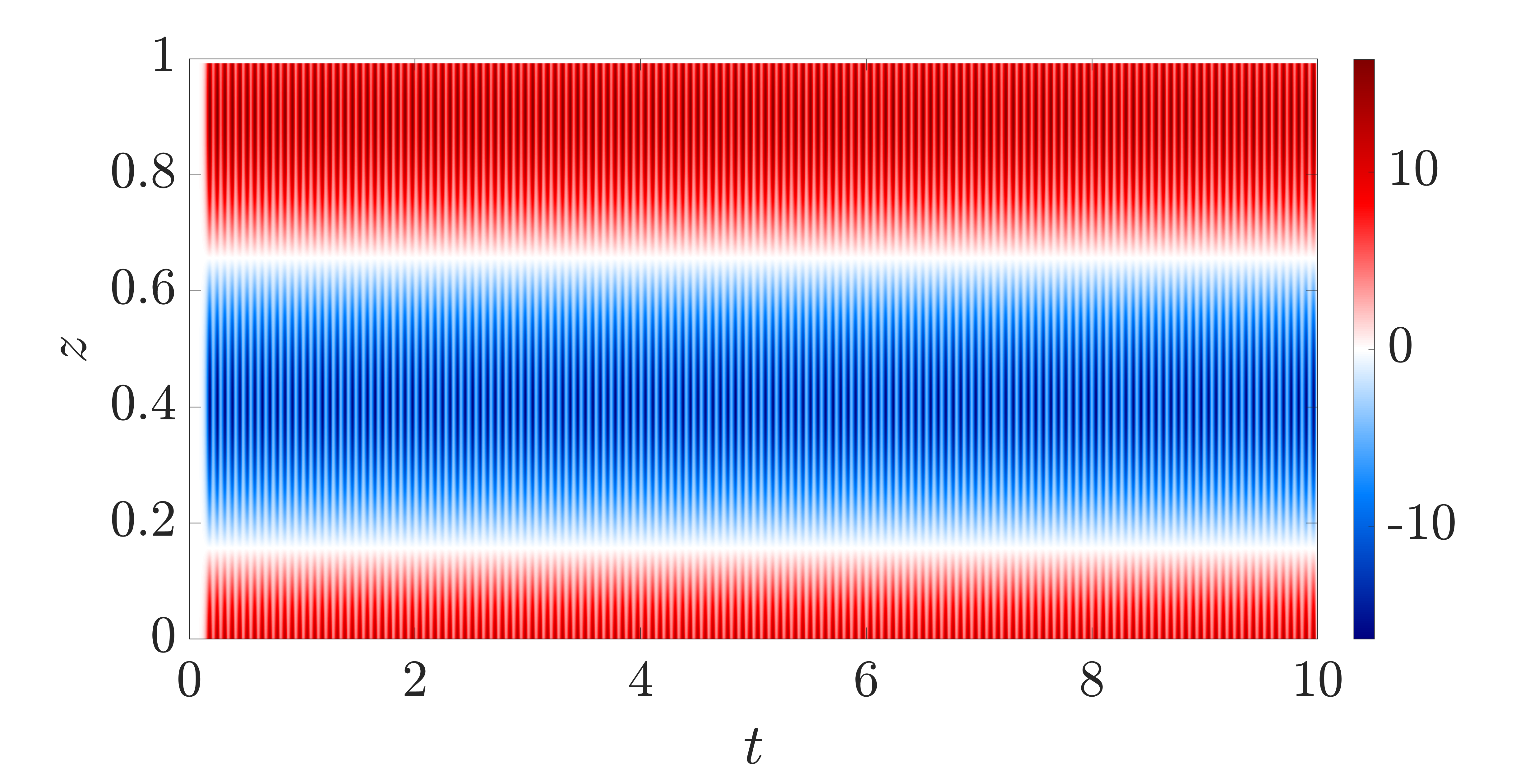}
    \includegraphics[width=0.49\textwidth]{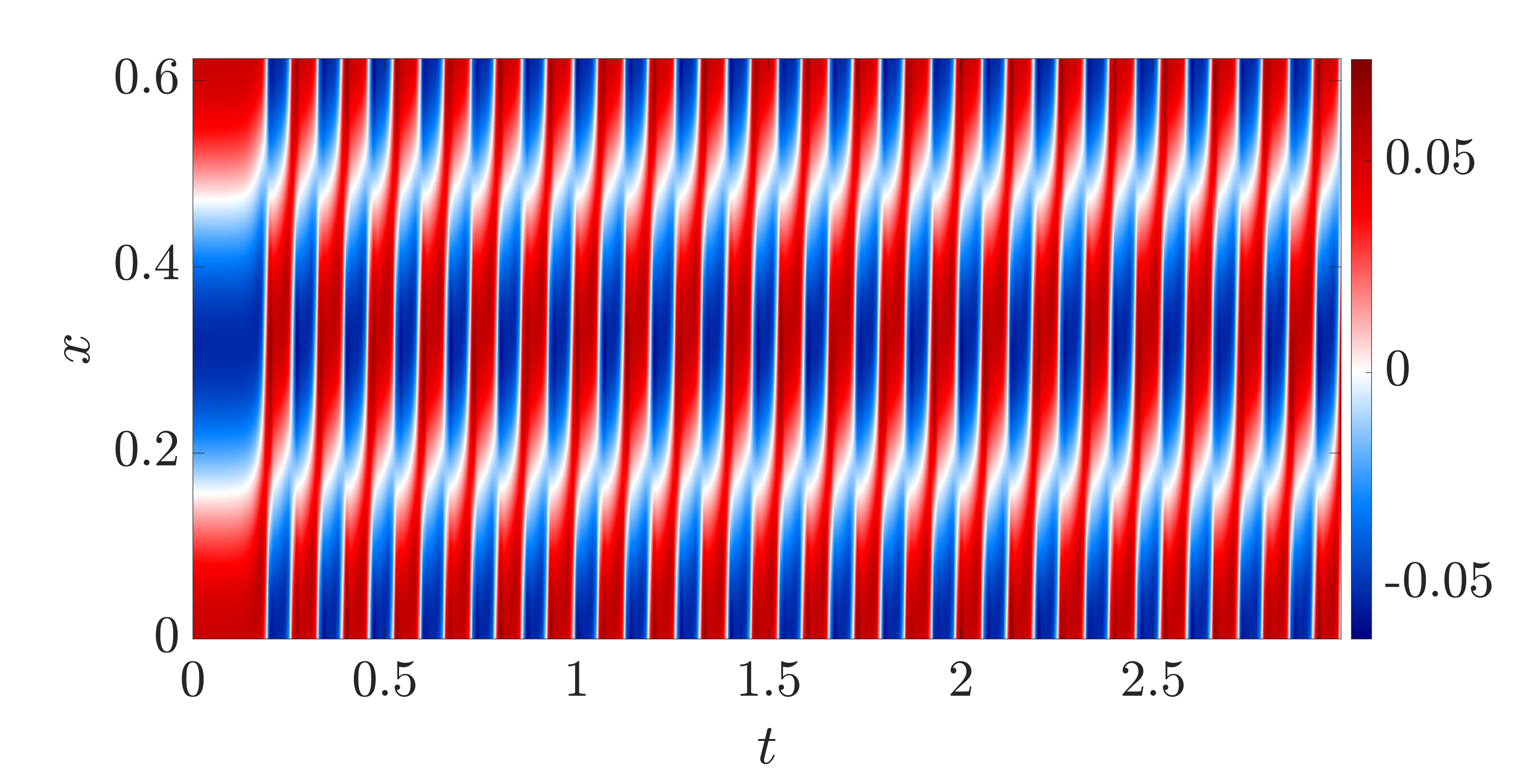}
    \caption{Same as figure \ref{fig:DNS_Ra_Tq_6e4_Pr_1} but obtained with $\beta=10^4$ and a random $Q(t=0)$. }
    \label{fig:DNS_Ra_Tq_6e4_Pr_1_beta_1e4}
\end{figure}

\begin{figure}
\hspace{0.05\textwidth} (a) $\langle u\rangle_{h}(z,t)$ \hspace{0.44\textwidth} (b)
\hspace{0.15\textwidth} (c)

    \centering
    \includegraphics[width=0.58\textwidth]{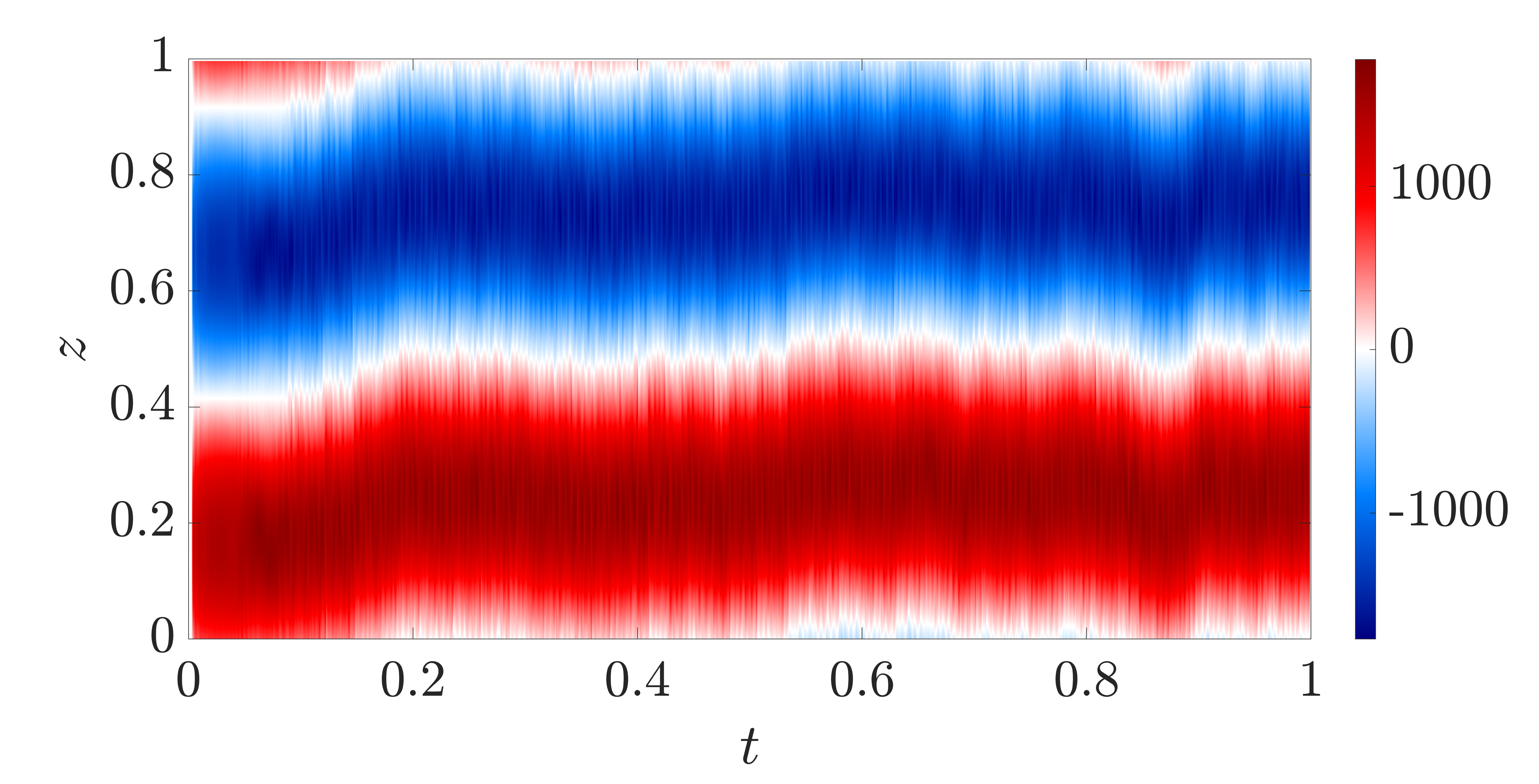}
    \includegraphics[width=0.33\textwidth,trim=-0 -0.25in 0 0]{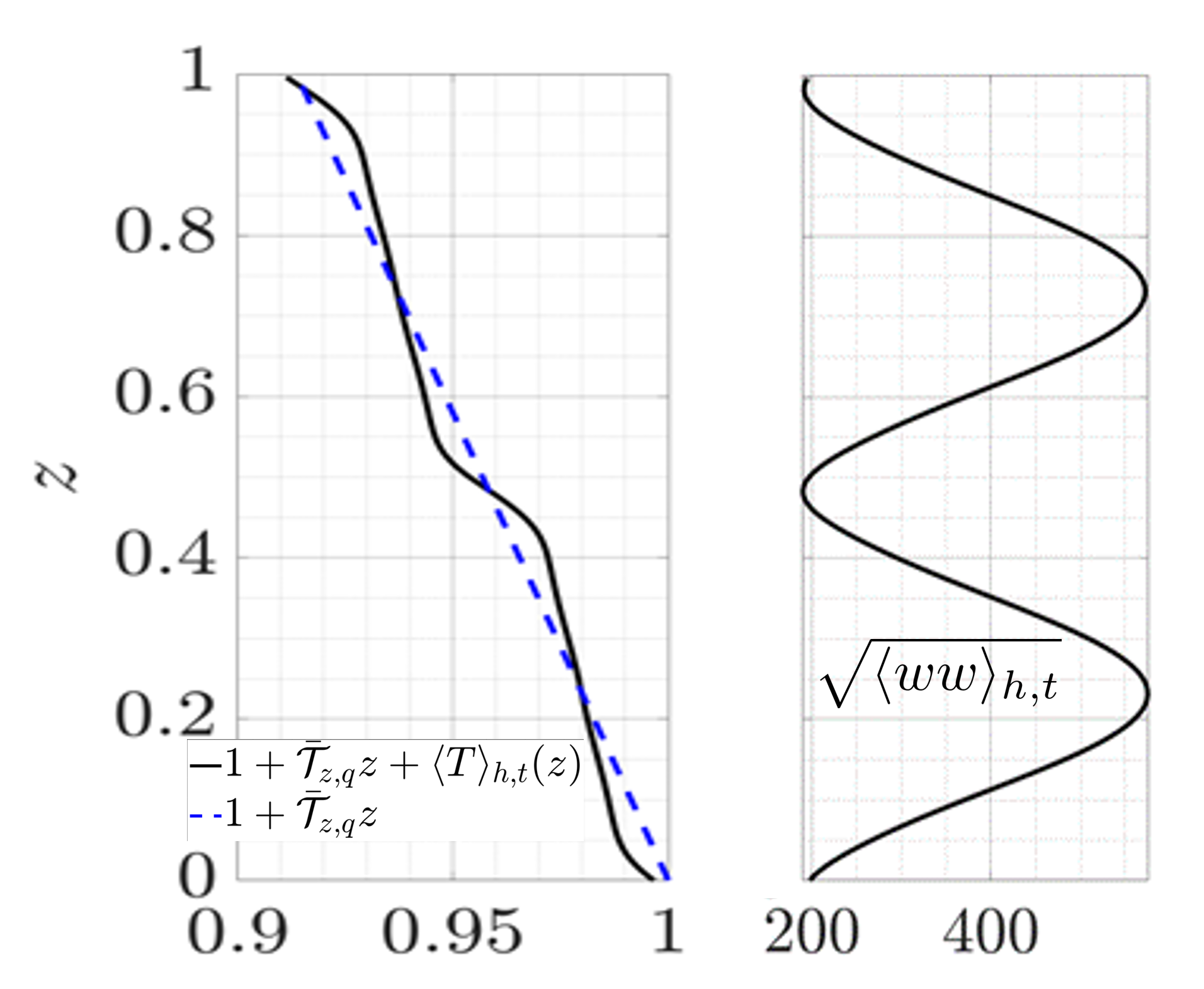}
    \caption{Same as figure \ref{fig:DNS_Ra_Tq_1e8_Pr_1} but obtained with $\beta=10^4$ and a random $Q(t=0)$. }
    \label{fig:DNS_Ra_Tq_1e8_Pr_1_beta_1e4}
\end{figure}

\begin{figure}
(a) \hspace{0.47\textwidth} (b)

    \centering
    \includegraphics[width=0.49\textwidth]{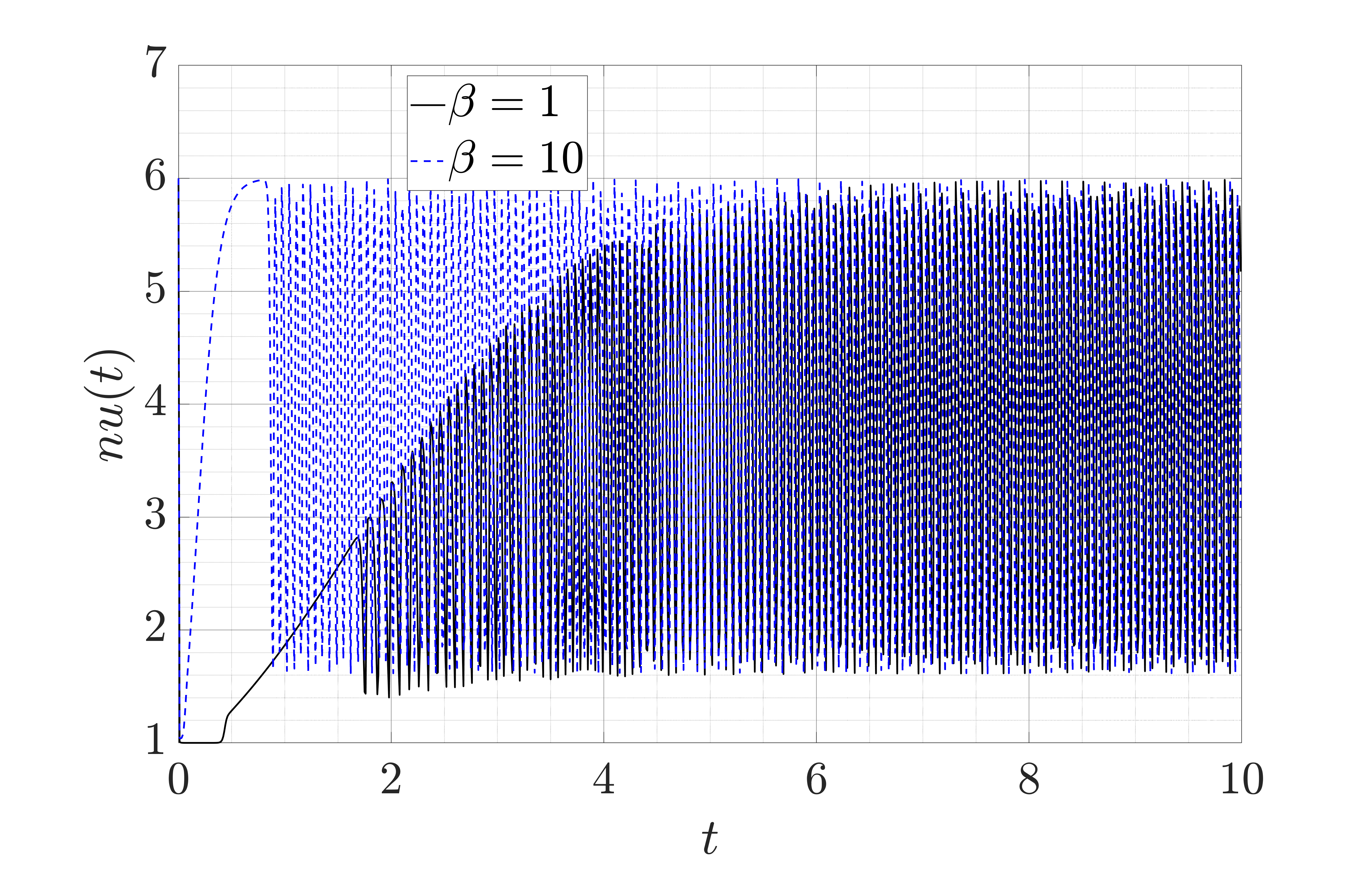}
    \includegraphics[width=0.49\textwidth]{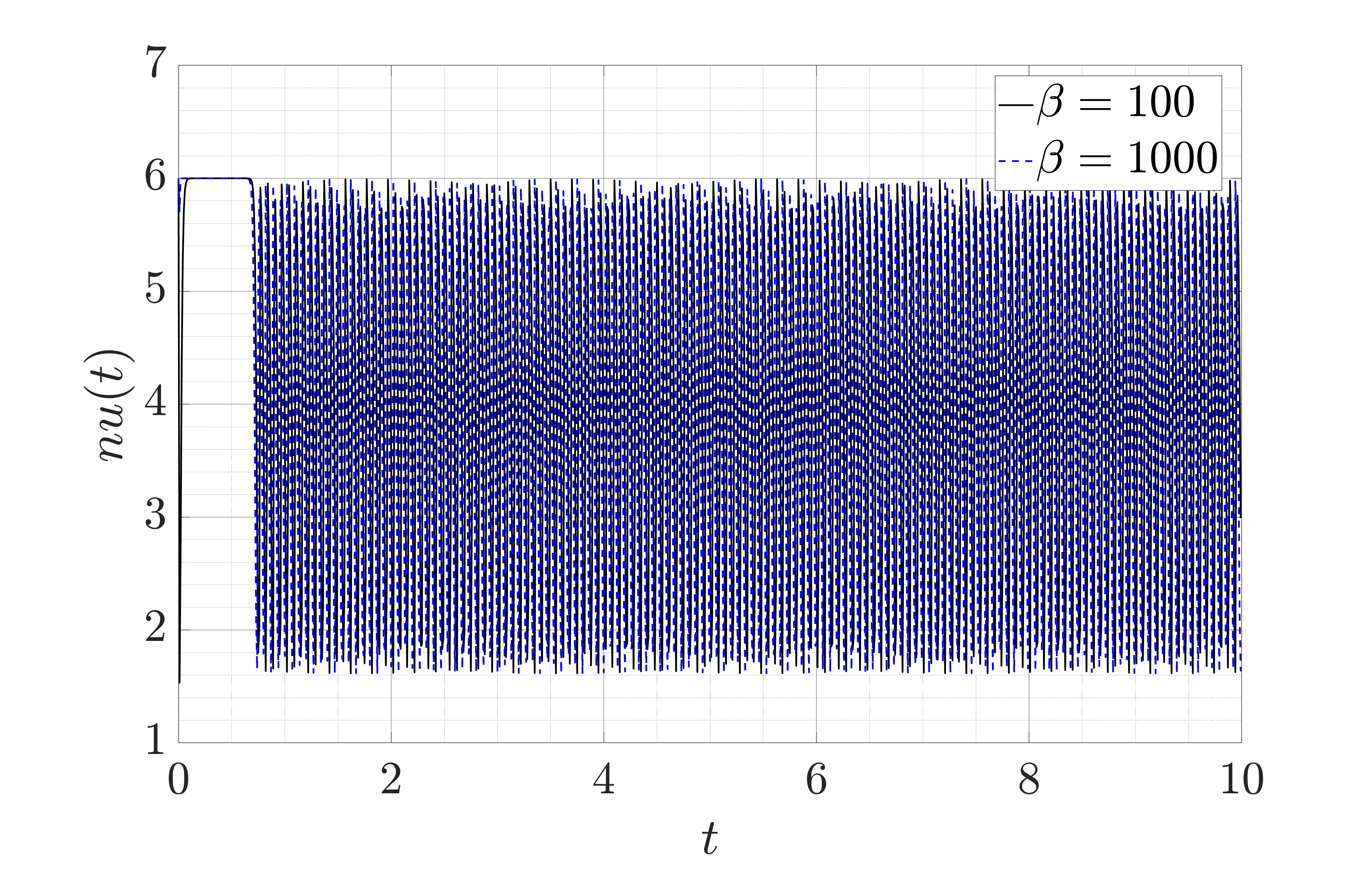}
    \caption{The instantaneous Nusselt number $nu(t)$ at $Ra_{T,q}=6\times 10^4$, $Pr=1$, $L_x=0.2\pi$ and $Q(t=0)=0$ with (a) $\beta=1$, 10 and (b) $\beta=100$, 1000. }
    \label{fig:DNS_Nu_Ra_Tq_6e4_Pr_1_beta_1_1e3}
\end{figure}

In this appendix, we present selected results obtained for a finite flux adjustment rate $\beta$ in \eqref{eq:Q} for comparison with the $\beta=\infty$ case studied in the previous sections. Figure \ref{fig:DNS_Ra_Tq_3e4_Pr_1_beta_1e4} displays a tilted elevator mode with $\beta=10^4$ and a random $Q(t=0)$ at $Ra_{T,q}=3\times 10^{4}$, revealing no substantial difference when compared with figure \ref{fig:DNS_Ra_Tq_3e4_Pr_1} for $\beta=\infty$. The only difference is in the length of the initial transient state (white region in panel (a)) and a shift in the vertical. Similarly, figure \ref{fig:DNS_Ra_Tq_6e4_Pr_1_beta_1e4} displays modulated traveling waves at $Ra_{T,q}=6\times 10^{4}$, $\beta=10^4$ that resemble, except for a vertical shift, those in figure \ref{fig:DNS_Ra_Tq_6e4_Pr_1} for $\beta=\infty$. This is so also for $Ra_{T,q}=10^8$ as figure \ref{fig:DNS_Ra_Tq_1e8_Pr_1_beta_1e4} also displays the same behavior as figure \ref{fig:DNS_Ra_Tq_1e8_Pr_1}.

We have also explored smaller values of $\beta$ when its effect will be more apparent. Figure \ref{fig:DNS_Nu_Ra_Tq_6e4_Pr_1_beta_1_1e3} shows $nu(t)$ for different $\beta$  associated with $Q(t=0)=0$ at $Ra_{T,q}=6\times 10^4$. For $\beta=1$, $nu(t)$ is reduced initially but then recovers to the level of the other $\beta$ cases at $t\approx 7$. Similar reduced Nusselt number also appears for $\beta=10$ but earlier. For $\beta=100$ and $1000$ in figure \ref{fig:DNS_Nu_Ra_Tq_6e4_Pr_1_beta_1_1e3}(b), $nu(t)$ quickly recovers to the value $nu(t)=6$ associated with elevator mode and then starts to oscillate. The late-time flow structures for these $\beta$ values display modulated traveling waves just as for $\beta=\infty$ (figure \ref{fig:DNS_Ra_Tq_6e4_Pr_1}) except for a vertical shift. 

We conclude that different values of the relaxation rate $\beta$ play no significant role in the late-time dynamics of the system.

\endgroup
\bibliography{main}
\bibliographystyle{jfm}

\end{document}